\documentclass[nonacm]{acmart}
\pdfoutput=1

\setcopyright{rightsretained}
\acmPrice{}
\acmDOI{10.1145/3622812}
\acmYear{2023}
\copyrightyear{2023}
\acmSubmissionID{oopslab23main-p183-p}
\acmJournal{PACMPL}
\acmVolume{7}
\acmNumber{OOPSLA2}
\acmArticle{237}
\acmMonth{10}
\received{2023-04-14}
\received[accepted]{2023-08-27}

\usepackage{subcaption} %% For complex figures with subfigures/subcaptions
\usepackage[nameinlink,noabbrev,capitalise]{cleveref}
\usepackage{listings}
\usepackage{graphicx}
\usepackage{enumitem}
\usepackage{mathtools}
\usepackage{tikz-cd}
\usepackage{tikz}
\usetikzlibrary{tikzmark}
\usepackage[normalem]{ulem}
\usepackage{mdframed}
\usepackage{mathpartir}
\usepackage{xspace}
\usepackage{wasysym}
\usepackage{threeparttable}
\usepackage{booktabs}
\usepackage{cprotect}
\usepackage{hyperref}

\renewcommand{\backrefalt}[4]{\textcolor{gray}{$\hookrightarrow$~{\ifcase #1 \or page\else pages\fi}~#2}} % for cute backrefs in bib

\usepackage[mathscr]{euscript}
\DeclareMathAlphabet\mathbfscr{U}{eus}{b}{n}

\definecolor{codegreen}{rgb}{0,0.6,0}
\definecolor{codegray}{rgb}{0.5,0.5,0.5}
\definecolor{codepurple}{rgb}{0.58,0,0.82}
\definecolor{backcolour}{rgb}{0.95,0.95,0.92}
\lstdefinestyle{mystyle}{
    backgroundcolor=\color{backcolour},
    commentstyle=\color{codegreen},
    keywordstyle=\color{magenta},
    numberstyle=\tiny\color{codegray},
    stringstyle=\color{codepurple},
    basicstyle=\ttfamily\footnotesize,
    breakatwhitespace=false,
    breaklines=true,
    captionpos=b,
    keepspaces=true,
    numbers=left,
    numbersep=5pt,
    showspaces=false,
    showstringspaces=false,
    showtabs=false,
    tabsize=2
}

\usepackage{syntax}

\usepackage[shorthand]{semantic} % for the math shorthands/ligatures such as |- for ⊢
\mathlig{=<}{\leq}
\mathlig{>=}{\geq}
\mathlig{|v|}{\Downarrow}
\mathlig{|}{\vee}
\mathlig{|_|}{\sqcup}
\mathlig{|-|}{\sqcap}
\mathlig{|=}{\vDash} % otherwise it defaults to \models, which is very ugly
\mathlig{[|}{\llbracket}
\mathlig{|]}{\rrbracket}
\mathlig{|->}{\mapsto}
\DeclareFontFamily{U}{skulls}{}
\DeclareFontShape{U}{skulls}{m}{n}{ <-> skull }{}
\newcommand{\skull}{\text{\usefont{U}{skulls}{m}{n}\symbol{'101}}}

\lstdefinestyle{snippet}{
  numbers=left,
  numberstyle=\tiny,
  frame=topbot,
  framexleftmargin=1.0em,
  framexrightmargin=0.1em,
  framextopmargin=0.6ex,
  framexbottommargin=0.6ex,
  backgroundcolor=\color{gray!6},
  numbersep=4pt, % Adjusted the space between line numbers and the content
}

\begin{document}

\makeatletter
\let\@authorsaddresses\@empty
\makeatother

\newcommand*{\Draft}{}%
\newcommand{\system}{$\textsf{HM}^{\ell}$\xspace}

\newcommand*{\TechReport}{}%

\ifdefined\TechReport
\newcommand{\report}[1]{#1}
\newcommand{\ifreport}[2]{#1}
\else
\newcommand{\report}[1]{}
\newcommand{\ifreport}[2]{#2}
\fi

\newcommand{\Cha}[1]{Chapter~{#1}}
\newcommand{\Sec}[1]{Section~\ref{#1}}
\newcommand{\Fig}[1]{Figure~\ref{#1}}
\newcommand{\Tbl}[1]{Table~\ref{#1}}
\newcommand{\Alg}[1]{Algorithm~\ref{#1}}
\newcommand{\App}[1]{Appendix~\ref{#1}}
\newcommand{\Def}[1]{Definition~\ref{#1}}

\newcommand{\Thm}[1]{Theorem~\ref{#1}}
\newcommand{\Lem}[1]{Lemma~\ref{#1}}

\newcommand{\aka}{\emph{a.k.a.\ }}
\newcommand{\ie}{\emph{i.e.,\ }}
\newcommand{\eg}{\emph{e.g.,\ }}
\newcommand{\cf}{\emph{cf.\ }}

\newcommand{\noi}{\noindent}
\newcommand{\inlitem}{$\bullet$~}

\newcommand{\qua}{\hspace{0.5em}}

\newcommand{\fname}[1]{\mathsf{#1}}
\newcommand{\dom}[1]{\fname{dom}(#1)}

\newcommand{\eqdef}{\,\stackrel{\mathclap{\normalfont\mbox{\scriptsize{def}}}}{~=\;~}\,}

\newcommand{\ruleName}[1]{\textsc{#1}}

\newcommand{\graydescr}[1]{\textcolor{darkgray}{ \textit{#1}}}

\newcommand{\cons}[1]{\fname{cons}(#1)}
\newcommand{\uni}[1]{\fname{uni}(#1)}
\newcommand{\type}[1]{\fname{type}(#1)}
\newcommand{\err}[1]{\fname{err}\,{#1}}
\newcommand{\neuter}[1]{\fname{neuter}_{#1}}

\newcommand{\kw}[1]{\textit{#1}}

\newcommand{\padd}[0]{\cdot}

\newcommand{\case}[5]{\fname{case}\ {#1}\ \fname{of}\ \{\, \iota_1({#2}) \Rightarrow {#3};\ \iota_2({#4}) \Rightarrow {#5}\, \}}

\newcommand{\betareducesto}{\triangleright_{\beta}}

\newcommand\eq{=}

\newcommand\al{\alpha}
\newcommand\be{\beta}
\newcommand\ga{\gamma}
\newcommand\de{\delta}

\newcommand\prov{p}
\newcommand\flow{Z}
\newcommand\noProv{\epsilon}
\newcommand{\loc}{\ensuremath{\ell}}
\newcommand\dead[1]{\skull(#1)}

\newcommand\pol{P}

\newcommand{\withP}[2]{\langle{#1,\, #2}\rangle}

\newcommand{\sub}{\ensuremath{\mathbin{<:}}}
\newcommand{\supp}{\ensuremath{\mathbin{:>}}}

\newcommand{\Pol}{\phi}
\newcommand{\notPol}{\neg\phi}
\newcommand{\Pos}{+}
\newcommand{\Neg}{-}

\newcommand{\rulesLabel}[1]{\text{\underline{\color{darkgray} #1}}\hfill\\}
\newcommand{\nextRules}[1]{\vspace{.5em}\\}
\newcommand{\trule}[1]{T-{#1}}
\newcommand{\crule}[1]{C-{#1}}
\newcommand{\urule}[1]{U-{#1}}
\newcommand{\srule}[1]{S-{#1}}

\newcommand{\truleName}[1]{\ruleName{\trule{#1}}}
\newcommand{\sruleName}[1]{\ruleName{\srule{#1}}}
\newcommand{\cruleName}[1]{\ruleName{\crule{#1}}}

\newcommand{\emptyCtx}{\epsilon}
\newcommand{\ctxPush}[1]{\cdot(#1)}

\newcommand{\rcd}[1]{\{\,#1\,\}}

\newcommand{\br}[1]{\{{#1}\}}
\newcommand{\setLit}[1]{\br{\,{#1}\,}}

\newcommand\LL[1]{{L\!\downarrow}\padd{#1}}
\newcommand\RL[1]{{R\!\downarrow}\padd{#1}}
\newcommand\LR[1]{{#1}\padd{L\!\uparrow}}
\newcommand\RR[1]{{#1}\padd{R\!\uparrow}}

\newcommand\Nprov[3]{\langle{#1}\rangle^{#2}_{#3}}
\newcommand\NL[2]{\langle{#1}\rangle^{#2}_{L}}
\newcommand\NR[2]{\langle{#1}\rangle^{#2}_{R}}
\newcommand\pFunL[1]{\NL{#1}{\rightarrow}}
\newcommand\pFunR[1]{\NR{#1}{\rightarrow}}
\newcommand\pSumL[1]{\NL{#1}{\tySum}}
\newcommand\pSumR[1]{\NR{#1}{\tySum}}
\newcommand\pProdL[1]{\NL{#1}{\tyProd}}
\newcommand\pProdR[1]{\NR{#1}{\tyProd}}
\newcommand\plb[1]{[#1]_{lb}}
\newcommand\pub[1]{[#1]_{ub}}
\newcommand\puni[1]{[#1]_{uni}}
\newcommand\punilst[1]{\overline{[#1]}_{uni}}
\newcommand\unify[3]{[#1 \sim #2,\, #3]}

\newcommand\seq{\,;\;}

\newcommand\ty{\tau}
\newcommand\tz{\delta}
\newcommand\pty{T}

\newcommand\tyOr{\sqcup}
\newcommand\tyAnd{\sqcap}
\newcommand\tyProd{\otimes}
\newcommand\tySum{\oplus}
\newcommand\bigtyOr{\bigsqcup}
\newcommand\bigtyAnd{\bigsqcap}
\newcommand\rec[2]{\mu{#1}.\,{#2}}
\newcommand\re[1]{\mu{#1}.\,}

\newcommand\tyCtor[1]{\mc{#1}}
\newcommand\tyPrim{\tyCtor{primitive}}
\newcommand\tyUnit{\mathbf{1}}
\newcommand\tyInt{\tyCtor{Int}}
\newcommand\tyNat{\tyCtor{Nat}}
\newcommand\tyBool{\tyCtor{Bool}}
\newcommand\tyStr{\tyCtor{Str}}

\newcommand\polyTy{\mathbb{T}}

\newcommand\true{\tyCtor{true}}
\newcommand\false{\tyCtor{false}}

\newcommand{\Int}{\mathsf{int}}
\newcommand{\Bool}{\mathsf{bool}}
\newcommand{\Unit}{\mathtt{unit}}
\newcommand{\True}{\mathtt{true}}
\newcommand{\False}{\mathtt{false}}
\newcommand{\ite}[3]{\ensuremath{\mathbf{if}\ #1\ \mathbf{then}\ #2\ \mathbf{else}\ #3}}

\newcommand{\mc}{\mathsf}
\newcommand{\tc}[1]{\textsf{#1}}

\newcommand{\st}{\mc{st}}
\newcommand{\vs}{\mc{vs}}
\newcommand{\mcl}{\mc{L}}
\newcommand{\mcr}{\mc{R}}

\newcommand{\lb}[2]{\mc{lb}_{#1}^\mc{#2}}
\newcommand{\ub}[2]{\mc{ub}_{#1}^\mc{#2}}

\newcommand{\I}{~|~}

\newcommand{\codeEmphasisColor}[0]{red!70!black}

\lstMakeShortInline[columns=fixed]@

\lstset{%
    % language=ML,
    language=Caml,
    % language=Haskell,
    basicstyle=\footnotesize,
    % literate={->}{{ $\rightarrow$ }}3,
    xleftmargin=3mm,
    moredelim=[is][\bfseries]{**}{**},
    moredelim=[is][\underbar]{__}{__},
    mathescape = true,
    literate=
      % {->}{{ $\rightarrow$ }}3{⟨}{{$\langle$}}1{⟩}{{$\rangle$}}1
      {╔}{}0
      {═}{==}1
      {─}{---}1
      {║}{}0
      {╠}{}0
      {╟}{}0
      {╙}{}0
      {•}{$\cdot$}1
      {▲}{$\blacktriangle$}1
      {▼}{$\blacktriangledown$}1
      {◉}{$\newmoon$}1
      {│}{|}1
    ,
    morekeywords={val},
    columns=fixed,
    basicstyle=\footnotesize\ttfamily,
    keywordstyle=\footnotesize\ttfamily\bfseries,
    commentstyle=\color{darkgray}\ttfamily,
    % morecomment=[f][\color{gray}][0]{--},
    morecomment=[f][\color{\codeEmphasisColor}][0]{--},
    % moredelim=**[is][\color{\codeEmphasisColor}\bfseries]{⌜}{⌝},
    % moredelim=**[is][\color{\codeEmphasisColor}]{^}{^},
    % morestring=*[b]~,
    % moredelim=**[is][\color{\codeEmphasisColor}]{⌜}{⌝},
    % moredelim=**[is][\color{\codeEmphasisColor}]{‹}{›},
    aboveskip=3pt,belowskip=3pt,lineskip=-0.2pt
}

\lstdefinelanguage{numsnippets}{%
    language=Caml,
    basicstyle=\footnotesize,
    xleftmargin=3mm,
    moredelim=[is][\bfseries]{**}{**},
    moredelim=[is][\underbar]{__}{__},
    mathescape = true,
    morekeywords={build},
    columns=fixed,
    basicstyle=\footnotesize\ttfamily,
    keywordstyle=\footnotesize\ttfamily\bfseries,
    commentstyle=\color{darkgray}\ttfamily,
    numbers=left,
    % morecomment=[f][\color{gray}][0]{--},
    morecomment=[f][\color{\codeEmphasisColor}][0]{--},
    % moredelim=**[is][\color{\codeEmphasisColor}\bfseries]{⌜}{⌝},
    % moredelim=**[is][\color{\codeEmphasisColor}]{^}{^},
    % morestring=*[b]~,
    % moredelim=**[is][\color{\codeEmphasisColor}]{⌜}{⌝},
    % moredelim=**[is][\color{\codeEmphasisColor}]{‹}{›},
    aboveskip=3pt,belowskip=3pt,lineskip=-0.2pt
}
\lstset{moredelim=[is][\bfseries]{***}{***}}
\lstdefinelanguage{error}{
  basicstyle=\footnotesize\ttfamily\color{\codeEmphasisColor},
  aboveskip=3mm,
  belowskip=3mm,
  lineskip=0pt,
  xleftmargin=3mm,
  moredelim=[is][\bfseries]{**}{**},
  % moredelim=[is][\underline]{__}{__},
  moredelim=[is][\underbar]{__}{__},
  moredelim=[is][{\color{black}\underbar}]{_B_}{__},
  moredelim=[is][{\bfseries\color{black}\underbar}]{*B_}{*_},
  moredelim=[is][\color{black}]{@}{@},
  mathescape = true,
  morekeywords={let,in,val},
  literate=
    {╔}{}0
    {═}{==}1
    {─}{---}1
    {║}{}0
    {╠}{}0
    {╟}{}0
    {╙}{}0
    {•}{$\cdot$}1
    {▲}{$\blacktriangle$}1
    {▼}{$\blacktriangledown$}1
    {◉}{$\newmoon$}1
    {│}{|}1
    % {•}{*}0
    {•}{$\cdot$}1
    {‹}{{\color{black}}}0
    % {›}{{\color{\codeEmphasisColor}}}0
    {›}{{\color{\codeEmphasisColor}} }1
    % {›}{{\color{gray}}}0
  ,
}

\mdfdefinestyle{codebox}{
  backgroundcolor=box-background,
  linecolor=box-border,
  linewidth=1pt,
  roundcorner=2pt,
  innerbottommargin=2pt,
  innertopmargin=2pt,
  innerleftmargin=2pt,
  skipabove=3pt
}

% \begin{Error}{lang}
% ... type error ...
% \end{Error}
\newenvironment{Error}[1]
{
  \medskip
  \begin{mdframed}[style=codebox]
  {\vspace{0.2em}\hspace{0.5em}\footnotesize\textsf{#1}\vspace{-0.8em}}
}
{
  \end{mdframed}
  \vspace{4pt}
}

\definecolor{box-background}{RGB}{248, 248, 248}
\definecolor{box-border}{RGB}{229, 229, 229}
\definecolor{linenumber-color}{RGB}{170, 170, 170}

%%%%%%%%%%%%%%%%%% Comments/TODO/FIXME/notes %%%%%%%%%%%%%%%%%%

%%%%%%% Invisible notes env (notes for authors only) `shadow' %%%%%%%
\specialcomment{shadow}{\begingroup\color{orange}}{\endgroup}

% To make it visible, comment the following line
\excludecomment{shadow}

\ifdefined \Draft
\newcommand{\TODO}[1]{\noindent\textcolor{orange}{{TODO: #1}}}
\newcommand{\TODOit}{\noindent\textcolor{orange}{{[TODO]}}}
\newcommand{\FIXME}[1]{\noindent\textcolor{red}{{FIXME: #1}}}
\newcommand{\FIXMEit}{\noindent\textcolor{red}{{[FIXME]}}}
\newcommand{\lionel}[1]{\noindent\textcolor{magenta}{{[Lionel:] #1}}}
\newcommand{\jonathan}[1]{\noindent\textcolor{ACMGreen}{{[Jonathan:] #1}}}
\newcommand{\ishan}[1]{\noindent\textcolor{cyan}{{[Ishan:] #1}}}
\newcommand{\david}[1]{\noindent\textcolor{green}{{[David:] #1}}}
\else
\newcommand{\TODO}[1]{}
\newcommand{\TODOit}{}
\newcommand{\FIXME}[1]{}
\newcommand{\FIXMEit}{}
\newcommand{\lionel}[1]{}
\newcommand{\ishan}[1]{}
\newcommand{\jonathan}[1]{}
\fi

\newcommand{\cut}[1]{}
\newcommand{\cutForLater}[1]{}
\newcommand{\cutForSpace}[1]{}
\newcommand{\cutObsolete}[1]{}
\newcommand{\cutForDolanVersion}[1]{}
\newcommand{\cutRedundant}[1]{}

\newcommand{\TODOlater}[1]{}

%%------------------------------------------------------------------------
%% DEFINITION HELPERS
%%------------------------------------------------------------------------

\newcommand{\alt}{~~|~~}

\newcommand{\inlinexp}[1]{
{\footnotesize
\[\begin{array}{l}
#1
\end{array}\]}}

\newcommand{\inlinexpa}[2]{
{\footnotesize
\[\begin{array}{#1}
#2
\end{array}\]}}

\newcommand{\infr} [3] [] {\infer[\textsc{#1}]{#3}{#2}}
\newcommand{\iand}        {\qquad}

\newcommand{\Ctxt}       {\mathcal{E}}
\newcommand{\InCtxt} [1] {\Ctxt[#1]}

%%------------------------------------------------------------------------
%% REDUCTION RELATION MACROS
%%------------------------------------------------------------------------

\newcommand{\subst} [3]    {#3 [#2 / #1]}
\newcommand{\dstep} [2]    {#1 ~\Downarrow~ #2}

\newcommand{\ssosredex}        {\rightarrow}
\newcommand{\ctxtreduce}       {\mapsto}
\newcommand{\sstep}     [3] [] {#2 &\ssosredex&  #3 &\textsc{#1}}
\newcommand{\ctxtstep}  [3] [] {#2 &\ctxtreduce& #3 &\textsc{#1}}

%%------------------------------------------------------------------------
%% TYPE DEFINITION MACROS
%%------------------------------------------------------------------------

\newcommand{\funct} [2] {#1\nobreak\rightarrow\nobreak#2}
\newcommand{\boolt}     {\mathtt{bool}}

\newcommand{\typeEnv}         {\Gamma}
\newcommand{\entails}         {\vdash}
\newcommand{\judgment}   [3] {#1 \entails #2 : #3}
\newcommand{\envent}      [2] {\judgment{\typeEnv}{#1}{#2}}
\newcommand{\extenvent}   [4] {\judgment{\typeEnv, #1 : #2}{#3}{#4}}
\newcommand{\envlookup}   [3] {\infr{#1(#2) = #3}{\judgment{#1}{#2}{#3}}}

%%------------------------------------------------------------------------
%% EXPRESSION MACROS
%%------------------------------------------------------------------------

%% lambda
\newcommand{\lamdefe}  [2] {\lambda #1.~#2}
\newcommand{\lamdefea} [2] {\begin{array}{l}\lambda#1.\\\hspace*{.5em}#2\\\end{array}}

%% let
\newcommand{\letdefe}    [3] {\letbind{#1}{#2}~\letin{#3}}
\newcommand{\letrecdefe}    [3] {\letrecbind{#1}{#2}~\letin{#3}}
\newcommand{\letdefarre} [3] {\begin{array}{l}\letbind{#1}{#2}\\\letin{#3})\\\end{array}}

\newcommand{\letbind}  [2] {\mathsf{let}~\lbind{#1}{#2}}
\newcommand{\letrecbind}  [2] {\mathsf{let\ rec}~\lbind{#1}{#2}}
\newcommand{\letbindp} [2] {\mathsf{let}~(\lbind{#1}{#2})}
\newcommand{\lbind}    [2] {#1=#2}
\newcommand{\letin}    [1] {\mathsf{in}~#1}

%% if
\newcommand{\ife}      [3] {\ifline{#1}~\thenline{#2}~\elseline{#3}}
\newcommand{\ifea}     [3] {\begin{array}{l}\ifline{#1}\\\thenline{#2}\\\elseline{#3}\end{array}}

\newcommand{\ifop}         {\mathsf{if}}
\newcommand{\ifline}   [1] {\ifop~ #1}
\newcommand{\thenline} [1] {\mathsf{then}~#1}
\newcommand{\elseline} [1] {\mathsf{else}~#1}

%% opers
\newcommand{\binopdef}     {\mathit{binop}}
\newcommand{\unopdef}      {\mathit{unop}}
\newcommand{\binope}   [2] {\binopdef~#1~#2}
\newcommand{\unope}    [1] {\unopdef~#1}

\newcommand{\andop}        {\mathsf{and}}
\newcommand{\orop}         {\mathsf{or}}
\newcommand{\notop}        {\mathsf{not}}
\newcommand{\ande}     [2] {\mathsf{and}~#1~#2}
\newcommand{\ore}      [2] {\mathsf{or}~#1~#2}
\newcommand{\note}     [1] {\mathsf{not}~#1}

%% values
\newcommand{\falsev}     {\mathsf{false}}
\newcommand{\truev} {\mathsf{true}}

%% helper functions for rules
\newcommand\addlb{\fname{add}\text{-}\fname{lb}}
\newcommand\addub{\fname{add}\text{-}\fname{ub}}
\newcommand{\freshen}[3]{\text{freshen}(#1)^{#2}_{#3}}
\newcommand{\appsubst}[2]{\text{ty-subst}(#1)_{#2}}
\newcommand{\extrudeH}[3]{\text{extr}(#1)^{#2}_{#3}}
\newcommand{\extrude}[2]{\text{extr}(#1)^H_{#2}}
\newcommand{\lvl}[1]{\text{lvl}(#1)^\sigma}
\newcommand{\lvlqi}[2]{\text{lvl}(#1)^{#2}}
\newcommand{\newlvl}[3]{\text{new-lvl}(#1 \mapsto #2, #3)}

%% Title information
\title{Getting into the Flow}
\subtitle{Towards Better Type Error Messages for Constraint-Based Type Inference}

%% Authors

\author{Ishan Bhanuka}
\orcid{0000-0002-7000-6534}
\affiliation{
  \institution{HKUST}
  % \city{Hong Kong}
  \country{Hong Kong, China}
}
% \email{xxx@xxx.de}

\author{Lionel Parreaux}
\orcid{0000-0002-8805-0728}
\affiliation{
  \institution{HKUST}
  % \city{Hong Kong}
  \country{Hong Kong, China}
}
% \email{xxx@xxx.de}

\author{David Binder}
\orcid{0000-0003-1272-0972}
\affiliation{
  \institution{University of Tübingen}
  \city{Tübingen}
  \country{Germany}
}
% \email{david.binder@uni-tuebingen.de}

\author{Jonathan Immanuel Brachthäuser}
\orcid{0000-0001-9128-0391}
\affiliation{
  \institution{University of Tübingen}
  \city{Tübingen}
  \country{Germany}
}
% \email{xxx@informatik.uni-tuebingen.de}

\def\shortauthors{Bhanuka, Parreaux, Binder, and Brachthäuser}

%% CCS
  

%% Keywords

\begin{abstract}
  Creating good type error messages for constraint-based type inference systems is difficult.
  % A good error message should reflect the programmer's understanding.
  % A good message should reflect the programmer's level of understanding of the program.
  Typical type error messages reflect implementation details of the underlying constraint-solving algorithms
  rather than the specific factors leading to type mismatches.
  We propose using subtyping constraints that capture data flow to classify and explain type errors.
  Our algorithm explains type errors as faulty data flows,
  which programmers are already used to reasoning about,
  and illustrates these data flows as sequences of relevant program locations.
  We show that our ideas and algorithm are not limited to languages with subtyping,
  as they can be readily integrated with Hindley-Milner type inference.
  In addition to these core contributions,
  we present the results of a user study to evaluate the quality of our messages compared to other implementations.
  While the quantitative evaluation does not show that flow-based messages improve
  the localization or understanding of the causes of type errors,
  the qualitative evaluation suggests a real need and demand for flow-based messages.
  \footnote{Ishan Bhanuka, Lionel Parreaux, David Binder, Jonathan Immanuel Brachthäuser. \href{https://dl.acm.org/doi/10.1145/3622812}{“Getting into the Flow -
Towards Better Type Error Message for Constraint-Based Type Inference”}. Technical Report. 2024. }
\end{abstract}

\maketitle

\section{Introduction}
\label{sec:intro}

\begin{comment}
=============================================
EXAMPLE SOURCE:
=============================================

val parse_version: string -> string
val show_major: string -> string

let appInfo = ("My Application", 1.5)

let process (name, vers) =
  name ^ show_major (parse_version vers)

let test = process appInfo
\end{comment}

Much academic research has gone into producing better type error messages for functional programming languages,
dating back at least to \citet{Wand86:src-type-errs}.
Yet, one would be none the wiser by looking at the error messages produced by existing compilers,
including those compilers designed specifically with learning in mind, such as Helium \citep{Heeren2003}.
For example, consider the following OCaml program\footnote{We use OCaml syntax for all code examples because of its prevalence in error localization literature.}\!,
where operator @(^)@ stands for string concatenation:
\vspace{0.4em}
\begin{lstlisting}[style=snippet, firstnumber=4]
let appInfo = ("My Application", 1.5)
let process (name, vers) = name ^ show_major (parse_version vers)
let main() = process appInfo
\end{lstlisting}
Here, we assume the following definitions imported from some library:
\vspace{0.4em}
\begin{lstlisting}[style=snippet, firstnumber=1]
val show_major    : string -> string
val parse_version : string -> string
\end{lstlisting}
This program contains a type error.
\cref{fig:zero-example} shows the error reported by the OCaml compiler (v. 4.14.0):

\begin{figure}
\begin{Error}{OCaml}
\begin{lstlisting}[language=error]
║  l.6: 	‹**let** main() = process _B_appInfo__›
║Error: This expression has type string * float
║       but an expression was expected of type string * string
║       Type float is not compatible with type string
\end{lstlisting}
\end{Error}
  \caption{Simple example of OCaml compiler error message.}
  \label{fig:zero-example}
\vspace{-14pt}
\end{figure}

\noindent
% Clearly, it is hard to tell, just by looking at this %very terse error 
It is not immediately clear, just from looking at this
report,
what caused the problem and how to go about fixing it,
unless one
% already has the rest of the source code in mind.
is already familiar with the source code and has it fresh in their mind---note that the definitions in our little program above could be very far apart in a real-world scenario.
The error seems to provide exactly as much information
as the type inference engine had on hand at the time it encountered a problem
% and only little information about the context is provided,
and little to no contextual information is provided,
which could have been helpful.
% On the other hand,
% Errors reported
% And the error
Error reports produced
by most other %popular
existing compilers for functional programming languages
are not significantly different than this.

% %
% Here is the report produced by Helium:

% \begin{Error}{Helium}
% \begin{lstlisting}[language=error]
% [TODO: Helium's message]
% \end{lstlisting}
% \end{Error}

\subsection{Flow-Based Error Messages}
How come the wealth of previous ideas for improving ML type errors
has not yet permeated modern compiler design practice?
This could be for a number of reasons.
Perhaps the previously-proposed approaches were too difficult to implement
or to integrate into existing type systems;
or they were too unreliable and their heuristics too difficult to tune;
or perhaps the corresponding explanations were not actually helpful to real programmers.

In this paper, we set out to %definitively
start addressing these questions by:
\begin{itemize}
  \item
      proposing a
      \emph{straightforward}, \emph{heuristics-free}
      approach to recording and reproducing the information relevant to ML type errors in terms of \emph{data flows},
      a concept that we expect users can get used to quickly,
      because it relates to how programs evaluate;
      % are likely to understand intuitively \jonathan{we have to downplay this. The results of the study suggest it has a learning curve.};
      % \\
      and by
  \item
      performing a randomized quasi-experimental study to evaluate whether our approach does help programmers understand the type errors found in actual ML programs. We compare the error messages we produce
      to those generated by {\tt ocamlc} and Helium \citep{Heeren2003}\footnote{
         Unfortunately, bit-rot seems to have gotten the better of the vast majority
         of historical type error reporting systems
         from academia, % and industry,
         so we are only able to compare with these two.
      }\!.
\end{itemize}
\smallskip

The approach we propose,
which we dub \system{} (to be read as ``\emph{H-M-loc}''),
produces the error message in \Cref{fig:first-example}
when given the same program as above.
% 
% val parse_version: string -> string
% val show_major: string -> string
% let appInfo = ("My Application", 1.5)
% let process (name, vers) =
%   name ^ show_major (parse_version vers)
% let test = process appInfo
% 
% \begin{Error}{\system{}}
% \begin{lstlisting}[language=error]
% [ERROR] Type `float` does not match `string`
% This float literal has type `float`
%   l.11:‹   **let** appInfo = ("My Application", _B_1.5__)›
% This variable has type `string`
%   l.15:‹   **let** process (name, _B_vers__) =›
% This reference has type `string`
%   l.16:‹     name ^ show_major (parse_version _B_vers__)›
%  `string` comes from this type expression
%   l.6:‹    **val** parse_version: _B_string__ -> string›
% \end{lstlisting}
% \end{Error}
% 
\begin{figure}%[t!]
\begin{Error}{\system{}}
\begin{lstlisting}[language=error]
[ERROR] Type `float` does not match `string`

◉ (float) comes from
│  - l.4  let appInfo = ("My Application", 1.5)
│                                          ^^^
│  - l.5  let process (name, vers) =
│                            ^^^^
│  - l.9  name ^ show_major (parse_version vers)
▼                                          ^^^^
◉ (string) comes from
   - l.2  val parse_version : string -> string
                              ^^^^^^
\end{lstlisting}
\end{Error}
  \caption{Simple example of \system{} error message.}
  \label{fig:first-example}
\vspace{-14pt}
\end{figure}
This report adds several helpful bits of context to the error.
% 1. it highlights the parts of the %original
% actual and expected types
% which are mismatched (here @Double@ and @Str@);\footnote{
%   This is not achieved by running a post-hoc \emph{diff} of the type representations
%   (as is done in some compilers, which regularly %sometimes resulting
%   results in wrong highlights),
%   but by properly tracking, in the type system,
%   which parts of the types are problematic.
% }
% and 2. it displays three other most relevant source locations pertaining
% to these problematic type parts:
% the location where the actual type part comes from,
% the location where it is eventually used,
% and finally the location showing the cause of the mismatch.
%
% \ishan{intro is bit on the long side. Maybe this can be shortened a bit.
% or the contributions brough forward and merged into this para.}
Most importantly, instead of displaying a single erroneous location,
it presents, in a logical order, each location involved in the erroneous \emph{data flow}.
This report illustrates the flow of data from the right-hand-side component of
the @appInfo@ pair into the @vers@ parameter of function @process@, and the flow of this @vers@
parameter into function @parse_version@, which is imported from some library.
% 
% It becomes clear from the report that our type error is caused by this erroneous data flow.
Our type error is caused by this specific erroneous data flow.

% \ishan{this section on gast is not necessary here since these points are covered in the contributions section 1.3}
% Note that this idea is not completely novel,
% as \citet{Gast2005} proposed a very similar idea already in 2005.
% What differentiates our work from previous work on the topic is:
% \emph{(1)}~we formally categorize the different \emph{kinds} of ML type error messages,
% which helps untangle the complexity of these errors;
% \emph{(2)}~we provide a simple type inference extension to an existing type inference algorithm
% to reconstruct the relevant information related to these errors;
% and
% \emph{(3)}~we describe a systematic approach to display error reports
% based on our new understanding of these error kinds.

% Jonathan: I commented out the last part. It disriminates us from Gast, but "from previous work"
%  can be read really general and I would focus on the technical differences, here.
% and
% \emph{(4)}~we perform a qualitative analysis of the results
% as well as, in the future,
%conducting a user study to experimentally validate the helpfulness of this approach.

\smallskip

% why the hell does latex break without this
% \newpage

% Moreover,
Note that if the abbreviated report above is still not
enough for the user to resolve the issue at hand, a further detailed explanation can be obtained.
\system{} supports a verbose mode that lists the entire sequence of locations involved in a type mismatch,
including those locations that go through nested type constructors.
The verbose message for the current example is illustrated in \cref{sec:appendix-verbose-example}.

Verbose error messages can become very long and unwieldy.
A promising future direction for our work will be to enable interactive type error debugging
in integrated development environments (IDEs),
whereby users will be able to explore the data flows involved in a type error interactively.

\subsection{Complex Type Errors}
The above program is simplistic and the erroneous data flow is easy to understand.
In more complex scenarios, errors can arise from two types flowing \emph{into} or \emph{from} the same location,
which we refer to as \emph{confluence} errors. Data flows are further complicated
when types flow \emph{through constructors}.
Later in the paper, we detail how we propose to handle these more advanced typing errors
in our error reporting system (\cref{sec:type-confluence-errors}).
As an early example, consider the following linear algebra program:
\vspace{0.4em}
\begin{lstlisting}[style=snippet]
let move (x, y) = (x / 2, y / 2)
let dist (x, y) = x *. x +. y *. y
let move_closer pos =
   if dist pos < 25.0 then pos else move pos
\end{lstlisting}
The OCaml compiler gives the following error for this program,
which again lacks context:
\begin{Error}{OCaml}
\begin{lstlisting}[language=error]
File line 3, characters 59-62:
3 |   ‹**let** move_closer pos = if dist pos < 25.0 then pos else move _B_pos__›
Error: This expression has type float * float
      but an expression was expected of type int * int
      Type float is not compatible with type int
\end{lstlisting}
\end{Error}
On the other hand, \cref{fig:second-example} shows \system{}'s verbose error report,
where the flows of all values and how they pass through constructors is precisely described:
\begin{figure}
\begin{Error}{\system{}}
\begin{lstlisting}[language=error]
[ERROR] Type `float` does not match `int`

◉ (float) comes from
▲  - lib. let ( *. ): float -> float -> float
│                     ^^^^^
│  - l.2  let dist (x, y) = x *. x +. y *. y
│                           ^
│ 
◉ (?a) is assumed for
   - l.2  let dist (x, y) = x *. x +. y *. y
                    ^
  ◉ (?a * _) comes from
  ▲  - l.2  let dist (x, y) = x *. x +. y *. y
  │                  ^^^^^^
  │  - l.4     if dist pos < 25.0 then pos else move pos
  │                    ^^^
  │  - l.3  let movecloser pos =
  │                        ^^^
  │ 
  ◉ (?c) is assumed for
  │  - l.3  let movecloser pos =
  │                        ^^^
  │  - l.4     if dist pos < 25.0 then pos else move pos
  │                                                  ^^^
  ▼ 
  ◉ (?b * _) comes from
     - l.1  let move (x, y) = (x / 2, y / 2)
                     ^^^^^^
◉ (?b) is assumed for
│  - l.1  let move (x, y) = (x / 2, y / 2)
│                   ^
│  - l.1  let move (x, y) = (x / 2, y / 2)
│                            ^
▼ 
◉ (int) comes from
   - lib. let ( / ): int -> int -> int
                     ^^^
\end{lstlisting}
\end{Error}
\caption{Complex example of \system{} error message.}
\label{fig:second-example}
\vspace{-14pt}
\end{figure}

This error report introduces new syntax that we explain below.

\begin{itemize}
  \item The indented parts are those corresponding to \emph{nested constructor flows}---here, the @float@ value does not immediately flow into an @int@ position,
  but rather flows into the left-hand side of a pair, which itself has its own
  flow. This nested flow is important to understand the entire context of the error and is shown
  in the verbose error reports.
  \item The labels @?a@, @?b@, and @?c@ are placeholders for types.
  @?b@ is the label for the argument of the @move_closer@ function and it flows into
  two use sites. It is used as the argument to the @dist@ function which expects a pair.
  The left type argument of the pair is labelled @?a@ which flows into a location expecting a @float@.
  Label @?b@ is also used as the argument to the @move@ function which also expects a pair with the left type argument
  labelled as @?c@. Type @?c@ flows into a location expecting an @int@. The data flow, presented like this,
  shows how two incompatible types are being unified causing a type error.
\end{itemize}

This program represents a more realistic example where composing different functions
together can lead to erroneous data flows.
%Such errors can happen in large code bases,
%with multiple contributors making changes to different parts of the code base.
%
% JB: the previous sentence is a bit difficult to understand and adds not enough to justify the complication IMHO.

\smallskip

\subsection{Contributions}
Our new approach is based on a theory of type provenance tracking.
% which is itself rooted in the new \emph{algebraic subtyping} technique \cite{dolanThesis,dolanMycroftSubtyping2017}
% for designing type systems with support for constraint-based type inference.
%
% We will show how to track type error provenance information in a sound and complete way,
% in the sense that we pinpoint \emph{all} the %tightest
% locations of a program
% which contribute to an error, \ie where something might have gone wrong (completeness),
% and \emph{not more} (soundness).
% Furthermore, we show how to leverage the resulting raw type provenance information
% in order to construct user-friendly error messages.
%
A key observation is that we have to treat type equality constraints $\tau_1 \eq \tau_2$ as \emph{asymmetric}, since such type constraints are read as information about a value flowing from a source where it has been introduced with type $\tau_1$ to a usage site where it will be used at type $\tau_2$.
This asymmetry suggest to look for inspiration from constraint-based inference algorithms for \emph{subtyping} constraints of the form $\tau_1 \sub \tau_2$, which are naturally asymmetric.
In this paper, we use the algebraic subtyping approach and algorithms developed by \citet{dolanMycroftSubtyping2017, dolanThesis}
and simplified by \citet{simpleEssence2020}, and combine it with the idea of \citet{Gast2005} to use data flow for explaining error messages.
While we use subtype inference to improve the quality of error messages,
our approach targets the familiar type theory of Hindley, Damas and Milner as the user-facing type system.

Specifically, we make the following contributions:

\begin{itemize}
  \item A \emph{classification system for unification errors} based on data flow, where each unification error is assigned a numeric level (\cref{sec:classification}).
     The classification allows us to speak about Level-$n$ errors, and to craft error messages specific for each level.
     We suggest that it is crucially important to use different textual explanations when explaining type errors of different levels.
  % \item \emph{The precise statement} of an observation in the algebraic subtyping community, namely that it is easier to create helpful error messages for a subtyping system than for a unification based system:
  %    A subtyping system only has to explain \emph{Level-0 errors}. \jonathan{I am not sure anymore, we want to mention this here. It is impossible to understand for the readers without reading the paper. Maybe make part of section 2?}
  %    \TODO{merge with point below}
  % \item \emph{a formal description of \system{}} which enriches terms with locations and types with provenances (\cref{subsec:formal:type-inference}).
  \item A \emph{subtyping constraint solving algorithm} which reports data flow-based error messages for \emph{Level-0 errors},
     which is very close to the one used in algebraic subtyping
     but additionally tracks the provenances of types and flows in the program (\cref{sec:formal}).
     This demonstrates an observation of the algebraic subtyping community that it is easier to create helpful error messages from a subtype-inference-based system rather than from a unification-based one.
  \item An \emph{equality-constraint solving algorithm} which reports data-flow-based error messages for \emph{Level-n errors} (where  $n \geq 1$).
     This algorithm is close to unification-based algorithms, but also tracks provenances of types and flows (\cref{sec:type-confluence-errors}).
  % \item \emph{an implementation} of \system{} as an extension of SimpleSub \cite{simpleEssence2020}, which typecheck OCaml programs and provides high quality flow errors. \jonathan{Should we have a short section on the implementation?}
  \item \emph{A user study} to empirically evaluate our error messages and to help us guide further research into improving their quality (\cref{sec:user-study}).
  The experiment compares the effects of \system{}, {\tt ocamlc}, and Helium on programmers' ability to understand and localize type errors. While the quantitative evaluation does not show that \system{} provides any measurable improvement over the state-of-the-art, a qualitative analysis suggests the demand for flow-based errors in situations with complex type errors.
  % \jonathan{I commented the following out, since reviewers complained about it.}
    % To the best of our knowledge, this is %one of the (or \emph{the})
    % the first user study on \emph{unification type error messages},\footnote{
    %   Much previous research %was dedicated to
    %   focused on finding objective ways of rating the %single
    %   locations reported by various error reporting tools, but no study directly assessed
    %   how much these errors and locations helped actual users understand and locate errors.
    % }
    % after decades of research in the domain.
\end{itemize}

\noindent
We provide an implementation of \system{} as an extension of Simple-sub \cite{simpleEssence2020}.
Our system type checks a reasonable subset of OCaml features while providing high-quality error messages\footnote{
  Our implementation is permanently available on Zenodo \cite{oopsla2023artefact}\!.
  A web demo is available on the repository which hosts the latest version of our implementation.
  The repository is hosted at \href{https://github.com/hkust-taco/hmloc}{github.com/hkust-taco/hmloc}.
}.

% \section{Old Introduction (for reference)}
% \input{introduction.tex}

\section{Classifying Type Errors}
\label{sec:classification}
Not all unification errors are created equal. By treating them as equal, compiler engineers pass up an opportunity to improve the quality of the error messages that we can generate.
Independent of the constraint algorithm we use, eventually the type checker might come to a point where it has discovered enough information to conclude that two incompatible types, such as $\tyBool$ and $\tyInt$, should be equal, at which point it emits an error.
Following this line of thought, we might conclude there is only one essential kind of unification error, namely that two types are incompatible.
Accordingly, most type checkers only use one textual template to display these errors to the user.
The error messages might be enriched with additional information about the typing context in which the error arose, or about the source code region for which it was generated, but the underlying textual template often stays the same.

In this section, we argue that this uniform view holds us back if we want to create great error messages for the user.
To improve error messages that arise from type unification, as a first step, we realize that not all unification errors are the same and introduce a precise \emph{classification of unification errors}.
Based on this classification, as a second step, it is then possible to craft a textual error message for each kind of unification error, instead of using one fits-all template. As we will see, we classify different constraint solving errors using the \emph{direction of data flow in the program}.

% Here we first need to explain the idea that types flow through the program, before we talk about direction.
\subsection{Flow of Types}
\label{subsec:classifying:flow-of-types}
Let us assume that we typecheck the faulty expression @not 1@. Traditionally, one would generate a type constraint
expressing that $\tyInt$ (the type of the literal @1@) has to be equal to $\tyBool$ (the argument type of the @not@ function).
However, we can observe that this information is directed and closely corresponds to the data flow:
\usetikzlibrary{calc}%
\begin{center}
% \vspace{-1.7em}
\begin{tikzpicture}
\footnotesize
\node (not) at (0, 0) {$\underset{\tyBool \,\rightarrow\, \tyBool}{\textbf{\texttt{not}\strut}}$};
\node (one) at (1, 0) {$\underset{\tyInt}{\texttt{1}\strut}$};
\draw[->] (one.south) to [bend left=60] ($(not.south) + (-1em, 0) $);
\end{tikzpicture}
\end{center}
% \vspace{-.5em}

\noindent In analogy to the well-known concept of data flow, we argue that programmers can reason about the flow of types to understand faulty programs.
In the above example, we say that the argument type $\tyInt$ \emph{flows into} the parameter type $\tyBool$.
It would be incorrect to generate a constraint for this expression which says that $\tyBool$ flows into $\tyInt$.
Most standard unification algorithms discard this directionality information, since they make implicit use of the rule of symmetry to solve constraints.
In these algorithms, the type equality constraint $\ty_1 = \ty_2$ is considered equivalent to the constraint $\ty_2 = \ty_1$.
As a first technical insight, we thus recognize that we have to use non-symmetric constraints if we want to preserve directionality information during constraint solving.
Luckily, there already is a ready-made notion for non-symmetric constraints: \emph{subtyping constraints} $\ty_1 \sub \ty_2$ which express that $\ty_1$ has to be a subtype of $\ty_2$.
We will see that we can equivalently interpret these subtyping constraints as expressing that a value of type $\ty_1$ flows into a context which expects a value of type $\ty_2$, and that this reading is independent of whether we consider a system with subtyping or without.

\subsection{Change of Direction}
\label{subsec:classifying:change-direction}
The flow of types provides us with a different explanation model for type errors.
In our example above, the flow was excessively short.
In realistic programs, the distance between the point where a type is introduced (like type $\tyInt$) and the point where it collides with a different expected type (type $\tyBool$ in our example) can be arbitrarily large.
Furthermore, type errors are not only introduced when one type flows directly into another incompatible one, but also if two incompatible types flow into a single location.
This is the case in the following example, where both $\tyInt$ and $\tyStr$ flow into the result type of the conditional expression:
\begin{center}
\vspace{-0.5em}
\begin{tikzpicture}
\footnotesize
\node (iff) {$\texttt{\textbf{if} true \textbf{then}}\strut$};
\node at ($(iff.east) + (0.5ex, 0)$) (fiv) {$\texttt{ 5 }\strut$};
\node at ($(fiv.east) + (1ex, 0)$) (els) {$\texttt{ \textbf{else} }\strut$};
\node at ($(els.east) + (1ex, 0)$) (hi) {$\texttt{"hi"}\strut$};
\draw[->] ($(fiv.south) + (0, 0.5ex) $) to [bend left=40] ($(iff.south west) + (1em, 0.5ex) $);
\draw[->] ($(hi.north) + (0, -0.3ex) $) to [bend left=-30] ($(iff.north west) + (1em, -0.3ex) $);
\end{tikzpicture}
\vspace{-0.8em}
\end{center}

\noindent We refer to these type errors as \emph{confluence} errors. When type checking a program like the one above, we would gather the two constraints $\tyStr \sub \alpha_{\textit{ret}}$ and $\tyInt \sub \alpha_{\textit{ret}}$, where $\alpha_{\textit{ret}}$ is a unification variable corresponding to the result of the conditional.
While there is an obvious type error in the program, just given the constraints we cannot immediately derive an inconsistency.
In order to do so, we would have to invoke the rule of symmetry.
As a second technical insight, we observe that invoking symmetry corresponds to a change of direction in the flow of types: $\tyStr \sub \alpha_{\textit{ret}} \supp \tyInt$
following the type flow from $\tyStr$ to $\tyInt$, we can notice that it reverses direction once.
Generalizing this observation, we present the following classification of type errors.

\begin{definition}
In a Level-$n$ unification error, the derivation of the contradiction has the form of a chain of subtyping constraints $\tau <:> \ldots <:> \tau'$ (with $\tau \neq \tau'$ and $<:>$ denoting either $<:$ or $:>$), where the direction of the subtyping constraints changes $n$ times.
Each change of direction corresponds to a reversal of the data flow which has to be explained to the user.
While the rule of symmetry allows HM type inference algorithms to ignore this information about data flow, retaining it is important to properly explain the cause of the type error.
\end{definition}

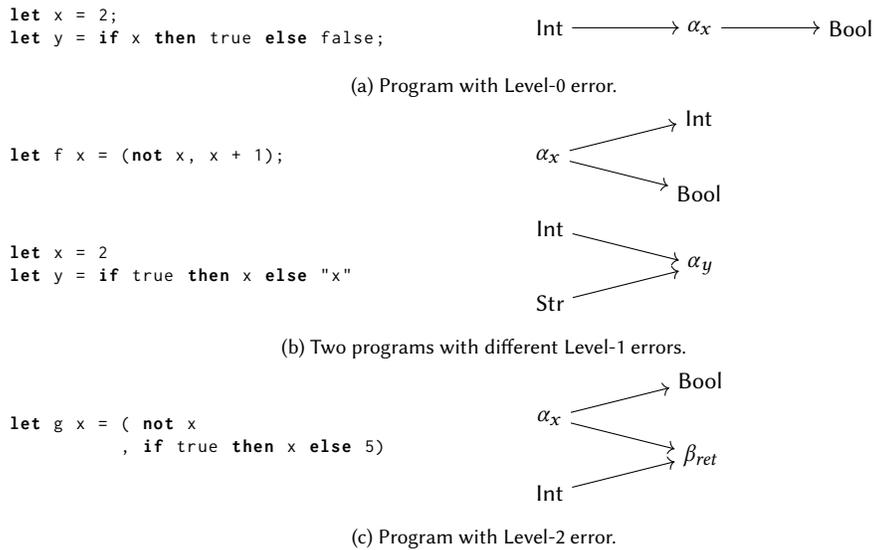
\begin{figure}[t!]
  %%
  %% Type 0 Errors
  %%
  \begin{subfigure}{\textwidth}
    \centering
    \begin{minipage}{0.4\textwidth}
\begin{lstlisting}
let x = 2;
let y = if x then true else false;
\end{lstlisting}
    \end{minipage}
    \hspace{1cm}
    \begin{minipage}{0.4\textwidth}
      \begin{tikzpicture}
        \node (A) at (0,0) {$\tyInt$};
        \node (B) at (2,0) {$\alpha_x$};
        \node (C) at (4,0) {$\tyBool$};
        \draw[->] (A) -- (B);
        \draw[->] (B) -- (C);
      \end{tikzpicture}
    \end{minipage}
    \caption{Program with Level-0 error.}
    \label{fig:classification-examples:level-zero}
  \end{subfigure}
  %%
  %% Type 1 Errors
  %%
  \begin{subfigure}{\textwidth}
    \centering
    \begin{minipage}{0.4\textwidth}
\begin{lstlisting}
let f x = (not x, x + 1);
\end{lstlisting}
    \end{minipage}
    \hspace{1cm}
    \begin{minipage}{0.4\textwidth}
      \begin{tikzpicture}
        \node (A) at (0,0) {$\alpha_x$};
        \node (B) at (2,0.5) {$\tyInt$};
        \node (C) at (2,-0.5) {$\tyBool$};
        \draw[->] (A) -- (B);
        \draw[->] (A) -- (C);
      \end{tikzpicture}
    \end{minipage}

    \begin{minipage}{0.4\textwidth}
\begin{lstlisting}
let x = 2
let y = if true then x else "x"
\end{lstlisting}
    \end{minipage}
    \hspace{1cm}
    \begin{minipage}{0.4\textwidth}
      \begin{tikzpicture}
        \node (A) at (0,0.5) {$\tyInt$};
        \node (B) at (0,-0.5) {$\tyStr$};
        \node (C) at (2,0) {$\alpha_{y}$};
        \draw[->] (A) -- (C);
        \draw[->] (B) -- (C);
      \end{tikzpicture}
    \end{minipage}
    \caption{Two programs with different Level-1 errors.}
    \label{fig:classification-examples:level-one}
  \end{subfigure}
  %%
  %% Type 2 Errors
  %%
  \begin{subfigure}{\textwidth}
    \centering
    \begin{minipage}{0.4\textwidth}
\begin{lstlisting}
let g x = ( not x
          , if true then x else 5)
\end{lstlisting}
    \end{minipage}
    \hspace{1cm}
    \begin{minipage}{0.4\textwidth}
      \begin{tikzpicture}
        \node (A) at (0,-0.5) {$\alpha_x$};
        \node (B) at (0,-1.5) {$\tyInt$};
        \node (C) at (2,0) {$\tyBool$};
        \node (D) at (2,-1) {$\beta_{\textit{ret}}$};
        \draw[->] (A) -- (C);
        \draw[->] (A) -- (D);
        \draw[->] (B) -- (D);
      \end{tikzpicture}
    \end{minipage}
    \caption{Program with Level-2 error.}
    \label{fig:classification-examples:level-two}
  \end{subfigure}
  \caption{Examples of faulty programs and their corresponding constraint graphs.}
  \vspace{-12pt}
    \label{fig:classification-examples}
\end{figure}

\noindent To see why this classification is useful, in the remainder of this section, we will consider concrete examples containing errors of various levels in \cref{fig:classification-examples} and the corresponding error messages that we generate.

\subsubsection{Level-0 Errors}
The snippet in \cref{fig:classification-examples:level-zero} contains a Level-0 error.
In this example, we have to introduce one unification variable $\alpha_x$ for the let-bound program variable $x$, and two constraints $\tyInt \sub \alpha_x$ and $\alpha_x \sub \tyBool$.
The first constraint expresses that the type $\tyInt$, introduced by the literal $2$, flows into the variable $x$, while the second constraint expresses that the type of the variable $x$ flows into the condition of the if-then-else expression which expects a boolean.
These constraints are presented in the corresponding graph as arrows, with the direction of the arrow corresponding to the flow of data through the program.
From these two constraints we can deduce the inconsistency $\tyInt \sub \tyBool$ \emph{without} having to reverse the data flow in the constraints.
This means that we can directly explain the error as the flow from one type to the other as shown in \cref{fig:level-0-example}.
\begin{figure}
\begin{Error}{\system{}}
\begin{lstlisting}[language=error]
[ERROR] Type `int` does not match `bool`

◉ (int) comes from
│  - l.1  let x = 2;
▼                 ^
◉ (bool) comes from
   - l.2  let y x = if x then true else false
                       ^
\end{lstlisting}
% \begin{lstlisting}[language=error]
% [ERROR] Type `int` does not match `bool`
% This integer literal has type `int`
%   l.2:‹ 	**let** x = _B_2__;›
% This if-then-else condition has type `int`
%   l.3:‹ 	**let** y = if _B_x__ then true else false;›
% \end{lstlisting}
\end{Error}
\caption{Level-0 error.}
\label{fig:level-0-example}
\vspace{-10pt}
\end{figure}
Level-0 errors allow for a good textual explanation in error messages, since we can point to a location in the program where a value of a certain type was introduced, follow it through intermediate bindings and point to a position in the code where a different type was expected.
Programs containing such Level-0 errors should always be rejected by a typechecker, since executing such programs would result in type mismatch errors at runtime (\ie non-value terms getting stuck and not reducing further).
Such programs are therefore rejected by both systems which support subtyping and systems which do not.

\subsubsection{Level-1 Errors}
In \cref{fig:classification-examples:level-one} we have two snippets which both exhibit Level-1 errors.
In the first of these snippets we have an if-then-else expression with incompatible cases for the if and else branch.
During constraint generation we would generate a unification variable $\alpha_{\textit{ret}}$ for the return type of the if-then-else expression, and two constraints: The constraint $\tyInt \sub \alpha_{\textit{ret}}$ for the if branch and the constraint $\tyStr \sub \alpha_{\textit{ret}}$ for the else branch.
But taken together, these constraints are only contradictory if we reverse the data flow once in the chain $\tyInt \sub \alpha_{\textit{ret}} \supp \tyStr$.
\cref{fig:level-1-example-converge} shows the error message.
% \pagebreak
\begin{figure}
\begin{Error}{\system{}}
\vspace{0.2cm}
\begin{lstlisting}[language=error]
[ERROR] Type `int` does not match `string`

        (int) ---> (?a) <--- (string)

◉ (int) comes from
│  - l.1  let x = 2
│                 ^
│  - l.2  let y = if true then x else "x"
▼                              ^
◉ (?a) is assumed here
▲  - l.2  let y = if true then x else "x"
│                 ^^^^^^^^^^^^^^^^^^^^^^^
◉ (string) comes from
   - l.2  let y = if true then x else "x"
                                      ^^^
\end{lstlisting}
% \begin{lstlisting}[language=error]
% [ERROR] Type `int` does not match `string`

%         int ---> ?a <--- string

% This `then` branch has type `int` and it flows into `?a`
%   l.2:‹ 	if true then _B_5__ else "hi"›
% This if-then-else expression has type `?a`
%   l.2:‹ 	_B_if true then 5 else "hi"__›
% This `else` branch has type `string` and it flows into `?a`
%   l.2:‹ 	if true then 5 else _B_"hi"__›
% \end{lstlisting}
\end{Error}
\caption{Level-1 ``confluence'' error with convergent flows}
\label{fig:level-1-example-converge}
\vspace{-10pt}
\end{figure}

In a system with subtyping and union and intersection types it would be possible to assign the type $\tyInt \sqcup \tyStr$ to the expression.
This shows that it is not strictly necessary to reject this expression, since the evaluation of this expression cannot lead to type unsoundness in itself, as long as its context can handle both an integer and a string.

The second example in \cref{fig:classification-examples:level-one} exhibits a different Level-1 error.
Here we have to generate a unification variable $\alpha_{x}$ for the lambda-bound variable $x$, and two constraints for the two different uses of $x$ in either side of the tuple.
The error is presented in \cref{fig:level-1-example}.
\begin{figure}
\begin{Error}{\system{}}
\vspace{0.2cm}
\begin{lstlisting}[language=error]
[ERROR] Type `int` does not match `bool`

        (int) <--- (?a) ---> (bool)

◉ (int) comes from
▲  - lib. let (+): int -> int -> int
│                  ^^^
│  - l.1  let f x = (not x , x + 1)
│                            ^
◉ (?a) is assumed here
│  - l.1  let f x = (not x , x + 1)
│               ^
│  - l.1  let f x = (not x , x + 1)
▼                        ^
◉ (bool) comes from
   - lib. let not: bool -> bool
                   ^^^^
\end{lstlisting}
% \begin{lstlisting}[language=error]
% [ERROR] Type `bool` does not match `int`

%         bool <--- ?a ---> int

%  `bool` comes from this type expression
%   builtin:56:‹ 	**let** not: _B_bool__ -> bool›
% This reference has type `bool` and it flows from `?a`
%   l.2:‹ 	**let** f x = (not _B_x__, x + 1);›
% This variable has type `?a`
%   l.2:‹ 	**let** f _B_x__ = (not x, x + 1);›
% This reference has type `?a` and it flows into `int`
%   l.2:‹ 	**let** f x = (not x, _B_x__ + 1);›
% This type reference has type `int`
%   builtin:16:‹ 	**let** (+): _B_int__ -> int -> int›
% \end{lstlisting}
\end{Error}
\caption{Level-1 error with divergent flows.}
\label{fig:level-1-example}
\vspace{-15pt}
\end{figure}
A system with support for subtyping could assign the type $(\tyInt \sqcap \tyBool) \to (\tyBool, \tyInt)$ to the expression, \ie a function which can only be called with a value which can act as both an integer and a boolean.

\subsubsection{Level-2 Errors}
If we combine the two snippets from \cref{fig:classification-examples:level-one} we obtain the example from \cref{fig:classification-examples:level-two} which exhibits a Level-2 error.
Here two unification variables, $\alpha_{x}$ and $\beta_{\textit{ret}}$ have to be generated, together with three constraints, and we have to change the direction of data flow twice to obtain the inconsistency between $\tyBool$ and $\tyInt$.
We conjecture that these kind of errors will already be quite rare in practice, even more so for errors of even higher levels, and even for a human it is no longer clear how the best error message should look like in this case.
But the algorithm that we present is still able to provide an explanation, mentions all the essential information, and is much more informative than what other implementations provide.
We show the error message for this example in \cref{fig:level-2-example}.
\begin{figure}
\begin{Error}{\system{}}
\vspace{0.2cm}
\begin{lstlisting}[language=error]
[ERROR] Type `bool` does not match `int`

        (bool) <--- (?a) ---> (?b) <--- (int)

◉ (bool) comes from
▲  - lib. let not: bool -> bool
│                  ^^^^
│  - l.1  let g x = ( not x
│                         ^
◉ (?a) is assumed here
│  - l.1  let g x = ( not x
│               ^
│  - l.2            , if true then x else 5)
▼                                  ^
◉ (?b) is assumed here
▲  - l.2            , if true then x else 5)
│                     ^^^^^^^^^^^^^^^^^^^^^
◉ (int) comes from
   - l.2            , if true then x else 5)
                                          ^
\end{lstlisting}
% \begin{lstlisting}[language=error]
%   TODO Add error message for Snippet 1c here. (Cannot be produced currently)
% \end{lstlisting}
\end{Error}
\caption{Level-2 error.}
\label{fig:level-2-example}
\vspace{-20pt}
\end{figure}

\subsection{Reporting Different Levels}

How do different type inference algorithms deal with these different errors?

Algebraic subtyping algorithms like MLsub usually only report Level-0 errors,
and not any of the higher-level-errors ($n > 0$), as these are not considered errors in this typing discipline.
This is why it has been remarked, in the subtyping literature,
that it should be easier to generate good error messages for a system based on subtyping---the error messages only have to explain a linear and obviously problematic data flow of information through the program.

Standard unification algorithms, on the other hand, have to account for Level-$n$ errors for arbitrary $n$, since symmetry is always valid for equality constraints.
However, since these algorithms usually do not track the reversal of the direction of data flow in the constraint solving process, the same textual explanation is used for all unification errors, regardless of their level.

Our classification now allows us to design detailed and specific error messages for both systems with and without subtyping.
If we have a system with subtyping, we only recognize Level-0 errors as proper errors, and display and explain them accordingly.
We describe how to do this in \cref{sec:formal}.
If, on the other hand, we are interested in a system which recognizes
the same errors as a standard unification algorithm, then we have to recognize and explain errors for all levels.
In \cref{sec:type-confluence-errors}, we extend the algorithm from \cref{sec:formal} to emulate a standard unification algorithm, and recognize errors for all the levels.
However unlike standard unification algorithm, we keep track of when the direction of data flow is reversed and report the full data flow for an error.

\section{Formalization}
\label{sec:formal}
In this section, we make the intuitions described in the previous sections formally precise.
There are two important properties of type inference in ML-style languages we omit in this section: let-generalization and the occurs-check.
We expect that both features will integrate well with the approach we have described in this paper, but leave the details for future work.

%%%%%%%%%%%%%%%%%%%%%%%%%%%%%%%%%%%%%%%%%%%%%%%%%%%%%%%%%%%%%%%%%%%%%%%%%%%%%%%
%% Subsection: Terms and Locations
%%%%%%%%%%%%%%%%%%%%%%%%%%%%%%%%%%%%%%%%%%%%%%%%%%%%%%%%%%%%%%%%%%%%%%%%%%%%%%%
\subsection{Terms and Locations}
\label{subsec:formal:terms-and-locs}

\Fig{fig:type-syntax} defines the syntax of terms $e$. The presentation is fairly standard,
but since we want to track the flow of information through the program, we need a way to refer to the \emph{locations} of subexpressions within the program.
For this reason, every subexpression and every binding site of a variable is annotated with a program location $\loc$.
In this article, we do not commit to any actual representation for program locations, but in our implementation we choose one based on line and column number ranges.
We also omit these locations in examples and explanations, or whenever they are not necessary.

\paragraph{Syntax of Terms}
Terms themselves consist of variables $x$, the unit constructor $\Unit$, integer literals $\overline{n}$ and integer addition $e + e$.
Booleans are constructed with literals $\True$ and $\False$, and eliminated using the conditional $\ite{e}{e}{e}$.
Functions are introduced using lambda abstraction $\lambda x.e$ and eliminated using function application $e\ e$.
Pairs are constructed using the pairing constructor $[e,e]$ and deconstructed using projections $\pi_1(e)$ and $\pi_2(e)$.
Sums are constructed using injections $\iota_1(e)$ and $\iota_2(e)$, and deconstructed using the pattern matching construct $\case{e}{x}{e}{x}{e}$.

%%%%%%%%%%%%%%%%%%%%%%%%%%%%%%%%%%%%%%%%%%%%%%%%%%%%%%%%%%%%%%%%%%%%%%%%%%%%%%%
%% Subsection: Types and Provenances
%%%%%%%%%%%%%%%%%%%%%%%%%%%%%%%%%%%%%%%%%%%%%%%%%%%%%%%%%%%%%%%%%%%%%%%%%%%%%%%
\subsection{Types and Provenances}
\label{subsec:formal:types-and-provs}

Where terms are annotated with locations, types are annotated with \emph{provenances} (defined in \Fig{fig:type-syntax}).
These provenances $p$ explain \emph{why} a certain type is used at a specific point in the program, and they are recorded and recombined during the type inference process.
Provenances are also used to report errors; in that case they explain the flow of information through the program that led to the mismatch of two types.
A provenance records a linear path through the program, so we have an operation $\cdot$ to concatenate two paths, and its unit $\noProv$ standing for the empty path.
Provenance concatenation is taken to be an associative operation where the \emph{absent provenance} $\noProv$ is taken to be the empty element.
Therefore, for example, $(\prov_0 \padd \prov_1) \padd \prov_2$ is the same as $\prov_0 \padd (\prov_1 \padd \prov_2)$ and is simply written $\prov_0 \padd \prov_1 \padd \prov_2$.
Similarly, $\noProv \padd \prov_0 \padd \noProv \padd \prov_1 \padd \noProv$, is the same as $\prov_0 \padd \prov_1$.
We also use locations $\loc$ in provenances, to record specific points in the flow of information through the program.
We will introduce and motivate the remaining syntactic forms of provenances in \cref{subsubsec:formal:subconstraints}, where they are used in the constraint solving process.

\paragraph{Syntax of Types}
The type forms are standard.
We have type variables $\alpha$, the unit type $\tyUnit$, and primitive types $\tyInt$ and $\tyBool$.
We have three binary type constructors referred to as $\odot$: the function type $\to$, the product type $\tyProd$ and the sum type $\tySum$.
As mentioned above, these are all annotated with provenances $p$.
Just as with terms, we will sometimes omit these provenances in examples and explanations.
In order to show how the different parts of an annotated type correspond to different parts of the information flow, we consider a very simple example.

\begin{example}
  \label{ex:simple-inferred-type}
  The inferred type of the term $[5^{\loc_1},\Unit^{\loc_2}]^{\loc_3}$ is $\tyInt^{\loc_1} \tyProd^{\loc_3} \tyUnit^{\loc_2}$.
\end{example}
The above example shows that in the inferred type, the top-level provenance $\loc_3$ only contains the information of the flow explaining the outermost type constructor $\_ \tyProd \_$.
It does not contain the information about the provenance of its arguments.
These provenances are annotated in the arguments to $\_ \tyProd \_$, namely $\tyInt^{\loc_1}$ and $\tyUnit^{\loc_2}$.

%%%%%%%%%%%%%%%%%%%%%%%%%%%%%%%%%%%%%%%%%%%%%%%%%%%%%%%%%%%%%%%%%%%%%%%%%%%%%%%
%% Figure: Syntax of terms and types.
%%%%%%%%%%%%%%%%%%%%%%%%%%%%%%%%%%%%%%%%%%%%%%%%%%%%%%%%%%%%%%%%%%%%%%%%%%%%%%%
\begin{figure*}[ht]
    {\small
    \begin{align*}
        %
        % Terms
        %
        \graydescr{Location}   && \loc   & \Coloneqq\ {\text{program location}} \\
        \graydescr{Term}       && e      & \Coloneqq\ {x^\loc} \mid {\Unit^\loc} \mid {\overline{n}^{\,\loc}} \mid {\True^\loc} \mid {\False^\loc} \mid (\ite{e}{e}{e})^\loc \mid {e +^\loc e} \mid {(\lambda x^\loc.\ e)^\loc} \mid {(e\ e)^\loc} \\
                               &&        & \mid {[e, e]^\loc} \mid {\pi_1(e)^\loc} \mid {\pi_2(e)^\loc} \mid {\iota_1(e)^\loc} \mid {\iota_2(e)^\loc} \mid {\case{e}{x^\loc}{e}{x^\loc}{e}^\loc} \\[0.3cm]
        %
        % Types
        %
        \graydescr{Constructor}&& \odot      & \Coloneqq\ {->} \mid {\tyProd} \mid {\tySum} \\
        \graydescr{Provenance} && \prov  & \Coloneqq\ {\prov \padd \prov} \mid {\noProv} \mid {\loc} \mid {\NL{\prov}{\odot}} \mid {\NR{\prov}{\odot}} \\
        \graydescr{Type}       && \ty,\tz    & \Coloneqq\ {\al^\prov} \mid {\tyUnit^\prov} \mid {\tyInt^\prov} \mid {\tyBool^\prov} \mid {\ty \odot^\prov \ty} \\
        \graydescr{Constraint} && Q      & \Coloneqq\ \ty \sub \ty \\[0.3cm]
        %
        % Context, State and Polarity
        %
        \graydescr{Context}    && \Gamma & \Coloneqq\ {\emptyCtx} \mid {\Gamma\ctxPush{x: \al}}\\
        \graydescr{State}      && \sigma & \Coloneqq\ \{\ \mathtt{bounds}: {\,\overline{\overline{\tau} \sub \alpha \sub \overline{\tau}}}
                                                       ,\ \mathtt{errors}:  {\,\overline{\prov}}
                                                      \ \}\\
    \end{align*}}
    \vspace{-15pt}
    \caption{Syntax of terms and types.}
    \vspace{-15pt}
        \label{fig:type-syntax}
\end{figure*}

%%%%%%%%%%%%%%%%%%%%%%%%%%%%%%%%%%%%%%%%%%%%%%%%%%%%%%%%%%%%%%%%%%%%%%%%%%%%%%%
%% Subsection: Type Inference
%%%%%%%%%%%%%%%%%%%%%%%%%%%%%%%%%%%%%%%%%%%%%%%%%%%%%%%%%%%%%%%%%%%%%%%%%%%%%%%
\subsection{Type Inference}
\label{subsec:formal:type-inference}

Our type inference algorithm is an extension of a particular implementation of algebraic subtyping, due to \citet{simpleEssence2020}.
For readers who want to grasp the full extent of the algebraic subtyping technique and the associated proofs of correctness,
we suggest referring to the relevant literature \cite{dolanMycroftSubtyping2017,dolanThesis}.
Note that locations and provenances do not affect type inference;
in the algorithm presented in the following sections, it is easy to see that erasing all mentions of locations and type provenance tracking from our algorithm results in the algorithm by \citeauthor{simpleEssence2020}.
Since algebraic subtyping accepts strictly more programs than Hindley-Milner type inference, it follows that all programs rejected by the algorithm in this section are also rejected by Hindley-Milner. However, programs which only exhibit level-n errors (with $n \geq 1$) are accepted by the algorithm in this section, but rejected by Hindley-Milner type inference. Since the goal of our approach is to improve the quality of error messages, and not to change the set of accepted programs, we add an additional phase described in \cref{sec:type-confluence-errors} so that the same programs are accepted by our algorithm and the standard unification based algorithms.

The type inference algorithm consists of three parts:
The algorithmic inference rules, discussed in \cref{subsubsec:formal:algorithmic-inference-rules}, the constraint solving algorithm, discussed in \cref{subsubsec:formal:constraint-solving}, and the computation of subconstraints, discussed in \cref{subsubsec:formal:subconstraints}.

%
% Subsubsection: Algorithmic Inference Rules
%
\subsubsection{Algorithmic Inference Rules}
\label{subsubsec:formal:algorithmic-inference-rules}

The type inference judgement $\sigma \shortmid \Gamma |- e : \ty \shortmid \sigma'$ is specified in \cref{fig:typ-rules-2}.
It takes a type inference state $\sigma$, a typing context $\Gamma$, and a term $e$, and returns a type $\ty$ along with a new type inference state $\sigma'$.
This type inference state consists of the lower and upper bounds for each type variable, and a list of errors that were generated during type inference.
To focus on the essential aspects of a rule, we fade out the state $\sigma$ when it is only threaded through.

The top-level judgement $\emptyset \shortmid \emptyCtx |- e : \ty \shortmid \sigma$,
also written $|- e : \ty \shortmid \sigma$,
tells us whether a term $e$ is well-typed:
if there exists a $\prov$ such that $\err\prov \in \sigma$,
then we say that $e$ is ill-typed and $\prov$ is a type provenance chain highlighting one of its type collision errors;
otherwise, we say that $e$ is well-typed.

We rely on the usual informal notion of freshness for type variables---$\al$ is ``fresh'' if it does not appear anywhere in the previous values of $\sigma$,
nor in the %function's arguments
previous parts of the premise of a typing rule.

\newcommand{\shadeout}[1]{\textcolor{lightgray}{#1}}
\newcommand{\tyInfJdgmt}[5]{\shadeout{#1 \shortmid}\ #2 \vdash #3 : #4\ \shadeout{\shortmid #5}}
\newcommand{\consJdgmt}[3]{\shadeout{#1 \shortmid}\ \cons{#2}\ \shadeout{\shortmid #3}}

%%%%%%%%%%%%%%%%%%%%%%%%%%%%%%%%%%%%%%%%%%%%%%%%%%%%%%%%%%%%%%%%%%%%%%%%%%%%%%%
%% Figure: Algorithmic Type Inference Rules
%%%%%%%%%%%%%%%%%%%%%%%%%%%%%%%%%%%%%%%%%%%%%%%%%%%%%%%%%%%%%%%%%%%%%%%%%%%%%%%
\begin{figure}[ht]
  {\small
  \begin{mathpar}
    \boxed{\strut \sigma \shortmid\ \Gamma \vdash e : \ty\ \shortmid \sigma}
    \hfill
    \\
    %%%%%%%%%%%%%%%%%%%%%%%%%%%%%%%%%%%%%%%%%%%%%%%%%%%%%%%%%%%%%%%%%%%%%%%%%%%%%%%
    %% T-Unit
    %%%%%%%%%%%%%%%%%%%%%%%%%%%%%%%%%%%%%%%%%%%%%%%%%%%%%%%%%%%%%%%%%%%%%%%%%%%%%%%
    \inferrule[\trule{Unit}]{
    }{
      \tyInfJdgmt{\sigma}{\Gamma}{\Unit^\loc}{\tyUnit^\loc}{\sigma}
    }

    %%%%%%%%%%%%%%%%%%%%%%%%%%%%%%%%%%%%%%%%%%%%%%%%%%%%%%%%%%%%%%%%%%%%%%%%%%%%%%%
    %% T-Lit
    %%%%%%%%%%%%%%%%%%%%%%%%%%%%%%%%%%%%%%%%%%%%%%%%%%%%%%%%%%%%%%%%%%%%%%%%%%%%%%%
    \inferrule[\trule{Lit}]{
    }{
      \tyInfJdgmt{\sigma}{\Gamma}{\overline{n}^{\,\loc}}{\tyInt^\loc}{\sigma}
    }

    %%%%%%%%%%%%%%%%%%%%%%%%%%%%%%%%%%%%%%%%%%%%%%%%%%%%%%%%%%%%%%%%%%%%%%%%%%%%%%%
    %% T-Plus
    %%%%%%%%%%%%%%%%%%%%%%%%%%%%%%%%%%%%%%%%%%%%%%%%%%%%%%%%%%%%%%%%%%%%%%%%%%%%%%%
    \inferrule[\trule{Plus}]{
      \tyInfJdgmt{\sigma_0}{\Gamma}{e_0}{\ty_0}{\sigma_1}
      \\
      \consJdgmt{\sigma_1}{\ty_0 \sub \tyInt^\loc}{\sigma_2}
      \\\\
      \tyInfJdgmt{\sigma_2}{\Gamma}{e_1}{\ty_1}{\sigma_3}
      \\
      \consJdgmt{\sigma_3}{\ty_1 \sub \tyInt^\loc}{\sigma_4}
    }{
      \tyInfJdgmt{\sigma_0}{\Gamma}{e_0 +^\loc e_1}{\tyInt^\loc}{\sigma_4}
    }

    %%%%%%%%%%%%%%%%%%%%%%%%%%%%%%%%%%%%%%%%%%%%%%%%%%%%%%%%%%%%%%%%%%%%%%%%%%%%%%%
    %% T-Var
    %%%%%%%%%%%%%%%%%%%%%%%%%%%%%%%%%%%%%%%%%%%%%%%%%%%%%%%%%%%%%%%%%%%%%%%%%%%%%%%
    \inferrule[\trule{Var}]{
      \shadeout{\sigma_0 \shortmid}\
      % \textit{refresh}(\Gamma(x)) = \ty
      \Gamma(x) = \ty
      \ \shadeout{\shortmid \sigma_1}
    }{
      \tyInfJdgmt{\sigma_0}{\Gamma}{x^\loc}{\ty^\loc}{\sigma_1}
    }\ x\in\dom\Gamma

    %%%%%%%%%%%%%%%%%%%%%%%%%%%%%%%%%%%%%%%%%%%%%%%%%%%%%%%%%%%%%%%%%%%%%%%%%%%%%%%
    %% T-Lam
    %%%%%%%%%%%%%%%%%%%%%%%%%%%%%%%%%%%%%%%%%%%%%%%%%%%%%%%%%%%%%%%%%%%%%%%%%%%%%%%
    \inferrule[\trule{Lam}]{
      \al\ \kw{fresh}
      \\
      \tyInfJdgmt{\sigma_0}{\Gamma\ctxPush{x:\al}}{e}{\ty}{\sigma_1}
    }{
      \tyInfJdgmt{\sigma_0}{\Gamma}{(\lambda x^{\loc_x}.\ e)^\loc}{\al^{\loc_x} ->^\loc \ty}{\sigma_1}
    }

    %%%%%%%%%%%%%%%%%%%%%%%%%%%%%%%%%%%%%%%%%%%%%%%%%%%%%%%%%%%%%%%%%%%%%%%%%%%%%%%
    %% T-App
    %%%%%%%%%%%%%%%%%%%%%%%%%%%%%%%%%%%%%%%%%%%%%%%%%%%%%%%%%%%%%%%%%%%%%%%%%%%%%%%
    \inferrule[\trule{App}]{
      \al\ \kw{fresh}
      \\
      \tyInfJdgmt{\sigma_0}{\Gamma}{e_0}{\ty_0}{\sigma_1}
      \\
      \tyInfJdgmt{\sigma_1}{\Gamma}{e_1}{\ty_1}{\sigma_2}
      \\
      \consJdgmt{\sigma_2}{\ty_0 \sub \ty_1 -> \al^\loc}{\sigma_3}
    }{
      \tyInfJdgmt{\sigma_0}{\Gamma}{(e_0\ e_1)^\loc}{\al^\loc}{\sigma_3}
    }

    %%%%%%%%%%%%%%%%%%%%%%%%%%%%%%%%%%%%%%%%%%%%%%%%%%%%%%%%%%%%%%%%%%%%%%%%%%%%%%%
    %% T-True
    %%%%%%%%%%%%%%%%%%%%%%%%%%%%%%%%%%%%%%%%%%%%%%%%%%%%%%%%%%%%%%%%%%%%%%%%%%%%%%%
    \inferrule[\trule{True}]{
    }{
      \tyInfJdgmt{\sigma}{\Gamma}{\True^\loc}{\tyBool^\loc}{\sigma}
    }

    %%%%%%%%%%%%%%%%%%%%%%%%%%%%%%%%%%%%%%%%%%%%%%%%%%%%%%%%%%%%%%%%%%%%%%%%%%%%%%%
    %% T-False
    %%%%%%%%%%%%%%%%%%%%%%%%%%%%%%%%%%%%%%%%%%%%%%%%%%%%%%%%%%%%%%%%%%%%%%%%%%%%%%%
    \inferrule[\trule{False}]{
      }{
        \tyInfJdgmt{\sigma}{\Gamma}{\False^\loc}{\tyBool^\loc}{\sigma}
    }\\

    %%%%%%%%%%%%%%%%%%%%%%%%%%%%%%%%%%%%%%%%%%%%%%%%%%%%%%%%%%%%%%%%%%%%%%%%%%%%%%%
    %% T-IfThenElse
    %%%%%%%%%%%%%%%%%%%%%%%%%%%%%%%%%%%%%%%%%%%%%%%%%%%%%%%%%%%%%%%%%%%%%%%%%%%%%%%
    \inferrule[\trule{IfThenElse}]{
      \al\ \kw{fresh} \\\\
      \tyInfJdgmt{\sigma_0}{\Gamma}{e_1}{\tau_1}{\sigma_1}\\
      \tyInfJdgmt{\sigma_1}{\Gamma}{e_2}{\tau_2}{\sigma_2}\\
      \tyInfJdgmt{\sigma_2}{\Gamma}{e_3}{\tau_3}{\sigma_3}\\\\
      \consJdgmt{\sigma_3}{\ty_1 \sub \tyBool}{\sigma_4}\\
      \consJdgmt{\sigma_4}{\ty_2 \sub \al^\loc}{\sigma_5}\\
      \consJdgmt{\sigma_5}{\ty_3 \sub \al^\loc}{\sigma_6}
    }{
      \tyInfJdgmt{\sigma_0}{\Gamma}{(\ite{e_1}{e_2}{e_3})^\loc}{\alpha^\loc}{\sigma_6}
    }\\

    %%%%%%%%%%%%%%%%%%%%%%%%%%%%%%%%%%%%%%%%%%%%%%%%%%%%%%%%%%%%%%%%%%%%%%%%%%%%%%%
    %% T-Prod
    %%%%%%%%%%%%%%%%%%%%%%%%%%%%%%%%%%%%%%%%%%%%%%%%%%%%%%%%%%%%%%%%%%%%%%%%%%%%%%%
    \inferrule[\trule{Prod}]{
      \tyInfJdgmt{\sigma_0}{\Gamma}{e_0}{\ty_0}{\sigma_1}
      \\
      \tyInfJdgmt{\sigma_1}{\Gamma}{e_1}{\ty_1}{\sigma_2}
    }{
      \tyInfJdgmt{\sigma_0}{\Gamma}{[e_0,\,e_1]^{\loc}}{\ty_0 \tyProd^\loc \ty_1}{\sigma_2}
    }
    \\
    %%%%%%%%%%%%%%%%%%%%%%%%%%%%%%%%%%%%%%%%%%%%%%%%%%%%%%%%%%%%%%%%%%%%%%%%%%%%%%%
    %% T-Lproj
    %%%%%%%%%%%%%%%%%%%%%%%%%%%%%%%%%%%%%%%%%%%%%%%%%%%%%%%%%%%%%%%%%%%%%%%%%%%%%%%
    \inferrule[\trule{Lproj}]{
      \tyInfJdgmt{\sigma_0}{\Gamma}{e}{\ty}{\sigma_1}
      \\
      \al,\be\ \kw{fresh}
      \\\\
      \consJdgmt{\sigma_1}{\ty \sub \al^\loc \tyProd^\loc \be^\noProv}{\sigma_2}
    }{
      \tyInfJdgmt{\sigma_0}{\Gamma}{\pi_1(e)^{\loc}}{\al^\loc}{\sigma_2}
    }

    %%%%%%%%%%%%%%%%%%%%%%%%%%%%%%%%%%%%%%%%%%%%%%%%%%%%%%%%%%%%%%%%%%%%%%%%%%%%%%%
    %% T-Rproj
    %%%%%%%%%%%%%%%%%%%%%%%%%%%%%%%%%%%%%%%%%%%%%%%%%%%%%%%%%%%%%%%%%%%%%%%%%%%%%%%
    \inferrule[\trule{Rproj}]{
      \tyInfJdgmt{\sigma_0}{\Gamma}{e}{\ty}{\sigma_1}
      \\
      \al,\be\ \kw{fresh}
      \\\\
      \consJdgmt{\sigma_1}{\ty \sub \be^\noProv \tyProd^\loc \al^\loc}{\sigma_2}
    }{
      \tyInfJdgmt{\sigma_0}{\Gamma}{\pi_2(e)^{\loc}}{\al^\loc}{\sigma_2}
    }
    \\
    %%%%%%%%%%%%%%%%%%%%%%%%%%%%%%%%%%%%%%%%%%%%%%%%%%%%%%%%%%%%%%%%%%%%%%%%%%%%%%%
    %% T-Linj
    %%%%%%%%%%%%%%%%%%%%%%%%%%%%%%%%%%%%%%%%%%%%%%%%%%%%%%%%%%%%%%%%%%%%%%%%%%%%%%%
    \inferrule[\trule{Linj}]{
      \al\ \kw{fresh}
      \\
      \tyInfJdgmt{\sigma_0}{\Gamma}{e}{\ty}{\sigma_1}
    }{
      \tyInfJdgmt{\sigma_0}{\Gamma}{\iota_1(e)^{\loc}}{\ty \tySum^\loc \al^\noProv}{\sigma_1}
    }

    %%%%%%%%%%%%%%%%%%%%%%%%%%%%%%%%%%%%%%%%%%%%%%%%%%%%%%%%%%%%%%%%%%%%%%%%%%%%%%%
    %% T-Rinj
    %%%%%%%%%%%%%%%%%%%%%%%%%%%%%%%%%%%%%%%%%%%%%%%%%%%%%%%%%%%%%%%%%%%%%%%%%%%%%%%
    \inferrule[\trule{Rinj}]{
      \al\ \kw{fresh}
      \\
      \tyInfJdgmt{\sigma_0}{\Gamma}{e}{\ty}{\sigma_1}
    }{
      \tyInfJdgmt{\sigma_0}{\Gamma}{\iota_2(e)^{\loc}}{\al^\noProv \tySum^\loc \ty}{\sigma_1}
    }

    %%%%%%%%%%%%%%%%%%%%%%%%%%%%%%%%%%%%%%%%%%%%%%%%%%%%%%%%%%%%%%%%%%%%%%%%%%%%%%%
    %% T-Case
    %%%%%%%%%%%%%%%%%%%%%%%%%%%%%%%%%%%%%%%%%%%%%%%%%%%%%%%%%%%%%%%%%%%%%%%%%%%%%%%
    \inferrule[\trule{Case}]{
      \al,\be,\ga\,\kw{fresh}
      \\
      \\
      \tyInfJdgmt{\sigma_0}{\Gamma}{e_0}{\ty_0}{\sigma_1}
      \\
      \tyInfJdgmt{\sigma_1}{\Gamma\ctxPush{x^{\loc_x}:\al^{\loc_x}}}{e_1}{\ty_1}{\sigma_2}
      \\
      \tyInfJdgmt{\sigma_2}{\Gamma\ctxPush{y^{\loc_y}:\be^{\loc_y}}}{e_2}{\ty_2}{\sigma_3}
      \\
      \consJdgmt{\sigma_3}{\ty_0 \sub \al^{\loc_x} \tySum^\loc \be^{\loc_y}}{\sigma_4}
      \\
      \consJdgmt{\sigma_4}{\ty_1 \sub \ga^\loc}{\sigma_5}
      \\
      \consJdgmt{\sigma_5}{\ty_2 \sub \ga^\loc}{\sigma_6}
    }{
      \tyInfJdgmt{\sigma_0}{\Gamma}{\case{e_0}{x^{\loc_x}}{e_1}{y^{\loc_y}}{e_2}^{\loc}}{\ga^\loc}{\sigma_6}
    }
  \end{mathpar}}
  \caption{Algorithmic type inference rules.}
  \vspace{-10pt}
    \label{fig:typ-rules-2}
\end{figure}

%
% Subsubsection: Constraint Solving Algorithm
%
\subsubsection{Constraint Solving Algorithm}
\label{subsubsec:formal:constraint-solving}

The constraint solving algorithm is specified in \cref{fig:constraining-rules}.
The type constraining function $\sigma \shortmid \cons{Q}^H \shortmid \sigma'$ takes a type inference state $\sigma$, a constraint $Q$,  and a set of current hypotheses $H$, and returns a new state $\sigma'$.

We introduce two helper functions `$\addlb$' and `$\addub$'
to add a type, respectively, to the upper and lower bounds of a type variable in the type inference state. For e.g. $\addub(\sigma_0, \al, \prov \cdot \ty^{\prov'}) = \sigma_1$ where $\sigma_1 = \sigma_0 \cup \{\al \sub \ty^{\prov \cdot \prov'}\}$. Similarly, we introduce helper functions `$\fname{lb}$' and `$\fname{ub}$'
to look up respectively the upper and lower bounds of a type variable in the type inference state.

\begin{description}
  \item[\textsc{\crule{Cache}}]
    This rule allows to immediately solve a constraint if it has already been encountered in the constraint solving process, and is therefore contained in the set of hypotheses $H$.
    This is necessary to avoid divergence in the presence of recursive types.\footnote{
      Whether the language is meant to support recursive types or not is an orthogonal concern,
      and it is easy to add a so-called ``\emph{occurs check}'' to make sure recursive types are rejected,
      if needed.
    }
    We define $\fname{reset}(\ty_0,\ty_1)$ as the substitution, in $(\ty_0,\ty_1)$, of all type provenances by $\noProv$.
    This is used to ensure that type provenances do not affect the memoization of the constraining function.
  \item[\textsc{\crule{Refl}}]
    A constraint between two equal types can be solved immediately.
    When we check for equality of types, we do not care for provenances.
    For this reason, we apply the $\fname{reset}$ function before we compare them.
  \item[\textsc{\crule{Var-L}}]
    (and similarly for \textsc{\crule{Var-R}})
    When we encounter a constraint $\alpha \sub \ty$ between a unification variable $\alpha$ and a type $\ty$ (which is not a unification variable) we have to do two things.
    We first add $\ty$ to the set of upper bounds of $\alpha$ in the state $\sigma$.
    Then we generate and solve one additional constraint between $\ty$ and all existing lower bounds for $\alpha$ in $\sigma$.
  \item[\textsc{\crule{Var-LR}}]
   When we encounter a constraint $\al \sub \al'$ between two unification variables, we add $\al$ to the lower bounds of $\al'$ and $\al'$ to the upper bounds of $\al$ before we generate the subconstraints to verify that the bounds are still consistent.
  \item[\textsc{\crule{Sub}}]
    When the constraint we have to solve is complex, i.e. neither of the two types is a unification variable, then we invoke the function $\fname{sub}$ in order to compute the subconstraints of the constraint.
    If this function returns with a set of new constraints, we solve them in turn.
  \item[\textsc{\crule{Error}}]
    If the function $\fname{sub}$ returns with an error, the returned $\sigma'$ is populated with $\err\prov$ elements containing the provenance chains $\prov$ corresponding to the error.
\end{description}

\begin{figure}[ht]
  {\small
  \begin{mathpar}
    \boxed{\strut\sigma \shortmid \cons{Q}^H  \shortmid \sigma }
    \hfill
    \\
    %
    % Cache
    %
    \inferrule[\crule{Cache}]{
      \fname{reset}(Q) \in H
    }{
      \shadeout{\sigma \shortmid}\ \cons{Q}^H \ \shadeout{\shortmid \sigma}
    }

    %
    % Refl
    %
    \inferrule[\crule{Refl}]{
      \fname{reset}(\ty_0) = \fname{reset}(\ty_1)
    }{
      \shadeout{\sigma \shortmid}\ \cons{{\ty_0} \sub {\ty_1}}^H\ \shadeout{\shortmid \sigma}
    }

    %
    % Var Left Right
    %
    \inferrule[\crule{Var-LR}]{
      \fname{add}\text{-}\fname{ub}(\sigma_0,\al,\al'^{\prov_0} \padd \prov_1) = \sigma_1\\
      \fname{add}\text{-}\fname{lb}(\sigma_1,\al',\prov_1 \padd \al^{\prov_0}) = \sigma_2\\
      \sigma_2 \shortmid \cons{[ \ty_{\al} \sub \al'^{\prov_1} \ | \ \ty_{\al}' \in \fname{lb}(\sigma_0,\alpha) ]}^{H\,\cup\,\fname{reset}(\al \sub \al')} \shortmid \sigma_3 \\
    }{
      \sigma_0 \shortmid \cons{ \al^{\prov_0} \sub \al'^{\prov_1} }^H \shortmid \sigma_3
    }

    %
    % Var Left
    %
    \inferrule[\crule{Var-L}]{
      \fname{add}\text{-}\fname{ub}(\sigma_0,\alpha,\ty^{\prov_0} \padd \prov_1) = \sigma_1\\
      \sigma_1 \shortmid \cons{[ \ty' \sub \ty^{\prov_1} \ | \ \ty' \in \fname{lb}(\sigma_0,\alpha) ]}^{H\,\cup\,\fname{reset}(\al \sub \ty)} \shortmid \sigma_2 \\
    }{
      \sigma_0 \shortmid \cons{ \al^{\prov_0} \sub \ty^{\prov_1} }^H \shortmid \sigma_2
    }

    %
    % Var Right
    %
    \inferrule[\crule{Var-R}]{
      \fname{add}\text{-}\fname{lb}(\sigma_0,\alpha,\prov_0 \padd \ty^{\prov_1}) = \sigma_1\\
      \sigma_1 \shortmid \cons{[ \ty^{\prov_0} \sub \ty' \ | \ \ty' \in \fname{ub}(\sigma_0,\alpha)]}^{H\,\cup\,\fname{reset}(\ty \sub \al)}
      \shortmid \sigma_1
    }{
      \sigma_0 \shortmid \cons{ \ty^{\prov_0} \sub \al^{\prov_1} }^H  \shortmid \sigma_2
    }

    %
    % Subconstraints
    %
    \inferrule[\crule{Sub}]{
      \fname{sub}(Q) = \overline{Q}
      \\
      \shadeout{\sigma_0 \shortmid}\ \cons{\overline{Q}} \ \shadeout{\shortmid \sigma_1}
    }{
      \shadeout{\sigma_0 \shortmid}\ \cons{Q} \ \shadeout{\shortmid \sigma_1}
    }

    %
    % Error
    %
    \inferrule[\crule{Error}]{
      \fname{sub}(Q) = \err{\prov}
    }{
      \shadeout{\sigma \shortmid}\ \cons{Q} \ \shadeout{\shortmid \sigma_0}
    }
  \end{mathpar}
  \begin{mathpar}
    \boxed{\strut\sigma \shortmid \cons{\overline{Q}}^H  \shortmid \sigma }
    \hfill
    \\

    %
    % Nil
    %
    \inferrule[\crule{Nil}]{
    }{
      \shadeout{\sigma \shortmid}\ \cons{[]} \ \shadeout{\shortmid \sigma}
    }

    \inferrule[\crule{Cons}]{
      \shadeout{\sigma_0 \shortmid}\ \cons{Q} \ \shadeout{\shortmid \sigma_1} \\
      \shadeout{\sigma_1 \shortmid}\ \cons{\overline{Q}} \ \shadeout{\shortmid \sigma_2}\\
    }{
      \shadeout{\sigma_0 \shortmid}\ \cons{Q::\overline{Q}} \ \shadeout{\shortmid \sigma_2}\\
    }
  \end{mathpar}}
  \caption{Constraint solving algorithm.}
    \vspace*{-15pt}
  \label{fig:constraining-rules}
\end{figure}

%
% Subsubsection: Computation of Subconstraints
%
\subsubsection{Computation of Subconstraints}
\label{subsubsec:formal:subconstraints}

The computation of subconstraints is defined in \cref{fig:subconstraints}.
The function $\fname{sub}(Q)$ takes a constraint $Q$ as input, and either computes a new list of constraints to be solved, or otherwise returns an error $\err{p}$ containing a provenance if the constraint cannot be solved.
We return new subconstraints if the types in the constraint are either both function types, both product types or both sum types.
In that case, we also have to recombine the provenances of the types which are involved in the constraint, in order to track how a data flow can be tracked through a constructor.
This is where the additional provenances which we introduced, but didn't explain, in \cref{subsec:formal:types-and-provs} come into play.
We  write $\NL\prov\odot$ and $\NR\prov\odot$ where the L and R indicate if the provenance comes from the left or right hand side of a constraint on a constructor type $\odot$.
We use the notations $\ty^{\prov_0} \padd \prov_1$ and $\prov_0 \padd \ty^{\prov_1}$ as shorthands for $\ty^{\prov_0 \padd \prov_1}$.
In every other case, that is, if the outermost types of the two sides of a subtyping constraint are not identical, the constraint is not solvable.

When we compute the subconstraints of two function types, we use the function $\fname{rev}(\prov)$ on provenances which yields a type provenance with the same contents, but in reverse order.
Reversal applies recursively, meaning that it also reverses the order of provenances nested inside constructors like ${\pFunL \prov}$.
The use of this function spells out a curious phenomenon:
passing through a function parameter \emph{reverses} the direction of a type flow,
switching from \emph{flowing into} to \emph{flowing from}, or \emph{vice versa}.
To illustrate this subtlety, consider
the following program annotated with the relevant locations ${\loc_{1..8}}$:
\begin{lstlisting}
let foo f$^{\loc_1}$  =  let g$^{\loc_2}$ = f$^{\loc_3}$ in g$^{\loc_4}$ "hello"$^{\loc_5}$
foo (fun x$^{\loc_6}$ -> x$^{\loc_7}$ + 1)$^{\loc_8}$
\end{lstlisting}
% let k = (fun x$^{\loc_6}$ -> x$^{\loc_7}$ + 1)$^{\loc_8}$
% foo k$^{\loc_9}$
%
The problematic flow here is
`$\loc_5 \padd \pFunL{\loc_4 \padd \loc_2 \padd \loc_3 \padd \loc_1 \padd \loc_8} \padd \loc_6 \padd \loc_7$'.
Here is how to understand it:
% , and how it could be reported in a (verbose-mode) user-readable error message:
\begin{quote}\textit{
The string literal at $\loc_5$ has type {\small\tt string};
it flows \textbf{\emph{into}} the \emph{parameter} of function {\small\tt g} at $\loc_4$
(hence the $L$, which denotes the \emph{left-hand-side} of a function type,
so we reverse the flow direction);
where {\small\tt g} itself flows
  \textbf{\emph{from}} let-bound identifier {\small\tt g} at $\loc_2$;
  \textbf{\emph{from}} parameter reference {\small\tt f} at $\loc_3$;
  \textbf{\emph{from}} parameter {\small\tt f} at $\loc_1$;
  \textbf{\emph{from}} the argument function at $\loc_8$;
  and (leaving the function type and reverting back to a forward flow)
  \textbf{\emph{into}} parameter {\small\tt x} at $\loc_6$;
  \textbf{\emph{into}} reference {\small\tt x} at $\loc_7$;
  where type {\small\tt int} is expected.
}\end{quote}
Notice how this flow starts as a normal \emph{forward} flows
but reverses to a \emph{backward} one upon entering the left-hand side of a function type
before going back to a forward flow as the flow leaves the function type.
Naturally, the \emph{complete} flow information described above is far too verbose to report directly to users.
We found that a good tradeoff (which we use in our tool)
is to only report the \emph{outer} flow `$\loc_5 \padd \ldots \padd \loc_6 \padd \loc_7$'
and reserve the full flow for the verbose mode and (in the future) for interactive type error exploration
and IDE integration.

\begin{figure}[ht]
  {\small
  \begin{align*}
    % Prod
    \fname{sub}({\ty_{00} \tyProd^{\prov_0} \ty_{01}} \sub {\ty_{10} \tyProd^{\prov_1} \ty_{11}}) &=
    [ {\ty_{00} \padd \pProdL{\prov_0 \padd \prov_1}} \sub \ty_{10}
     \,,\ {\ty_{01} \padd \pProdR{\prov_0 \padd \prov_1}} \sub \ty_{11} ]\\
    % Sum
    \fname{sub}({\ty_{00} \tySum^{\prov_0} \ty_{01}} \sub {\ty_{10} \tySum^{\prov_1} \ty_{11}}) &=
    [ {\ty_{00} \padd \pSumL{\prov_0 \padd \prov_1} } \sub \ty_{10}
     \,,\ {\ty_{01} \padd \pSumR{\prov_0 \padd \prov_1} } \sub \ty_{11} ]\\
    % Fun
    \fname{sub}({\ty_{00} ->^{\prov_0} \ty_{01}} \sub {\ty_{10} ->^{\prov_1} \ty_{11}}) &=
    [ {\ty_{10} \padd \pFunL{\prov_1 \padd \fname{rev}(\prov_0)}} \sub \ty_{00}
    \,,\ {\ty_{01} \padd \pFunR{\prov_0 \padd \prov_1}} \sub \ty_{11} ]\\
    % Error
    \fname{sub}(\ty_1^{p_0} \sub \ty_2^{p_1}) &= \err{\prov_0 \padd \prov_1} \\
  \end{align*}}
  \vspace{-20pt}
  \caption{Subconstraining rules.}
  \vspace{-5pt}
    \label{fig:subconstraints}
\end{figure}

% %%
% %% Subsection: Generating Error Messages from Provenances
% %%
% \subsection{Generating Error Messages from Provenances}
% \label{subsec:formal:generating-error-messages}

% Let us now verify that this algorithm does indeed track the provenances of types correctly, and can be used to generate error messages for Level-0 errors.
% For this, we take up the example  from \cref{fig:classification-examples:level-zero} and show how to infer a flow-based error for it.

% \begin{example}[Level-0 Unification Error]
%   Annotating the example from \cref{fig:classification-examples:level-zero} with locations results in the following program:
%   \begin{lstlisting}
%     let x$^{\loc_0}$ = 2$^{\loc_1}$
%     let y$^{\loc_2}$ = (if x$^{\loc_3}$ then true$^{\loc_4}$ else false$^{\loc_5}$)$^{\loc_6}$
%   \end{lstlisting}
%   The type inference algorithm yields an error with the provenance $\loc_1 \padd \loc_0 \padd \loc_3$ which describes precisely how the type $\tyInt$, which was introduced at the integer literal at $\loc_1$, flows through the intermediate binding at $\loc_0$ to the use site at $\loc_3$, where a boolean value is expected.
% \end{example}

Finally, notice that all the type constructors used in this section are \emph{variant}:
product and sum types are covariant in their components and functions are contravariant in their parameters
and covariant in their results.
In a system with subtyping, it is always possible to separate the covariant and contravariant uses of type parameters,
so that this pervasive variance is generally feasible.
However, in ML languages like OCaml, some type constructors are defined as invariant,
such as mutable references.
To handle these, we need a notion of \emph{non-directional unification}, which is studied in the next section.

\section{Type confluence errors}
\label{sec:type-confluence-errors}
The algorithm presented in \cref{sec:formal} only recognizes and reports Level-0 errors.
Now, we extend it to also report Level-n errors for $n \geq 1$, by tracking data flows described by constraints, as explained in \cref{subsec:classifying:flow-of-types}.

%%%%%%%%%%%%%%%%%%%%%%%%%%%%%%%%%%%%%%%%%%%%%%%%%%%%%%%%%%%%%%%%%%%%%%%%%%%%%%%
%% Figure: Extended syntax of terms and types.
%%%%%%%%%%%%%%%%%%%%%%%%%%%%%%%%%%%%%%%%%%%%%%%%%%%%%%%%%%%%%%%%%%%%%%%%%%%%%%%
\begin{figure*}[ht]
  {\small
  \begin{align*}
      \graydescr{Provenance} && \prov      & \Coloneqq\ ... \mid Z\\
      \graydescr{Relation} && \bullet      & \Coloneqq \sub^\prov 
        \mid \supp^\prov 
        \mid\ \sim^{\Nprov{Z}{\tyProd \mid \tySum \mid ->}{L \mid R}}\\
      \graydescr{Data flow} && Z      & \Coloneqq\ \ty \bullet \ty \mid Z \bullet \ty \mid \ty \bullet Z\\
  \end{align*}}
  \vspace{-20pt}
  \caption{Extended syntax with unification.}
    \vspace{-10pt}
  \label{fig:extended-type-syntax}
\end{figure*}

\paragraph*{Data flows}
We extend the syntax from \cref{fig:type-syntax} with the notion of data flows.
A data flow Z is a sequence of types related by either $\sub^\prov$, $\supp^\prov$, or $\sim^{\Nprov{Z}{\tyProd \mid \tySum \mid ->}{L \mid R}}$, and we abstract over these relations with the symbol $\bullet$.
We already discussed in \cref{subsec:classifying:flow-of-types} that constraints represent data flows;
we can therefore embed constraints into data flows:
constraint $\ty^{\prov_0} \sub \ty^{\prov_1}$, for example, corresponds to data flow  $\ty \sub^{\prov_0 \cdot \prov_1} \ty$.

If a data flow has the form $\ty_1 \bullet \ldots \bullet \ty_2$, then we use the shorthand $\ty_1\ \_\ \ty_2$ if we are only interested in the two outermost types of the data flow.
Unifying a data flow $Z = \ty_1\ \_\ \ty_2$ says that $\ty_1$ and $\ty_2$ must be equal, and the data flow for this unification is explained by the components of $Z$.
If the types $\ty_1$ and $\ty_2$ are not equal, then we have to generate a detailed unification error from Z, which we will explain in \cref{subsubsec:error-reports}.

We introduce $\sim$ to relate arguments of constructor types.
We also introduce nested data flows $\Nprov{Z}{\tyProd \mid \tySum \mid ->}{L \mid R}$ for arguments of constructor types.
Consider the data flow $\ty_{00} \sim^{\pProdL{Z}} \ty_{10}$, which has the nested data flow $Z = \ty_{00} \tyProd \ty_{01} \sub \al \sub \ty_{10} \tyProd \ty_{11}$.
We say that the types $\ty_{00}$ and $\ty_{10}$ are the left arguments of product types at the terminal ends of data flow $Z$.
A key intuition is that the types don't flow directly in $Z$ but are carried by the product types.
Similarly, sum and function type arguments have nested data flows too.

A data flow is \emph{valid} for a given a type inference state if all the individual relations in it are valid.
A sub-typing relation is valid if it is contained in the state $\sigma$, and the relation $\sim^{\Nprov{Z}{\odot}{L|R}}$ is valid if the nested data flow is valid and terminated by the correct constructor types.

{\small
\vspace{-10pt}
\begin{align*}
  \fname{valid}(Z)_\sigma &= \fname{valid}(\ty \bullet \ty')_\sigma\ \forall \ty, \ty'.\ (...\ \ty \bullet \ty'\ ...) \in Z\\
  \fname{valid}(\ty \sim^{\NL{Z}{\odot}} \ty')_\sigma &= \fname{valid}(Z)_\sigma \text{ where } Z = (\ty \odot \_)\ \_\ (\ty' \odot \_)\\
  \fname{valid}(\ty \sim^{\NR{Z}{\odot}} \ty')_\sigma &= \fname{valid}(Z)_\sigma \text{ where } Z = (\_ \odot \ty)\ \_\ (\_ \odot \ty')\\
  \fname{valid}(\ty \sub^\prov \ty')_\sigma &= \ty \sub^\prov \ty' \in \sigma\\
  \fname{valid}(\ty \supp^\prov \ty')_\sigma &= \ty' \sub^{\fname{rev}(\prov)} \ty \in \sigma\\
\end{align*}
\vspace{-30pt}}

%%%%%%%%%%%%%%%%%%%%%%%%%%%%%%%%%%%%%%%%%%%%%%%%%%%%%%%%%%%%%%%%%%%%%%%%%%%%%%%
%% Unification algorithm
%%%%%%%%%%%%%%%%%%%%%%%%%%%%%%%%%%%%%%%%%%%%%%%%%%%%%%%%%%%%%%%%%%%%%%%%%%%%%%%

\subsection{Unification Algorithm}
\label{subsubsec:formal:unification-algorithm}
\begin{figure}[ht]
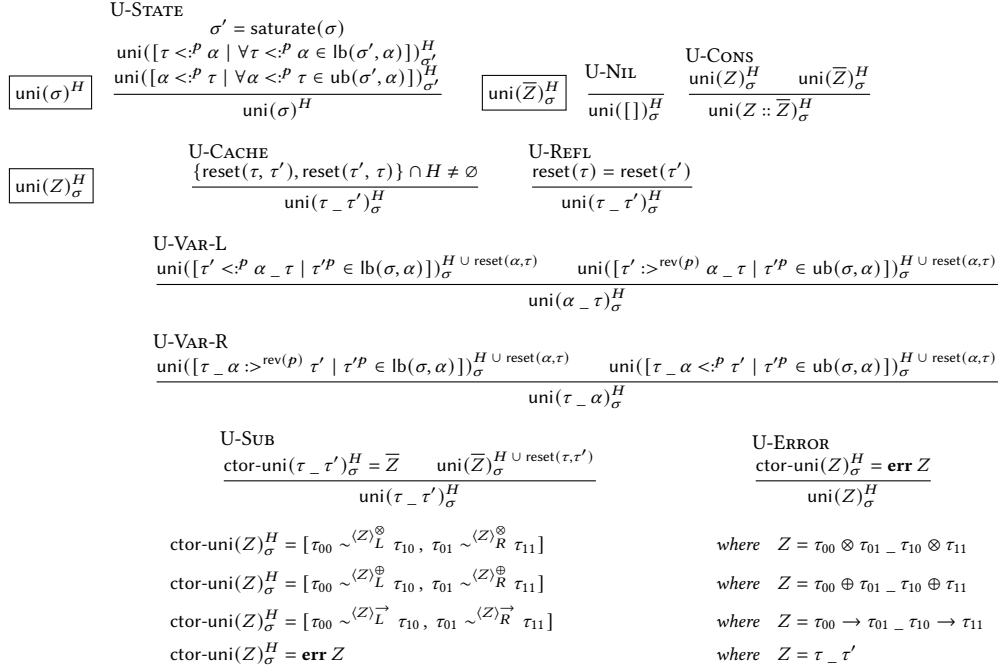

  {\footnotesize
  \begin{mathpar}
    \boxed{\uni{\sigma}^H}
    % \hfill
    % \\
    \quad
    %
    % Unify bounds in a state
    %
    \inferrule[\urule{State}]{
      \sigma' = \fname{saturate}(\sigma)
      \\\\
      \uni{[\ty \sub^\prov \al \mid \forall \ty \sub^\prov \al \in \fname{lb}(\sigma', \al)]}^H_{\sigma'}
      \\\\
      \uni{[\al \sub^\prov \ty \mid \forall \al \sub^\prov \ty \in \fname{ub}(\sigma', \al)]}^H_{\sigma'}
    }{
      \uni{\sigma}^H
    }
    % \\
    % \hfill
    \qquad
  % \end{mathpar}
  % \begin{mathpar}
  % \\
    \boxed{\uni{\overline{Z}}^H_\sigma}
    % \hfill
    % \\
    \quad
    %
    % Nil
    %
    \inferrule[\urule{Nil}]{
    }{
      \uni{[]}^H_\sigma
    }
    \quad
    \inferrule[\urule{Cons}]{
      \uni{Z}^H_\sigma\\
      \uni{\overline{Z}}^H_\sigma\\
    }{
      \uni{Z :: \overline{Z}}^H_\sigma\\
    }
    \hfill

  % \end{mathpar}
  % \begin{mathpar}
  \\
    \boxed{\uni{Z}^H_\sigma}
    % \hfill
    % \\

    %
    % Unify cache
    %
    \inferrule[\urule{Cache}]{
      \{
        \fname{reset}(\ty,\, \ty'),
        \fname{reset}(\ty',\, \ty)
      \} \cap H \neq \varnothing
    }{
      \uni{\ty\ \_\ \ty'}^H_\sigma
    }

    %
    % Unify with equal type
    %
    \inferrule[\urule{Refl}]{
      \fname{reset}(\ty) = \fname{reset}(\ty')
    }{
      \uni{\ty\ \_\ \ty'}^H_\sigma
    }
    \hfill
    \\

    %
    % Unify with type variable
    %
    \inferrule[\urule{Var-L}]{
      \uni{[\ty' \sub^\prov \al\ \_\ \ty \mid \ty'^\prov \in \fname{lb}(\sigma, \al)]}^{H\ \cup\ \fname{reset}(\al, \ty)}_\sigma\\
      \uni{[\ty' \supp^{\fname{rev}(\prov)} \al\ \_\ \ty \mid \ty'^\prov \in \fname{ub}(\sigma, \al)]}^{H\ \cup\ \fname{reset}(\al, \ty)}_\sigma\\
    }{
      \uni{\al\ \_\ \ty}^H_\sigma
    }

    %
    % Unify with type variable
    %
    \inferrule[\urule{Var-R}]{
      \uni{[\ty\ \_\ \al \supp^{\fname{rev}(\prov)} \ty' \mid \ty'^\prov \in \fname{lb}(\sigma, \al)]}^{H\ \cup\ \fname{reset}(\al, \ty)}_\sigma\\
      \uni{[\ty\ \_\ \al \sub^\prov \ty' \mid \ty'^\prov \in \fname{ub}(\sigma, \al)]}^{H\ \cup\ \fname{reset}(\al, \ty)}_\sigma\\
    }{
      \uni{\ty\ \_\ \al}^H_\sigma
    }

    %
    % Unify constructors
    %
    \inferrule[\urule{Sub}]{
      \fname{ctor\text{-}uni}(\ty\ \_\ \ty')^H_\sigma = \overline{Z}
      \\
      \uni{\overline{Z}}^{H\ \cup\ \fname{reset}(\ty, \ty')}_\sigma
    }{
      \uni{\ty\ \_\ \ty'}^H_\sigma
    }

    %
    % Unify unification error
    %
    \inferrule[\urule{Error}]{
      \fname{ctor\text{-}uni}(Z)^H_\sigma = \textbf{err}\ Z\\
    }{
      \uni{Z}^H_\sigma
    }
  \end{mathpar}
  \begin{align*}
    % Prod
    \fname{ctor\text{-}uni}(
      Z
    )^H_\sigma &=
      [
      \ty_{00} \sim^{\pProdL{Z}} \ty_{10}
      \,,\ 
       \ty_{01} \sim^{\pProdR{Z}} \ty_{11}
      ]
      &
      \textit{where}\quad
      Z &= \ty_{00} \tyProd \ty_{01}\ \_\ \ty_{10} \tyProd \ty_{11}
    \\
    % Sum
    \fname{ctor\text{-}uni}(
      Z
    )^H_\sigma &=
      [
      \ty_{00} \sim^{\pSumL{Z}} {\ty_{10}}
      \,,\ 
      \ty_{01} \sim^{\pSumR{Z}} \ty_{11}
      ]
      &
      \textit{where}\quad
      Z &= \ty_{00} \tySum \ty_{01}\ \_ \ \ty_{10} \tySum \ty_{11}
    \\
    % Fun
    \fname{ctor\text{-}uni}(
      Z
    )^H_\sigma &=
      [
      \ty_{00} \sim^{\pFunL{Z}} \ty_{10}
      \,,\ 
      \ty_{01} \sim^{\pFunR{Z}} \ty_{11}
      ]
      &
      \textit{where}\quad
      Z &= \ty_{00} -> \ty_{01}\ \_\ \ty_{10} -> \ty_{11}
    \\
    % Err
    \fname{ctor\text{-}uni}(
      Z
    )^H_\sigma &= \textbf{err}\ Z
      &
      \textit{where}\quad
      Z &= \ty\ \_\ \ty'
  \end{align*}}
  \vspace{-1em}
  \caption{Data-flow-tracking unification.}
    \label{fig:unification-rules}
\end{figure}

The unification algorithm is specified in \cref{fig:unification-rules}.
We start with function $\uni{\sigma}^H$, which takes a type inference state $\sigma$ and recurses over its
bounds through the
unification function $\uni{Z}^H_\sigma$, where $Z$ is a data flow and $H$ is the current set of hypotheses.
% , and $\sigma$ is the type inference state.
This function equates the first and last types of a data flow and terminates with an error if they are not equal.
A piece of global state could be threaded through the inference rules describing this function
to collect all incorrect data flows; however, we omit this to keep the algorithm's specification concise.
It is enough to see that, given a derivation of this function, we can gather all unification errors
by collecting all uses of the \ruleName{U-Error} rule.

We say that type inference state $\sigma$ is \emph{saturated} when
for all $\al$ and $\al'$ we have $\al \in \fname{lb}(\sigma,\,\al') <=> \al' \in \fname{ub}(\sigma,\,\al)$.
Helper function `$\fname{saturate}$' saturates its input state in the obvious way.

\begin{description}
  \item[\textsc{\urule{state}}]
    This rule is the entry point for unification.
    We first saturate the type inference state.
    Then, for each type variable $\al$ in state $\sigma$, we unify it with its upper and lower bound types $\ty$.
  \item[\textsc{\urule{Cache}}]
    A unification is solved trivially if it is already cached. We use `$\fname{reset}$' to erase the provenance of the types before looking up the cache, where $\fname{reset}(\ty, \ty') = (\fname{reset}(\ty), \fname{reset}(\ty'))$.
    This cache is necessary for the same reason as the cache in \cref{subsubsec:formal:constraint-solving},
    \ie to prevent divergence in the case of cyclic bounds.
    However, since unifying two types is a symmetric operation,
    we now look up both subtyping directions in the cache
    (\ie both $\fname{reset}(\ty,\, \ty')$ and $\fname{reset}(\ty',\, \ty)$).
  \item[\textsc{\urule{Refl}}]
    A unification between two types that are equal modulo their provenance is solved immediately.
    We use reset to erase the provenance information before comparing the types.
  \item[\textsc{\urule{Var-L}}]
    This rule unifies type variable $\al$ with $\ty$. It produces a set of data flows from the the upper and lower bounds of $\al$ to $\ty$. Since $\al$ is on the left of the relation, the new relations are concatenated to the left of the existing data flow $\flow$. This preserves the left to right continuity of the data flow.
  \item[\textsc{\urule{Var-R}}]
    This is similar to \urule{Var-L} except the type variable $\al$ is on the right of the relation. So the new relation gets concatenated on the right.
  \item[\textsc{\urule{Sub}}]
    When the unification involves constructed types (product, sum and function types), we invoke $\fname{ctor}\text{-}\fname{uni}$ to compute sub-unifications for their type arguments. If the function cannot equate the constructor types it returns an error.
  \item[\textsc{\urule{Error}}]
    If \urule{Sub} returns an error for an incorrect data flow, this rule terminates the algorithm.
    The actual algorithm can be implemented by threading through a state
    and collecting all such errors before reporting them.
\end{description}

%%%%%%%%%%%%%%%%%%%%%%%%%%%%%%%%%%%%%%%%%%%%%%%%%%%%%%%%%%%%%%%%%%%%%%%%%%%%%%%
%% Unification algorithm
%%%%%%%%%%%%%%%%%%%%%%%%%%%%%%%%%%%%%%%%%%%%%%%%%%%%%%%%%%%%%%%%%%%%%%%%%%%%%%%

Notice that because we look up \emph{both}
pairs $(\ty,\,\ty')$ and $(\ty',\,\ty)$ in the cache,
we will only ever traverse a given data flow between two variables
in a single direction only, instead of potentially traversing the data flow once in each direction,
which could otherwise happen in the presence of type variable cycles,
such as $\al \sub \be, \be \sub \al$.
Long cyclic chains could lead to a potentially exponential amount of erroneous data flows, which we avoid this way.
Moreover, our concrete implementation performs a breadth-first search
while following the algorithm of \Fig{fig:unification-rules}
in order to find the shortest erroneous paths,
and then stops the search without traversing further constraints.
From our own observations, this seems to make inference fast in practice.\footnote{This can be observed in the web demo, which updates the output quickly upon every keystroke, even for larger programs.}
We only report one erroneous path to the programmer to explain a given error, even when many possible paths (including several shortest ones) are available.
This choice can be considered \emph{arbitrary},
as there is no reason to consider that one particular path should be better at explaining the error than the others.
In the future, different approaches to report such equivalent paths could be studied:
We could report all paths, explain cyclic data flows separately to programmers,
or use heuristics to determine which path will be more understandable.

%%
%% Subsection: Detailed Error Reports
%%
\subsection{Detailed Error Reports}
\label{subsubsec:error-reports}
We create error reports from the data flows of failed unifications. Each data flow is transformed into a sequence of source locations separated by alternating data flow directions,
encoding the back-and-forth nature of higher-level error,
and possibly containing nested flows when the problematic unification is indirect
and goes through type constructor arguments.
Additional type information in the data flow is used to add helpful details to the report.  The data flow information can be used for interactive error reporting, interactive code exploration, code hints in IDEs to name a few.
Our current implementation only reports errors and we share examples in \cref{sec:appendix:examples}.

%%
%% Example: Level-0 Data Flow
%%
\begin{example}[Level-0 Data Flow]
  \label{ex:unification-provenance-0}
  We can visualize a data flow as a type flowing through a sequence of program locations.
  A constraint is the most basic data flow.
  We can translate the subtyping constraint $\al^{\prov_0} \sub \ty^{\prov_1}$ into the data flow $\al \sub^{\prov_0 \cdot \prov_1} \ty$.
  After unification, $\ty$ and $\al$ must be the same type, so we can say that the data represented by the type flows through the locations $\prov_0$ and $\prov_1$.
  This can be visualized by the following diagram:
  \begin{center}
    \begin{tikzpicture}
      \node (A) at (-1.5,0) {$\al$};
      \node (B) at (1.5,0) {$\ty$};
      \draw[->] (A) -- (B) node[midway, above]{$\prov_0 \ \ \ \prov_1$};
    \end{tikzpicture}
  \end{center}
\end{example}
\vspace{-.5em}

%%
%% Example: Level-1 Data Flow
%%
\begin{example}[Level-1 Data Flow]
  \label{ex:unification-provenance-1}
  We annotate the second program from \cref{fig:classification-examples:level-one}, which yields the following annotated program:
  \begin{lstlisting}
    let x$^{\loc_0}$ = 2$^{\loc_1}$
    let y$^{\loc_2}$ = (if true then x$^{\loc_4}$ else "x"$^{\loc_5}$)$^{\loc_3}$
  \end{lstlisting}
  For this program we generate the constraints $\tyInt^{\loc_4 \cdot \loc_0 \cdot \loc_1} \sub \al_{y}^{\loc_3 \cdot \loc_2}$ and $\tyStr^{\loc_5} \sub \al_{y}^{\loc_3 \cdot \loc_2}$.
  We get a unification error for $\tyInt \sub^{\prov_0} \al_y \supp^{\prov_1} \tyStr$ where $\prov_0 = \loc_1 \cdot \loc_0 \cdot \loc_4 \cdot \loc_3 \cdot \loc_2$ and $\prov_1 = \loc_2 \cdot \loc_3 \cdot \loc_5$.
  Notice that invoking symmetry for $\tyStr$ relation to creates a linear data flow.
  This flow looks like this:
  \begin{center}
    \begin{tikzpicture}
      \node (A) at (-3.0,0) {$\tyInt$};
      \node (B) at (1.8,0) {$\tyStr$};
      \node (C) at (0,0) {$\alpha_{y}$};
      \draw[->] (A) -- (C) node[midway, above]{$\loc_1 \ \ \ \loc_0 \ \ \ \loc_4 \ \ \ \loc_3$};
      \draw[->] (B) -- (C) node[midway, above]{$\loc_3 \ \ \ \loc_5$};
    \end{tikzpicture}
  \end{center}
\end{example}
  \vspace{-.5em}

%%
%% Example: Level-2 Data Flow
%%
\begin{example}[Level-2 Data Flow]
  \label{ex:unification-provenance-2}
  Similarly, we annotate the locations for the program in \cref{fig:classification-examples:level-two} which yields the following annotated program:
  \begin{lstlisting}
    let g x$^{\loc_1}$ = (not x$^{\loc_2}$, (if true then x$^{\loc_4}$ else 5$^{\loc_5}$)$^{\loc_3}$)
  \end{lstlisting}
  We generate the following constraints and unification error for $\al_{x}$ and $\al_{ite}$\footnote{$\al_{ite}$ is the type for the if-then-else expression.}.
  The constraints are $\tyInt^{\loc_5} \sub \al_{ite}^{\loc_3}$, $\al_{x}^{\loc_4 \cdot \loc_1} \sub \al_{ite}^{\loc_3}$ and $\al_{x}^{\loc_1} \sub \tyBool^{\loc_2}$.
  We visualized the unification error $\tyInt \sub^{\loc_5 \cdot \loc_3} \al_{ite} \supp^{\loc_3 \cdot \loc_4 \cdot \loc_1} \al_x \sub^{\loc_1 \cdot \loc_2} \tyBool$ in the following way:

  \begin{center}
    \begin{tikzpicture}
      \node (A) at (-3.0,0) {$\tyInt$};
      \node (B) at (-1,0) {$\al_{ite}$};
      \node (C) at (1.5,0) {$\al_x$};
      \node (D) at (3.5,0) {$\tyBool$};
      \draw[->] (A) -- (B) node[midway, above]{$\loc_5 \ \ \ \loc_3$};
      \draw[<-] (B) -- (C) node[midway, above]{$\loc_3 \ \ \ \loc_4 \ \ \ \loc_1$};
      \draw[->] (C) -- (D) node[midway, above]{$\loc_1 \ \ \ \loc_2$};
    \end{tikzpicture}
  \end{center}
\end{example}

%%
%% Example: Constructor Data Flow
%%
\begin{example}[Constructor Data Flow]
  \label{ex:unification-provenance-3}
  We introduce the following program to demonstrate a nested data flow:
  \begin{lstlisting}
    let x$^{\loc_0}$ = true$^{\loc_1}$
    let y$^{\loc_2}$ = (if true then (x$^{\loc_4}$, "true")$^{\loc_6}$ else ("false"$^{\loc_5}$,"false")$^{\loc_7}$)$^{\loc_3}$
  \end{lstlisting}
  We consider constraints $(\al_l^{\loc_4} \tyProd^{\loc_6} \_) \sub \al_{y}^{\loc_3 \cdot \loc_2}$, $(\tyStr^{\loc_5} \tyProd^{\loc_7} \_) \sub \al_{y}^{\loc_3 \cdot \loc_2}$, $\tyBool^{\loc_1 \cdot \loc_0} \sub \al_l^{\loc_4}$. Unification gives an error for $\tyBool \sub^{\loc_1 \cdot \loc_0 \cdot \loc_4} \al_l \sim^{\pProdL{Z}} \tyStr$ where $Z = (\al_l^{\loc_4} \tyProd \_) \sub^{\loc_6 \cdot \loc_3 \cdot \loc_2} \al_{y} \supp^{\loc_2 \cdot \loc_3 \cdot \loc_7} (\tyStr^{\loc_5} \tyProd \_) \sub \al_{y}$.
  We elide the right type argument and irrelevant provenances for clarity.
  It is visualized in the following diagram:
  \begin{center}
    \begin{tikzpicture}
      \node (A) at (-2.1,0) {$\al_l^{\loc_4} \tyProd \_$};
      \node (B) at (2.0,0) {$\tyStr^{\loc_5} \tyProd \_$};
      \node (C) at (0,0) {$\alpha_{y}$};
      \node (DA) at (-2.4, -0.2) {};
      \node (EB) at (1.7, -0.2) {};
      \node (F) at (-4.6, -1.5) {$\tyBool$};
      \node (D) at (-2.4,-1.5) {$\al_l$};
      \node (E) at (1.7, -1.5) {$\tyStr$};
      \draw[->] (A) -- (C) node[midway, above]{$\loc_6 \ \ \ \loc_3$};
      \draw[->] (B) -- (C) node[midway, above]{$\loc_3 \ \ \ \loc_5$};
      \draw[->] (D) -- (DA) node[midway, left, above, rotate=90]{$\loc_4$};
      \draw[->] (E) -- (EB) node[midway, right, above, rotate=-90]{$\loc_5$};
      \draw[->] (F) -- (D) node[midway, above]{$\loc_0 \ \ \ \loc_1 \ \ \ \loc_4$};
    \end{tikzpicture}
  \end{center}
\end{example}
  \vspace{-.2em}

Converting this representation into a textual error message is straightforward. The sequence of program locations are shown vertically betwe
en lines describing intermediate types. The constructor data flow is shown with a horizontal offset corresponding to the height of the nested
data flow in the diagram. Our layout is not prescriptive and future implementations can
experiment with other layouts.

Notice that in this section, we now unify constructor arguments using the non-directional symbol $\sim$
instead of the directional $\sub$.
This is appropriate because some type constructors can be invariant in OCaml,
and even variant constructors use unification semantics during type inference anyway.
%
% On the other hand,
But since product, sum, function, and other types can still be considered variant even in ML
(because their type parameters are used only at one polarity in their definitions),
this can still be used to construct properly directional flows in type error explanations.
For instance, notice that in the diagram above, we use directional arrows in all the edges of the graph,
because $\otimes$ is covariant in its first argument. We would use a non-directional edge if the type constructor was invariant, for example, if it had been a mutable reference.
We can reconstruct the directionality of these variant flows by inspecting the nature of the type constructor
in the nested $\sim^{\Nprov{Z}{\tyProd \mid \tySum \mid ->}{L \mid R}}$ unification forms.

We also include a one-line flow summary (or \emph{outline}) in the survey error messages.
It was omitted from the introductory examples 
% since it is a layout-specific detail.
as it is not essential.
At a glance, it shows the user a high-level overview of the erroneous flow by stripping away the location info and using ASCII symbols to show the flow direction. \cref{fig:flow-summary} shows flow summaries for the examples discussed above.

\begin{figure}[h]
  \centering
  \begin{subfigure}{0.45\textwidth}
    \lstset{xleftmargin=30pt}
    \begin{lstlisting}
      (?a) ---> ($\ty$)
    \end{lstlisting}
  \caption{\cref{ex:unification-provenance-0}}
  \end{subfigure}
  \hfill
  \begin{subfigure}{0.45\textwidth}
    \begin{lstlisting}
      (int) ---> (?a) <--- (str)
    \end{lstlisting}
  \caption{\cref{ex:unification-provenance-1}}
  \end{subfigure}

  \vspace*{0.25em}
  \begin{subfigure}{\textwidth}
  \centering
    \lstset{xleftmargin=80pt}
    \begin{lstlisting}
      (int) ---> (?a) <--- (?b) <--- (str)
    \end{lstlisting}
  \caption{\cref{ex:unification-provenance-2}}
  \end{subfigure}

  \vspace*{0.25em}
  \begin{subfigure}{\textwidth}
  \centering
    \begin{lstlisting}
      (bool) ---> (?a) ~~~> (?a * _) ---> (?b) <--- (str * _) <~~~ (str)
    \end{lstlisting}
  \caption{\cref{ex:unification-provenance-3}}
  \end{subfigure}
  \vspace*{0.01em}
  \caption{Flow summary for data flows shown in \cref*{subsubsec:error-reports}.}
  \vspace{-10pt}
  \label{fig:flow-summary}
\end{figure}

% \section{Qualitative Evaluation}
% \label{sec:evaluation}
% \input{qualitative-evaluation.tex}

% TODO after the ICFP/ECOOP deadline
% \section{Towards Interactive Type Error Debugging}
% \label{sec:interactive-errors}
% \input{interactive-errors.tex}

\section{Let Polymorphism}
\label{sec:let-poly}
In this section, we introduce the notion of polymorphism level to support let-polymorphism
in \system, following
\citet{Remy-inria92:extension-ml-sorted-eqn-thry-on-tys},
who created this technique for the Caml/OCaml compilers \cite{kiselyov:efficient-gen-levels}.\footnote{
  \citet{Parreaux20:simple-essence-alg-subt}
  generalized the idea to the context of algebraic subtyping in \emph{subtype extrusion}
  and later formalized it
  \cite{parreaux-superf-2024}.
  % However, 
  In this paper,
  we use the simpler algorithm that works for unification
  (not subtyping)
  for simplicity.
}

We extend the syntax in \cref{fig:extended-type-syntax-letpoly},
to include let bindings and \kw{Substitutions} which map type variables to type variables.
Polymorphism levels are natural numbers and the \kw{levels} field in the typing
state maps a type variable to a level. We extend the typing context to store a type $\ty$
of a polymorphism level $i$ as $\forall_i \ty$. We also introduce a few helper functions
on the typing state, namely:
\begin{itemize}
  \item $\newlvl{\al}{i}{\sigma} = \sigma'$: New typing state $\sigma'$ has $\al \mapsto i$ added
    to it or updated, if $\sigma$ already associated a level to $\al$.
  \item $\appsubst{S, \sigma}{i} = \sigma'$: New state $\sigma'$ has all
    substitutions applied to old state $\sigma$ as described in \cref{fig:helper-apply-substitution}.
    For each substitution $\al \mapsto \al'$, all the constraints of $\al$ are duplicated where $\al$
    is substituted with $\al'$
    while preserving provenance information. A level mapping $\al' \mapsto i$ is also added
    to the new state.
  \item $\lvl{\ty} = i$: The polymorphism level of the type $\ty$ is $i$ for a given state $\sigma$.
    The semantics is described in \cref{fig:helper-level}. A primitive type has
    level 0 and a constructor type's level is the maximum level of its constituent types.
    Type variables are associated with their own polymorphism levels. We use
    record-dot style syntax like $\sigma.\mathtt{bounds}$ to access the bounds
    associated with the state.
\end{itemize}

\begin{figure*}[ht]
  {\small
  \begin{align*}
        %
        % Terms
        %
        \graydescr{Term}       && e      & \Coloneqq ... \mid \letdefe{x^\loc}{e}{e} \mid \letrecdefe{x^\loc}{e}{e}\\
        %
        % Substitutions
        %
        \graydescr{Substitutions}       && S    & \Coloneqq\ \overline{\al \mapsto \al}\\
        %
        % Terms
        %
        \graydescr{Level}       && i      & \Coloneqq \text{natural number}\\
        \graydescr{Context}    && \Gamma & \Coloneqq\ {\emptyCtx} \mid {\Gamma\ctxPush{x: \al}} \mid {\Gamma\ctxPush{x: \forall_i\,\ty}}\\
        %
        % State
        %
        \graydescr{State}      && \sigma & \Coloneqq\ \{\ \mathtt{bounds}: {\,\overline{\overline{\tau} \sub \alpha \sub \overline{\tau}}}
                                                ,\ \mathtt{levels}:  {\,\overline{\al \mapsto i}}
                                                ,\ \mathtt{errors}:  {\,\overline{\prov}}
                                              \ \}\\
  \end{align*}}
  \caption{Extended syntax for let-polymorphism.}
    \label{fig:extended-type-syntax-letpoly}
\end{figure*}

\begin{figure*}[ht]
  {\small
\begin{align*}
  \appsubst{[], \sigma}{i} &= \sigma\\
  \appsubst{\al \mapsto \al' :: S, \sigma}{i} &= \appsubst{S, \sigma'}{i}
    &\text{where } \sigma &= \{\ \mathtt{bounds} : b,\ \mathtt{levels}: l,\ \mathtt{errors}: e\ \}\\
                &&b' &= [\ \al' \sub^\prov \ty \mid \ty^\prov \in \fname{ub}(\sigma, \al)\ ] ::: [\ \ty \sub^\prov \al' \mid \ty^\prov \in \fname{lb}(\sigma, \al)\ ] ::: b\\
                &&\sigma' &= \{\ \mathtt{bounds} : b',\ \mathtt{levels}: \al' \mapsto i :: l,\ \mathtt{errors}: e\ \}\\
\end{align*}}
\caption{Application of type substitution to typing state.}
  \label{fig:helper-apply-substitution}
\end{figure*}

\begin{figure*}[ht]
  {\small
\begin{align*}
  \lvl{\al} &= i &&\text{where } \al \mapsto i \in \sigma.\mathtt{bounds}\\\
  \lvl{\ty \odot \ty'} &= \text{max}(\lvl{\ty}, \lvl{\ty'})\\
  \lvl{\ty} &= 0 &&\text{where } \ty \in \{\ \tyUnit, \tyInt, \tyBool\ \}
\end{align*}}
\caption{Level of a type.}
  \label{fig:helper-level}
\end{figure*}

We extend the typing rules, with the notion of polymorphism level. All top level
let bindings and terms start with polymorphism level 0. \cref{fig:typ-rules-2-letpoly}
shows only the relevant rules.
\begin{description}
  \item[\truleName{PolyType}] This rule looks up a polymorphic type from the typing context
    and \kw{freshen}s it. \kw{freshen} collects a substitution $S$ which is
    applied to typing state $\sigma$. In practice, this operation only
    affect types whose polymorphism level is lesser or equal to the current level $i$.
  \item[\truleName{Let}] This is a new rule that types let expressions. It increments
    the polymorphism level before typing the let-bound expression.
  \item[\truleName{Let-Rec}] This is rule types let-rec expressions and is similar. It increments
    to the \truleName{Let} rule. The key difference is that it creates a fresh type
    variable $\al$ for the binding.
\end{description}

\begin{figure}[ht]
  {\small
  \begin{mathpar}
    \boxed{\strut \sigma \shortmid\ i, \Gamma \vdash e : \ty\ \shortmid \sigma}
    \hfill
    \\
    %%%%%%%%%%%%%%%%%%%%%%%%%%%%%%%%%%%%%%%%%%%%%%%%%%%%%%%%%%%%%%%%%%%%%%%%%%%%%%%
    %% T-Var
    %%%%%%%%%%%%%%%%%%%%%%%%%%%%%%%%%%%%%%%%%%%%%%%%%%%%%%%%%%%%%%%%%%%%%%%%%%%%%%%
    \inferrule[\trule{PolyType}]{
      \Gamma(x) = \forall_j\ \ty\\
      (\ty', S) = \freshen{\ty, []}{\sigma_0}{j}\\
      \sigma_1 = \appsubst{S, \sigma_0}{i}
    }{
      \sigma_0 \shortmid i, \Gamma \vdash x^\loc : \ty'^\loc \shortmid \sigma_1
    }\ x\in\dom\Gamma

    %%%%%%%%%%%%%%%%%%%%%%%%%%%%%%%%%%%%%%%%%%%%%%%%%%%%%%%%%%%%%%%%%%%%%%%%%%%%%%%
    %% T-Let
    %%%%%%%%%%%%%%%%%%%%%%%%%%%%%%%%%%%%%%%%%%%%%%%%%%%%%%%%%%%%%%%%%%%%%%%%%%%%%%%
    \inferrule[\trule{Let}]{
      \tyInfJdgmt{\sigma_0}{i + 1, \Gamma}{e_1}{\ty}{\sigma_1}
      \\
      \tyInfJdgmt{\sigma_3}{i, \Gamma\ctxPush{x:\forall_i\ty}}{e_2}{\ty'}{\sigma_4}
    }{
      \sigma_0 \shortmid i, \Gamma \vdash \letdefe{x^\loc}{e_1}{e_2}: \ty' \shortmid \sigma_4
    }

    %%%%%%%%%%%%%%%%%%%%%%%%%%%%%%%%%%%%%%%%%%%%%%%%%%%%%%%%%%%%%%%%%%%%%%%%%%%%%%%
    %% T-Let-Rec
    %%%%%%%%%%%%%%%%%%%%%%%%%%%%%%%%%%%%%%%%%%%%%%%%%%%%%%%%%%%%%%%%%%%%%%%%%%%%%%%
    \inferrule[\trule{Let-Rec}]{
      \al\ \kw{fresh}
      \\
      \sigma_1 = \newlvl{\al}{i + 1}{\sigma_0}
      \\
      \tyInfJdgmt{\sigma_1}{i + 1, \Gamma\ctxPush{x:\al}}{e_1}{\ty}{\sigma_2}
      \\
      \consJdgmt{\sigma_2}{\ty \sub \al^\loc}{\sigma_3}
      \\
      \tyInfJdgmt{\sigma_3}{i,\Gamma\ctxPush{x:\forall_i\al}}{e_2}{\ty'}{\sigma_4}
    }{
      \sigma_0 \shortmid i, \Gamma \vdash \letrecdefe{x^\loc}{e_1}{e_2}: \ty' \shortmid \sigma_4
    }
  \end{mathpar}}
  \caption{Extended type inference rules for let-polymorphism.}
    \label{fig:typ-rules-2-letpoly}
\end{figure}

\begin{comment}

-- at level 0:
let x = t1 in
let y = t2 x x in
...

\end{comment}

The semantics for \kw{freshen} are described in \cref{fig:helper-freshen}.
Freshening a type means traversing it and mapping all the type variables
of lesser or equal polymorphism level to fresh type variables.
This creates a substitution $S$ which when used like $\appsubst{S, \sigma}{i}$
specializes the type up to polymorphism level $i$.

\begin{figure*}[ht]
  {\small
\begin{align*}
  %
  % Create fresh variable
  %
  \freshen{\al, S}{\sigma}{i} &= (\beta, \al \mapsto \beta :: S)
    &\text{where } &\al \notin \fname{dom}(S),\ \al \mapsto i' \in \sigma.\mathtt{bounds},\ i' \le i,\ \beta\ \kw{fresh}\\
  %
  % Apply mapping
  %
  \freshen{\al, S}{\sigma}{i} &= (\beta, S)
    &\text{where } &\al \in \fname{dom}(S),\ S(\al) = \beta\\
  %
  % Constructor type
  %
  \freshen{\ty \odot \ty', S}{\sigma}{i} &= (\delta \odot \delta', S'')
    &\text{where } &(\delta, S') = \freshen{\ty, S}{\sigma}{i}\\
    &&           &(\delta', S'') = \freshen{\ty', S'}{\sigma}{i}\\
  %
  % No action needed for all other cases
  %
  \freshen{\ty, S}{\sigma}{i} &= (\ty, S)
    &\text{where } &\ty \in \{{\ \tyInt, \tyBool, \tyUnit\ }\}
\end{align*}}
\caption{Freshening of types above a polymorphism level.}
  \label{fig:helper-freshen}
\end{figure*}

Most constraining rules are unaffected by polymorphism level.
However, we add a refinement in \cref{fig:constraining-rules-extended-letpoly}.
\begin{description}
  \item[\cruleName{R-Extr}] The type variable is constrained with a type of higher polymorphism
    level, as its upper bound. The type is extruded to the level of the
    type variable before proceeding with the constraint.
  \item[\cruleName{L-Extr}] 
    This rule is similar to \cruleName{R-Extr} but for type variable lower bounds instead of upper bounds.
\end{description}

A type behaves polymorphically only when its polymorphism level
is higher than that of the typing context. In such cases, freshen clones
the type upto the level of the typing context. However, we don't want it to
behave polymorphically below its level. To ensure this, we use \kw{extrusion} described in \cref{fig:helper-extrude}.
It reduces the polymorphism of a type variable upto a level. It also does this
for all its upper and lower bound types, including those reachable transitively.
The extrude function also maintains a cache for visited type variables to break cycles.

\begin{figure}[ht]
  {\small
  \begin{mathpar}
    \boxed{\strut\sigma \shortmid i, \cons{Q}^H  \shortmid \sigma }
    \hfill
    \\
    %
    % Var Left Right
    %
    \inferrule[\crule{R-Extr}]{
      i_\al = \lvlqi{\al}{\sigma_0}\\
      \extrudeH{\ty, \sigma_0}{\{\}}{i_\al} = \sigma_1\\
      \shadeout{\sigma_1 \shortmid\ } i, \text{cons}(\al^{\prov_0} \sub \ty^{\prov_1}) \shadeout{\ \shortmid \sigma_2}
    }{
      \sigma_0 \shortmid i, \cons{ \al^{\prov_0} \sub \ty^{\prov_1} }^H \shortmid \sigma_2
    }{
      \ \lvlqi{\al}{\sigma_0} < \lvlqi{\ty}{\sigma_0}
    }
    
    \inferrule[\crule{L-Extr}]{}{(Similar)}
  \end{mathpar}
  }
  \caption{Extension to the constraint solving algorithm.}
    \label{fig:constraining-rules-extended-letpoly}
\end{figure}

\begin{figure*}[ht]
  {\small
\begin{align*}
  % no types to extrude
  \extrude{[], \sigma}{i} &= \sigma
  \\
  % type variable already extruded
  \extrude{\al :: \overline{\ty}, \sigma}{i} &= \extrude{\overline{\ty}, \sigma}{i}
    &\text{where }
      & \al \in H
  \\
  % type variable not extruded but lower than level i
  \extrude{\al :: \overline{\ty}, \sigma}{i} &= \extrude{\overline{\ty}, \sigma}{i}
    &\text{where }
      & \lvl\al \le i
  \\
  \extrude{\al :: \overline{\ty}, \sigma}{i} &= \extrudeH{\overline{\ty}'', \sigma'}{H\ \cup\ \{\al\}}{i}
    &\text{where }
      &\overline{\ty}' = [\al' \mid \al' \in \fname{lb}({\sigma, \al})] ::: \overline{\ty}\\
    &&&\overline{\ty}'' = [\al''' \mid \al'' \in \fname{ub}({\sigma, \al})] ::: \overline{\ty}'\\
    &&&\sigma' = \newlvl{\al}{i}{\sigma}
  \\
  \extrude{\ty \odot \ty' :: \overline{\ty}, \sigma}{i} &= \extrude{\overline{\ty}, \sigma'}{i}
    &\text{where }
      &i < \lvl{\ty \odot \ty'}\\
    &&&\sigma' = \extrude{\ty}{\sigma}\\
    &&&\sigma'' = \extrude{\ty'}{\sigma'}
  \\
  \extrude{\ty :: \overline{\ty}, \sigma}{i} &= \extrude{\overline{\ty}, \sigma}{i}
    &\text{where }
      &\ty \in {\{\ \tyInt, \tyBool, \tyUnit\ \}}
\end{align*}}
\caption{Extrusion of a higher level type to a lower level.}
  \label{fig:helper-extrude}
\end{figure*}

\subsubsection{Extrusion example}
With these changes programs with both \emph{top-level} and \emph{local} let-polymorphism
type check. The program in \cref{fig:extrusion-example} demonstrates how polymorphism is lost by
extrusion.

The function $g$, initially, appears polymorphic and the program appears valid.
However, in the body of $g$ the parameter $y$ is applied to $x$ which is
introduced by $f$. When typing the expression $x y$, we encounter a constraint like
$\consJdgmt{\sigma_0}{ \al_x \sub (\al_y -> \al')}{\sigma_1}$ where ${\al_x \mapsto 0, \al_y \mapsto 1} \in \sigma_0$,
$\al'$ is some fresh type variable, and all other symbols have their usual meanings. This constraining rule
extrudes $\al_y$ to polymorphism level 0 with \crule{R-Extr} rule.
Now $\al_y$ is no longer polymorphic at level 1 and the program is no longer valid.
In \cref{fig:extrusion-example-error}, we see the \system for the program.

\begin{figure}
\begin{lstlisting}[style=snippet]
let f x =
  let g y =
    let h = x y
    in y
  in let k = g 1 in g "hi"
\end{lstlisting}
\caption{Extrusion example.}
\label{fig:extrusion-example}
\end{figure}

\begin{figure}
\begin{Error}{\system{}}
\begin{lstlisting}[language=error]
[ERROR] Type `string` does not match `int`

        (string) ---> (?a) <--- (int)

◉ (string) comes from
│  - l.5    in let k = g 1 in g "hi"
▼                               ^^^^
◉ (?a) is assumed for
▲  - l.2    let g y =
│                 ^
◉ (int) comes from
   - l.5    in let k = g 1 in g "hi"
                         ^
\end{lstlisting}
\end{Error}
\caption{Error message for incorrect usage of function $g$.}  
\label{fig:extrusion-example-error}
\end{figure}

\section{Experimental Evaluation}
\label{sec:user-study}
In this paper, we hypothesize that flow-based type error messages are more effective than location-based type error messages in helping programmers understand errors.
Traditional error messages only provide %information on
\emph{one} of the possible locations related to each error,
which can be useful to understand \emph{where} a program goes wrong
but is often insufficient to understand the full context of the error.
In contrast, our proposed system presents the flow of types, which intends to help understand
\emph{how} a program is goes wrong.
Traditional systems thus focus on locality, while we attempt to facilitate causal reasoning,
which we believe is essential to effectively understand and repair errors.

We conduct a randomized quasi-experimental study to validate our hypothesis. The experiment compares the understanding of programmers using flow-based type error messages with those using traditional location-based error messages.

In the user study, we ask participants to understand and describe program errors, measuring \emph{a)} the perceived satisfaction of the participants with the provided error messages and \emph{b)} whether the participant sufficiently understood the error.

% One canonical language (today) that implements Hindley-Milner type
% inference is OCaml. It is widely used in industry and academia.
% We have implemented DFEs as an alternative to the type inference for OCaml
% programs.
% To investigate the research questions, we want to compare the performance of
% programmers that we assign into one of the following groups:

% 1. treatment group (A): programmers are shown DFEs and have to answer questions / solve tasks.
% 2. control groups (B, C, D): programmers are shown traditional OCaml error messages (B) (or other state-of-the-art systems (C) and (D)) and have to answer questions / solve tasks.

% Problem: what if reviewers will see the data collection prompt on Twitter etc.?

\subsection{Experiment Design}
The experiment has been conducted in form of an online survey (using lab.js \cite{henninger2021lab} and hosted on Open Lab \cite{shevchenko2022open}). After a demographic questionaire and a short introduction to OCaml,
participants were presented with a series of errornous OCaml programs. Provided with the program and an error message (side-to-side), participants were invited to understand the error using the provided error message
followed by a series of questions:
\begin{itemize}
    \item[Q1] ``In your own words: what is the problem in the program above?''
    \item[Q2] ``How much did the error message help you to locate the problem''
    \item[Q3] ``How much did the error message help you to understand the problem''
\end{itemize}
The first question (Q1) was asked as free form text asking participants to keep the answer short. The remaining two questions (Q2 and Q3) were answered on a five-point Likert scale ranging from ``Not helpful'' to ``Very helpful''.
The survey ended with a single optional free form text field asking users ``Is there anything you want to tell us?''.
% We submit an offline version of the user study as part of supplementary material.
%\jonathan{What do we do for the camera ready there?}

\paragraph{Conditions}
When starting the survey, each participant was randomly assigned (drawn without replacement) to one of three conditions:

\begin{itemize}
    \item[(A)] \textit{\system{}} -- our implementation of \system{}, extending SystemSub with flow-based type errors.
    \item[(B)] \textit{OCaml} -- as a first control group, we compare against the standard OCaml compiler.
    \item[(C)] \textit{Helium} -- as a second control group, we compare against Helium \citep{Heeren2003}.
\end{itemize}

Helium is a compiler for Haskell, not OCaml,  but the subset of programs we consider can be easily translated from OCaml to Haskell.
We therefore translated the OCaml examples to Haskell by hand, used Helium to generate an error message, and translated the types contained in the resulting error message back to use OCaml Syntax.

\paragraph{Selection of programs}
We prepared ten different ill-typed example programs that we manually ranked as ``easy'' (3), ``medium'' (4), or ``hard'' (3), see \cref{sec:appendix:examples} for full details of the example programs.
Programs labeled as ``easy'' were constructed specifically for the study, while examples labeled ``medium'' or ``hard'' were
selected from the datasets shared by \cite{Seidel2017}.
From each of the three categories, each participant was presented with two randomly sampled example programs; that is, six programs in total.
While the order of examples within each category was random; the categories itself have always been presented in ascending complexity.

\subsection{Participants}
We shared the online survey with professionals and researchers by posting it on relevant online platforms
% \jonathan{do we need to be more specific here?}
(such as Reddit and Twitter).
%To control for programmer expertise, participants were asked for their professional status,
Participants were asked to self-estimate their experience in programming in general, functional programming, statically typed programming, and programming in OCaml on a 5-point Likert scale  \cite{feigenspan2012measuring}.
From a total of 455 participants, 318 started and 119 concluded the survey (40 \system{}, 39 OCaml, 40 Helium). We manually excluded two participants (both in the OCaml condition) since they visibly did not invest enough effort to answer the questions.
Of the 117 non-excluded participants, 70 assigned themselves an expertise $>= 4$ for ``functional programming'' or ``OCaml''.
Only 14 participants did assign themselves an expertise $\le{} 3$ for all categories.

% \jonathan{Do we need to report that we lost data on professional status?}
% \lionel{I don't think so.}

\subsection{Evaluation}
The study sets out to analyse whether the error-message condition (\system{}, OCaml, or Helium) significantly influences \emph{a)} the perceived usefulness of the error messages, as well as \emph{b)} the understanding of the programming error.
We first discuss the evaluation of perceived usefulness (Q2 and Q3) before discussing the evaluation of the open question (Q1).

After data collection, we realized a mistake made while hand-translating Helium error message for \texttt{hard2} back to OCaml types. It left a Haskell formatted list type, which might have been confusing to participants and led to poor responses. Thus we exclude example \texttt{hard2} from our evaluation.

\begin{figure*}[t]
  \small
  \centering
  \emph{Q2: ``How much did the error message help you to locate the problem?''}
  \includegraphics[width=\textwidth]{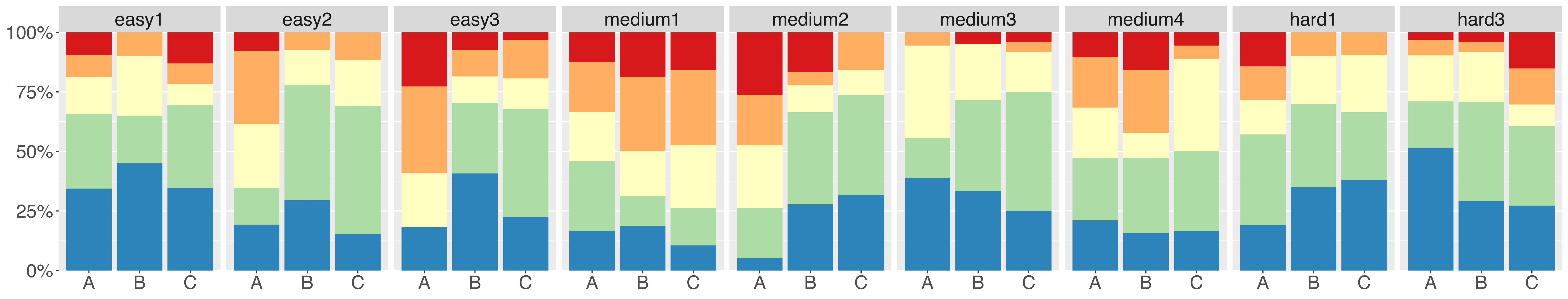}
  \emph{Q3: ``How much did the error message help you to understand the problem?''}
  \includegraphics[width=\textwidth]{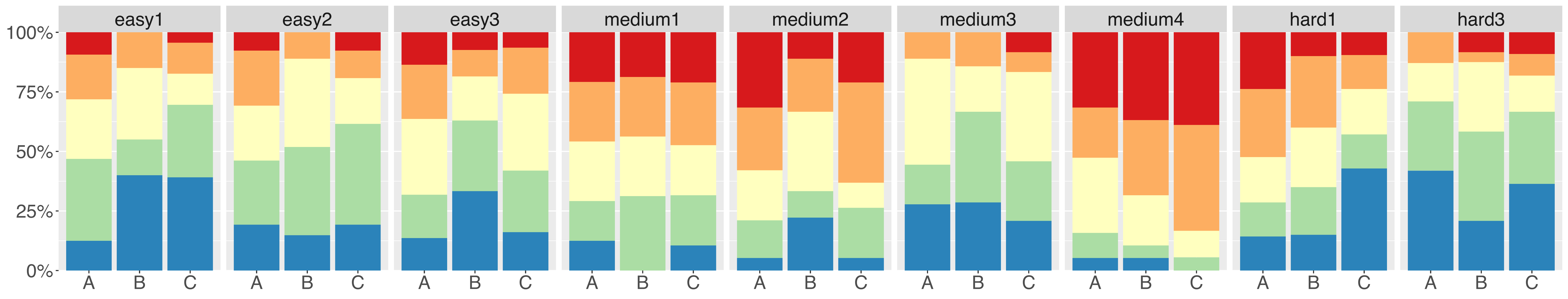}
  \caption{Participants answering the respective question on a five-point Likert scale from ``Not helpful'' (top, red) to ``Very helpful'' (bottom, blue). We compare conditions \system{} (A), OCaml (B), and Helium (C). \label{fig:results:perceived}}
\end{figure*}

\subsubsection{Perceived Usefulness (Q2 and Q3)}
For each of the ten individual example programs, and each of the three conditions (A), (B), and (C), Figure \ref{fig:results:perceived} presents the results for the perceived usefulness reported by participants.
The five-point Likert-scaled data is visualized as stacked bar charts, where the lowest (dark-blue) component corresponds to the answer ``Very helpful'', and the highest (dark-red) component corresponds to the answer ``Not helpful''.

\paragraph{Locating the problem (Q2)}
Performing a Kruskal-Wallis rank sum test \citep{kruskall1995ranks} for each of the ten tasks,
we can find significant differences for \texttt{easy2} ($p=0.0119,\chi^2=8.87$), \texttt{easy3} ($p=0.0025,\chi^2=11.96$), \texttt{medium2} ($p=0.0067,\chi^2=10.015$).
Individual test results are available in \cref{sec:appendix:examples}.
\cref{table:1} lists the full results for all performed Kruskal-Wallis  rank sum tests. All participants correctly described the error for program \texttt{easy1}.
\begin{table}
\caption{Results of all Kruskal-Wallis tests per program.}
  \centering
  \begin{threeparttable}
  \small
  \begin{tabular}{lrrrrrr}
      \toprule
                                   & \multicolumn{2}{c}{\textsf{\textbf{Q1 Describe}}} & \multicolumn{2}{c}{\textsf{\textbf{Q2 Locate}}} & \multicolumn{2}{c}{\textsf\textsf{\textbf{Q3 Understand}}}  \\
     \textsf{\textbf{Program}}   & $p$ & $\chi^2$                                       & $p$ & $\chi^2$                                     & $p$ & $\chi^2$  \\ \midrule
{\tt easy1} & N/A & N/A & $0.7748$ & $0.51$ & $0.0805$ & $5.04$ \\
{\tt easy2} & $0.8808$ & $0.25$ & $0.0119$ & $8.87$ & $0.6259$ & $0.94$ \\
{\tt easy3} & $0.3015$ & $2.40$ & $0.0025$ & $11.95$ & $0.0860$ & $4.91$ \\
{\tt medium1} & $0.8943$ & $0.22$ & $0.5009$ & $1.38$ & $0.9988$ & $0.00$ \\
{\tt medium2} & $0.7996$ & $0.45$ & $0.0067$ & $10.02$ & $0.1727$ & $3.51$ \\
{\tt medium3} & $0.4535$ & $1.58$ & $0.9386$ & $0.13$ & $0.5235$ & $1.29$ \\
{\tt medium4} & $0.9042$ & $0.20$ & $0.6442$ & $0.88$ & $0.3943$ & $1.86$ \\
{\tt hard1} & $0.8190$ & $0.40$ & $0.2410$ & $2.85$ & $0.0632$ & $5.52$ \\
% {\tt hard2} & $0.0010$ & $13.84$ & $0.0007$ & $14.65$ & $0.0000$ & $43.66$ \\
{\tt hard3} & $0.4928$ & $1.42$ & $0.1257$ & $4.15$ & $0.3650$ & $2.02$ \\
  \bottomrule
  \end{tabular}
  \end{threeparttable}

\label{table:1}
\end{table}

To determine which groups are different, we perform a post-hoc Dunn test \citep{dunn1964multiple} for each of the significant example programs.
A Bonferonni adjustment of the $p$-value is used to account for the error rate introduced by the pair-wise comparison.
% ; that is, $p$-values indicate singificant differences when $\alpha / 2 = 0.05 / 2 = 0.0025$.
For \texttt{easy2}, we have significant differences for $A\!-\!B$ with $-2.934363$ ($p=0.0050$). That is, participants reported messages of \system{} to help less with locating than those of OCaml.
For \texttt{easy3}, we have significant differences for $A\!-\!B$ with $-3.2893$ ($p=0.0015$) and
$A\!-\!C$ with $-2.7449$ ($p=0.0091$). That is, participants reported messages of \system{} to help less with locating than those of OCaml or Helium.
For \texttt{medium2}, we have significant differences $A\!-\!C$ with $-3.018982$ ($p=0.0038$). That is, participants reported messages of \system{} to help less with locating than those of Helium.

\begin{figure*}
  \small
  \centering
  \emph{Q1: ``In you own words: what is the problem in the program above?''}
  \includegraphics[width=\textwidth]{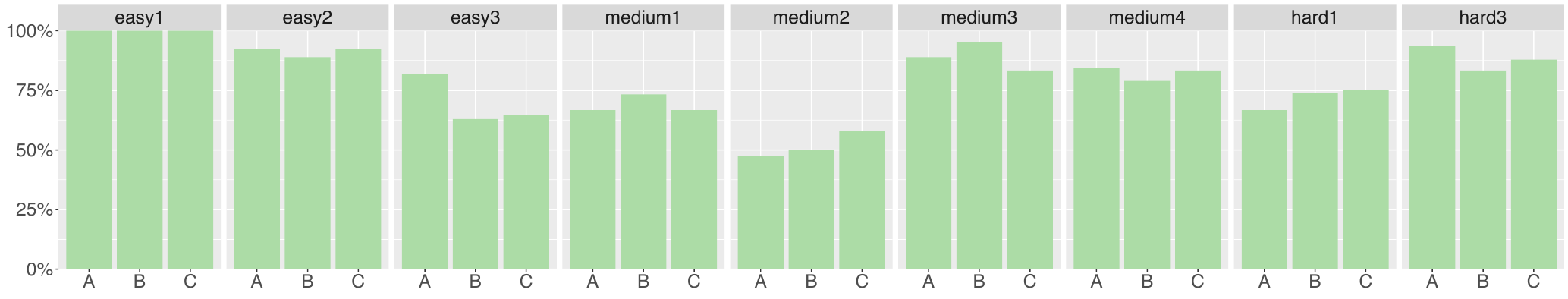}
  \caption{Percentage of participants that correctly described the problem. We compare conditions \system{} (A), OCaml (B), and Helium (C). \label{fig:results:measured}}
  \vspace{-10pt}
\end{figure*}

\paragraph{Understanding the problem (Q3)}
Similar to Q2, we perform a Kruskal-Wallis test, however there is no significant difference between the results of the three systems.

% The Bonferroni Adjustment: Adjusted p-value = p*m
%   where:
%     p: The original p-value
%     m: The total number of comparisons being made

\subsubsection{Understanding Errors (Q1)}
To evaluate the open question Q1, we manually assigned a binary grade to the provided textual answers, judging whether the participant understood the underlying problem or not. Emphasis was given on participants' understanding of erroneous program expressions and the fixes they suggested.
To avoid biases, the manual coding was performed blind---that is, the condition of a participant during evaluation was hidden to the grader. The same grader graded the responses from all participants.
Figure \ref{fig:results:measured} reports the results for the three conditions, again grouped per example program.

\subsection{Interpretation}
Our experiment yielded little statistically significant difference between respondents using different tools to locate and understand errors. In general, the study cannot show that \system{} improves understanding or localization of problems.

In a few cases, other systems even performed significantly better than \system{}.
For Q2, which measured the perceived value of the error message in locating the error, \texttt{easy2}, \texttt{easy3}, and \texttt{medium2} showed significant results. In all three cases, participants reported that \system{} error messages helped less in locating the problem. All three programs (especially \texttt{easy2} and \texttt{easy3}) are comparatively small. Constructing (and consuming) a detailed data flow explanation that is longer than the programs might not pay off in complexity.
For Q3, which measured the perceived value of the error message in understanding the error, no results were significant.

A qualitative analysis of the freeform feedback provided by participants reveals interesting insight into why \system{} might have performed worse than OCaml and Helium in \emph{locating} errors.

\paragraph{Errors are too verbose or unnecessary for small programs}
Many participants reported the \system{} errors to be excessively verbose, which was detrimental to understanding errors for \emph{small} programs.
\begin{quote}\it
%hcpibnpalfkb5oqy
The earliest examples were so trivial I could've figured them out without error messages.
\end{quote}
%
% %
% \begin{quote}\it
% %hcpibnpalfk2pqyn
% Your error messages seemed very verbose, though that seems to be the style these days.
% \end{quote}
However, some participants also reported that they were indeed useful for longer problems.
\begin{quote}\it
%hcpibnpalfjx8sri
I felt that the error messages were more helpful on the longer problems, only because I didn't need to look at them on the shorter problems.
\end{quote}

\noindent The problem of errors that are too verbose could be remedied by designing better layouts or using heuristics to hide some of the extra locations
for simple programs.
\noindent Furthering the previous point, in many of the simple cases (in \emph{easy} and \emph{medium} programs),
  the respondents were frustrated by the comparatively large number of locations reported by \system{}
  and instead just wanted to see \emph{``the'' location where the error happened}.
  We conjecture that it would be helpful to combine our work with previous research on identifying single error locations and provide an ``\emph{incipit}'' of sorts before the precise data flow---reporting a single location, which is often enough to understand simple errors.
  This way, programmers would only look at the precise data flow when they need to obtain more context
  and understanding for the root causes of the error.
  In an interactive setting the additional information provided by flow errors could be presented on request of the user.
While the participant's concerns are important and it would be interesting to investigate more compact representations, real-world programs
are almost always larger than the programs surveyed in our study.

\paragraph{Error message layout and notation}
Respondents also complained about the error message layout as well as other aspects of the used notation, including the data-flow arrows, line number formatting, ellipsis for large code and so on.
\begin{quote}\it
%hcpibnpalfkb5oqy
Many of the error messages I was shown seemed upside-down, in that they showed what I believe to be the source of the error at the bottom, with a stack of more-removed sources of incompatible constraints extending upwards. This might look better coming from a CLI, but in this format it was weird and unhelpful.
\end{quote}

\noindent However, other partipants appreciated the structure of \system{}'s error messages:
\begin{quote}\it
%hcpibnpalfkte6to
In general, I really like the detail and consistency in the error messages. This really helps with solving more subtle errors, but it also adds a lot of noise for more simple issues. I feel like this isn't a real problem since even a little experience will allow you to immediately identify the issues at a glance of the error message by looking at the right things.
One possible improvement could be to mention (and preview) the offending statement with the relevant parts marked before going in depth on the breakdown of the type interpretation. As it is now, one part of the breakdown hides the other conflicting part of the statement
\end{quote}

\noindent
There is ample opportunity to improve all aspects of the presentation of error messages, including the data-flow summaries, the textual explanation, and the graphical layout. For each of the components, the usefulness is still unclear, the concrete design can be improved, and the
interplay between the different aspects needs to be studied.
More research and testing is needed to develop effective error message layouts for data-flow style reasoning.

\noindent
One factor that might have contributed to the confusion of participants is that many of our respondents were experienced practitioners who were accustomed to message layout from standard tools.
We also conjecture that experienced programmers already developed a deep understanding for \emph{how} type-checking in OCaml proceeds and thus learnt to infer information from existing error messages. It naturally requires some time to adapt to the new format of error messages, as also one participant described:
\begin{quote}\it
%hcpibif0lfooolxp
The error message has all the information, it takes some time to get used to it though.
\end{quote}
Similarly, many respondents found the unification variable shown in \system{} messages unhelpful. The unexplained ``\lstinline+?a+'' notation used for unification variables combined with the new concept of data flows was confusing to these participants.
We conjecture that a gentle introduction to \system{}'s notation,
  as well as familiarity built over time, could potentially remedy these issues.

\paragraph{Flow-based reasoning}
The work presented in this paper builds upon the assumption that the underlying mental model (also called ``notional machine'' by \citet{boulay1981notional}) of flow-based reasoning is natural and can help programmers to understand and locate error messages. A possible interpretation of the below user feedback is that this assumption is false.
\begin{quote}\it
I honestly find the `\lstinline+int --> ?a <-- bool+' notation quite confusing. It is useful in some cases where there is no obvious expected or actual type, but in cases where the unification variable is unnecessary, it adds quite a bit of unnecessary mental overhead.
\end{quote}
Maybe understanding how values (and types) flow through a program does not contribute to the understanding of type errors. Maybe guiding users along the data flow is not helpful afterall, since they could also follow the data flow themselves without the overhead of processing verbose error messages with positions marked by unification variables that are not part of the original program text. Our work aims to \emph{support} users in this process, which only makes sense if the process itself is practically useful.

\paragraph{No specialized messages}
OCaml and Helium had specialized messages for certain errors. Messages like ``function applied to too many arguments'' (\texttt{hard1}) and ``expected type \ldots'' seem more helpful than just data flow information. Of course, \system{} could be specialized to also add such helpful text to error messages, which are orthogonal to the idea of presenting data flow.

\subsubsection{The Need for Detailed Explanations}
While some participants remarked that the messages of \system{} are too verbose, those in the control groups often remarked the opposite
about messages generated by OCaml and Helium.

\begin{quote}\it
%hcpibkllg41h672
I think in many of the examples it would be helpful for the errors to explain where the constraint was introduced that we are hitting [...].
\end{quote}
Another participant recognized the difficulty of extensive inference:
\begin{quote}\it
%hcpibnpalfkh4hqg
Presumably some of the harder ones were caused by extensive inference throughout the code. It might help to show and look at multiple errors in such cases. Perhaps there's a way to group them, but this is hypothetical.
\end{quote}
Yet another participant would like to see more type information of the different components that constitute a problematic call.
\begin{quote}\it
%hcpibnpalfl2sw5u
The error messages are not at all clear about what the expected and what the found type is. Also, it is not clear why it believes the found type is the found type.
Maybe it could show the different elements that are applied to the function separately as well?
\end{quote}
Participants were also aware of the difficulty to trade-off concise error messages and sufficiently detailed information.
\begin{quote}\it
%hcpibnpalfkuvg9f
Almost every type error in a program is an accumulation of multiple (smaller) erroneous parts. Any such error that states "something happened exactly here and here" is incomplete, because there's far more context that should be included to fully understand the error.
Some languages show the entire chain of type unification that led to the error, but that's rather verbose. I hope one can find a solution that shows just enough context to be perfectly helpful.
\end{quote}

% hcpibdzslg0pnret
% I feel like the example programs were pretty hard to understand in the first place, so it's not surprising that errors were hard to understand too.

The above quotes are only a few examples of the feedback that we received by the control groups.
In many cases, respondents want to know source expressions for the two conflicting types,
instead of the singular location where the error was triggered. They also want to see surrounding source code in the error message to gain context.

\system{}'s error messages address exactly these two concerns.
We interpret this as support our hypothesis that data-flow-style error messages could be useful for programmers.

Overall, the quantitative analysis cannot show that \system{} improves over {\tt ocamlc} or Helium. The qualitative analysis suggests that this might be due to the error message layout and notations or the verbosity. However, some respondents found the verbosity and context helpful as well, particularly for large programs and subtle errors.
\system{}'s contribution is the detailed data flow information, not the specific error message layout.
However, \system{} in its current form is certainly not perfect and respondents point out a few shortcomings, such as the exact textual representation of errors.
We believe, in future work can well utilize it to design better error message layouts, code exploration tools, and IDE intellisense features.

% Besides the usual descriptive statistics, for each of questions ($Q1-Q5$), we will perform a multi-variate analysis to test for differences between the three groups (\system{}, OCaml, and Helium) and knowledge level of participants (novices vs. experts).

\subsection{Threats to Validity}
\paragraph{Internal validity}
For most of the tasks, the results were insignificant. Potentially, programs in the used corpus were not complex enough to measure differences between the systems.
The study was conducted remotely with no control over the context. Participants might have been distracted while answering, spent different amount of time ($\textrm{mean}=32\textrm{min}, \textrm{sd}=22\textrm{min}$), and have used different devices.
We used the browser's \emph{user agent} to identify mobile devices.
We were initially concerned that mobile users may have a harder time taking the survey and may
subsequently provide a worse quality of responses, but all open-text answers provided by the 14 participants on mobile devices were of a high quality, so we decided to keep them.
Participants might not know enough OCaml to understand the errors, which we tried to address by presenting each participant with a one-page introduction into the relevant OCaml concepts.
Participants might recognize the different styles of error messages from different systems.
We tried to limit the influence of this bias by fixing one participant to one specific condition. However, experienced OCaml programmers reportedly recognized the errors generated by Helium and our system to be non-standard.
Participants might be biased in favour of our system, since it is socially desirable: firstly, it is always interesting to see a new tool that could turn out an improvement and secondly, because participants might be fellow researchers and practitioners that want to support research in the field.
Participants were not trained on how to read \system{} error messages. Unknown notation and unconventional layout of error messages might have confused participants.

\paragraph{External validity}
Our example programs were sourced from university students solving programming assignments \cite{Seidel2017}.
These may not be indicative of the broader range of programming practices prevalent at large.
The results of the study might not carry over to languages other than OCaml. Our approach can be used with all languages that implement Hindley-Milner-style type inference. However, we chose OCaml because of extensive empirical analysis already done on OCaml error localization by other researchers \cite{Geng2022, Zhang2014}
and because of the relative simplicity of the language's base constructs.
For instance, compared to Haskell, the lack of type classes in OCaml makes it easier to quickly introduce the language to beginner and novice programmer before starting the survey.
Conveniently, previous research made available large datasets of ill-typed OCaml programs ranging from low to high complexity.

% \section{Ongoing and Future work}
% \label{sec:futureWork}
% \input{futurework.tex}

\section{Related Work}
\label{sec:relatedWork}
\paragraph*{Algebraic subtyping}
Type inference for systems with subtyping and parametric polymorphism is a known hard problem.
We build upon the algebraic subtyping approach developed in \cite{dolanThesis} and \cite{dolanMycroftSubtyping2017}. More precisely, we build upon the more recent publications \cite{simpleEssence2020} and \cite{parreaux2022}.

\paragraph*{Explaining type errors with data flow}
The approach to explaining type errors using the data flow of the program is very reminiscent of the similar approach by \citet{Gast2005}. It describes an algebra based on subtyping constraints and defines a "consistency" relation between types. This is similar to how we unify the bounds of type variables to find flow errors where incompatible types flow into or flow from the same type variable. \citet{Neubauer2003} use sum types to encode all the types an expression can be and flows sets to track the locations each type can flow through. Our work builds on this by formally categorizing different kinds of data flows and describing a systematic approach to display error reports for them. We also provide an implementation of our algorithm that integrates with existing type systems and supports let polymorphism.
% \ishan{this section on gast is not necessary here since these points are covered in the contributions section 1.3}
% Note that this idea is not completely novel,
% as \citet{Gast2005} proposed a very similar idea already in 2005.
% What differentiates our work from previous work on the topic is:
% \emph{(1)}~we formally categorize the different \emph{kinds} of ML type error messages,
% which helps untangle the complexity of these errors;
% \emph{(2)}~we provide a simple type inference extension to an existing type inference algorithm
% to reconstruct the relevant information related to these errors;
% and
% \emph{(3)}~we describe a systematic approach to display error reports
% based on our new understanding of these error kinds.
% \ishan{Not clear how to say our approach is better or addresses some of its shortcomings. Most notably they don't provide a unification algorithm}
% \lionel{weirdly, \citet{Gast2005} mentions this work but doesn't really discuss its shortcomings and how he improves on it}

\paragraph*{Algorithmic error localization}
Previous work on type errors focus on finding the program expression most likely causing the error. \citet{Zhang2014} demonstrate a constraint based system to identify expressions that create unsatisfiable constraints. Using heuristics they pick the simplest explanation for the error. \citet{Loncaric2016} also demonstrate a constraint based system that can integrate with existing type systems to produce error reports efficiently. Our system directly integrates provenance tracking with constraint solving allowing us to track detailed information. Heuristics based error localization can complement detailed data flow errors.
% Previous work on type errors focus on finding the program expression most likely causing the error. \cite{Zhang2014} demonstrate a constraint based system to identify expressions that create unsatisfiable constraints. Using heuristics they pick the simplest explanation for the error. \cite{Loncaric2016} also demonstrate a constraint based system that can integrate with existing type systems to produce error reports efficiently. \ishan{We extend upon this line of reasoning and do constraint solving and error reporting directly in the type system. This also allows us to track detailed provenance information and show error messages that explain the how type flow is causing errors. This will be very helpful in understanding errors that occur far from where the mismatching types originate.}

\paragraph*{Data driven error localization}
More recent work by \citet{Geng2022,Seidel2017} leverages language models and supervised learning techniques to localize errors. They use large datasets with pairs of ill-typed and fixed programs to train models, which can then predict the likely location for the fix with high accuracy. However these techniques are limited to identifying a program expression and cannot create error messages which explain the flow that causes the error.

\paragraph*{Improving compiler error messages}
There's been considerable research on type errors messages \cite{heeren05:top} and their role in programmer experience.
\citet{Becker2019} mention that type error messages play an important role in helping the programmer fix the error. \citet{Marceau2011} argue that reporting all type errors and mapping error messages back to source code are crucial for effective error messages. Furthermore, \citet{Marceau2011b} recommends not to highlight specific fixes as they may be incorrect. Techniques from \citet{Wrenn2017} can be used to evaluate and improve data flow style error messages. Finally, \citet{Kochhar2016} surveyed software engineering practitioners to find that respondents prefer general solutions that can integrate with existing tooling and IDEs, furthermore they should scale to large codebases. Our system addresses the key challenge of mapping an error back to source code locations and existing tools and IDEs can be instrumented with the detailed data provenance information for interactive debugging.
% \cite{Becker2019} mentions that type error messages play an important role in helping the programmer fix the error. \cite{Marceau2011} argue that reporting all type errors and mapping error messages back to source code are crucial for effective error messages. Finally \cite{Kochhar2016} surveyed software engineering practitioners to find that respondents prefer general solutions that can integrate with existing tooling and IDEs, furthermore they should scale to large codebases. \ishan{Our system addresses many of the concerns posed by practitioners. It can extend any ML based type system with better error messages. Moreover it collects of flow information that can be used to support rich interactive error debugging experience in IDEs.}

% ~\\
% \TODO{relate with some of the extensive work on control-flow analysis. For example:
% \citet{Rehof-popl01:ty-based-flow-anal-from-poly-subt-to-cfl-reach} make the following very relevant observation:}
% \begin{quote}
%   \textit{Interestingly, the flow relation of [FRD00b] without subtyping emerges by collapsing all M-productions to e, corresponding to an equational interpretation of the subtype relation. The flow relation thereby becomes regular, and flow queries become linear time in the size of the flow graph. Our present results show that the addition of subtyping corresponds to a passage from regular to context-free flow.}
% \end{quote}

\section{Conclusion and Future Work}
\label{sec:conclusion}
We now conclude and suggest directions for future work.

\subsection{Conclusion}

If we want powerful type inference techniques to become broadly accepted in mainstream programming languages, we have to generate excellent error messages when type inference goes wrong.
In this article, we laid some foundations to improve one important class of error messages: type error messages arising from constraint solving for both subtyping and equality constraints.
Our main insight is that these constraints contain information about the data flow that led to the error, and that we can use this information to generate more informative error messages.
We carried out a user study and compared our error messages to those of {\tt ocamlc} and Helium.
The study suggests that the additional information can potentially be overwhelming, so we have to carefully consider under what circumstances we use it and and how much of it we present to the user.
While the empirical part of the study could not quantitatively show that \system{} improved over the state of the art, we also received encouraging feedback by participants which suggest that the general approach of flow-based error messages is worthwhile.

\subsection{Future Work}

We see two important directions of future work related to the
extension of our method to model additional features of common type systems.

% we discuss this in \cref{subsec:futurework:rankn-polymorphism,subsec:futurework:gadts,subsec:futurework:constrained-types,subsec:futurework:letpolymorphism,subsec:futurework:modalities}.
% \TODO{Which of the possible directions do we want to include here?}
% The second direction is the empirical evaluation of the error messages which we presented, and which is already ongoing.
% This is described in \cref{subsec:empirical-evaluation:user-study}.

% \paragraph*{Let polymorphism}
% \label{subsec:futurework:letpolymorphism}

% The formalism that we presented does not include let bindings which are polymorphically generalized, even though this is a standard feature of both the Hindley-Damas-Milner algorithms and algebraic subtyping.
% Our implementation supports both \emph{top-level} and \emph{local} let-polymorphism. We are yet to investigate its formalization, although we do not expect any particular difficulty.
% However, authors such as \citet{vytiniotis2010let} argue that local let bindings usually need not be generalized.

%
% We cannot yet anticipate whether let polymorphism integrates well with our approach, or whether we will find another reason why let should not be generalized~\cite{vytiniotis2010let}.

\paragraph*{Occurs check}
\label{subsec:futurework:occurs-check}
We have not implemented the \emph{occurs check} in our prototype yet.
The occurs check is a standard feature of Hindley-Milner type inference that catches cycles in constraint graphs.
One of the most significant results in our study is that occurs-check failures
are much harder to understand for users than unification failures,
and so there is much space for improvement.
We hope to significantly improve these error messages using our approach based on data flows.
Running the occurs-check separately also has algorithmic complexity advantages \cite{Remy-inria92:extension-ml-sorted-eqn-thry-on-tys}.

% Consider the example of self-application.
% \begin{lstlisting}
%     \x. x x
% \end{lstlisting}
% \TODO{Finish this example}
% \begin{Error}{\system{}}
% \begin{lstlisting}[language=error]
% [ERROR] Occurs check failed.
%            |--> ?a --|
%            |---------|
% .....
% \end{lstlisting}
% \end{Error}

\paragraph*{More advanced type system features}

We would like to investigate how flow-based reasoning scales to more advanced type system features,
where type error messages can often become even more confusing than traditional unification error.
For example, we are particularly interested in studying support for higher-rank and first-class polymorphism
\cite{jones2007practical},
especially since these approaches could benefit from subtyping \cite{Lebotlan03:MLF-ICFP}.
Other important and tricky type system features include generalized algebraic data types,
constrained types, modalities, and linear types.

% \subsection{Higher Rank Polymorphism}
% \label{subsec:futurework:rankn-polymorphism}

% \cite{jones2007practical}

% \subsection{Generalized Algebraic Data Types}
% \label{subsec:futurework:gadts}

% \subsection{Constrained Types}
% \label{subsec:futurework:constrained-types}

% \subsection{Modalities}
% \label{subsec:futurework:modalities}

\begin{acks}
  We want to thank the anonymous reviewers for their comments and for their help in improving the paper.
  We would also like to thank Volker Franz for feedback on the study design and Marlen Brachthäuser for help with the study design and the empirical evaluation.
  Ji\v{r}\'{i} Bene\v{s} contributed the idea to replicate the flow overview in a gutter on the left of our error messages.
\end{acks}

\section*{Data-Availability Statement}
The implementation of the system \system{} described in this paper is permanently available on Zenodo \cite{oopsla2023artefact}.
The latest implementation, which may contain slight changes and features not described in this paper, can be found at \href{https://github.com/hkust-taco/hmloc}{github.com/hkust-taco/hmloc}.
An offline version of the user survey, the raw collected data, as well as the processing scripts are available on request.

\clearpage

%% Bibliography
\bibliography{bib}

\appendix

\section{Further Examples}
\label{sec:appendix:further-examples}

\subsection{Verbose Version of Example From Introduction}
\label{sec:appendix-verbose-example}

Following is the additional result of running our type checker in \emph{verbose} mode:

\begin{Error}{\system{} (verbose mode)}
\begin{lstlisting}[language=error]
[ERROR] Type `float` does not match `string`

        (float) ~~~> (string * float) ---> (_ * ?a) ~~~> (?a) ---> (string)

◉ (float) comes from
   - l.4  let appInfo = ("My Application", 1.5)
                                           ^^^
  ◉ (string * float) comes from
  │  - l.4  let appInfo = ("My Application", 1.5)
  │                       ^^^^^^^^^^^^^^^^^^^^^^^
  │  - l.9  let test = process appInfo
  │                            ^^^^^^^
  ▼ 
  ◉ (_ * ?a) comes from
     - l.6  let process (name, vers) =
                        ^^^^^^^^^^^^
◉ (?a) is assumed for
│  - l.6  let process (name, vers) =
│                            ^^^^
│  - l.7    name ^ show_major (parse_version vers)
│                                            ^^^^
▼ 
◉ (string) comes from
   - l.1  val parse_version: string -> string
                             ^^^^^^
\end{lstlisting}
\end{Error}
Given these error reports,
we now understand exactly what the cause of the error is,
even without having seen the rest of the source code:
% the problem is that
version information passed into the @process@
function are expected to be strings,
since they are passed to the @parse_version@ function,
but the user provided a @float@ number instead while defining @appInfo@.
%

% \section{Study: Test Results}
% \label{sec:appendix:results}
% \cref{table:results} lists the full results for all performed Kruskal-Wallis  rank sum tests. All participants correctly described the error for program \texttt{easy1}.
% \begin{table}
% \input{results-table.tex}
% \caption{\label{table:results}Results of all Kruskal-Wallis tests per program.}
% \end{table}

\section{Study: Example Programs}
\label{sec:appendix:examples}
In our user study (cf. \cref{sec:user-study}) we presented participants with error messages.
Each participant was assigned to one of three different systems, and the programs were selected from ten different examples.
This section contains all the examples, together with the error messages generated by each of the three systems, and the empirical results for each example.

%%
%% Easy 1
%%
\subsection{Easy 1}
\label{subsec:appendix:easy1}

\begin{center}
  \includegraphics[width=6cm]{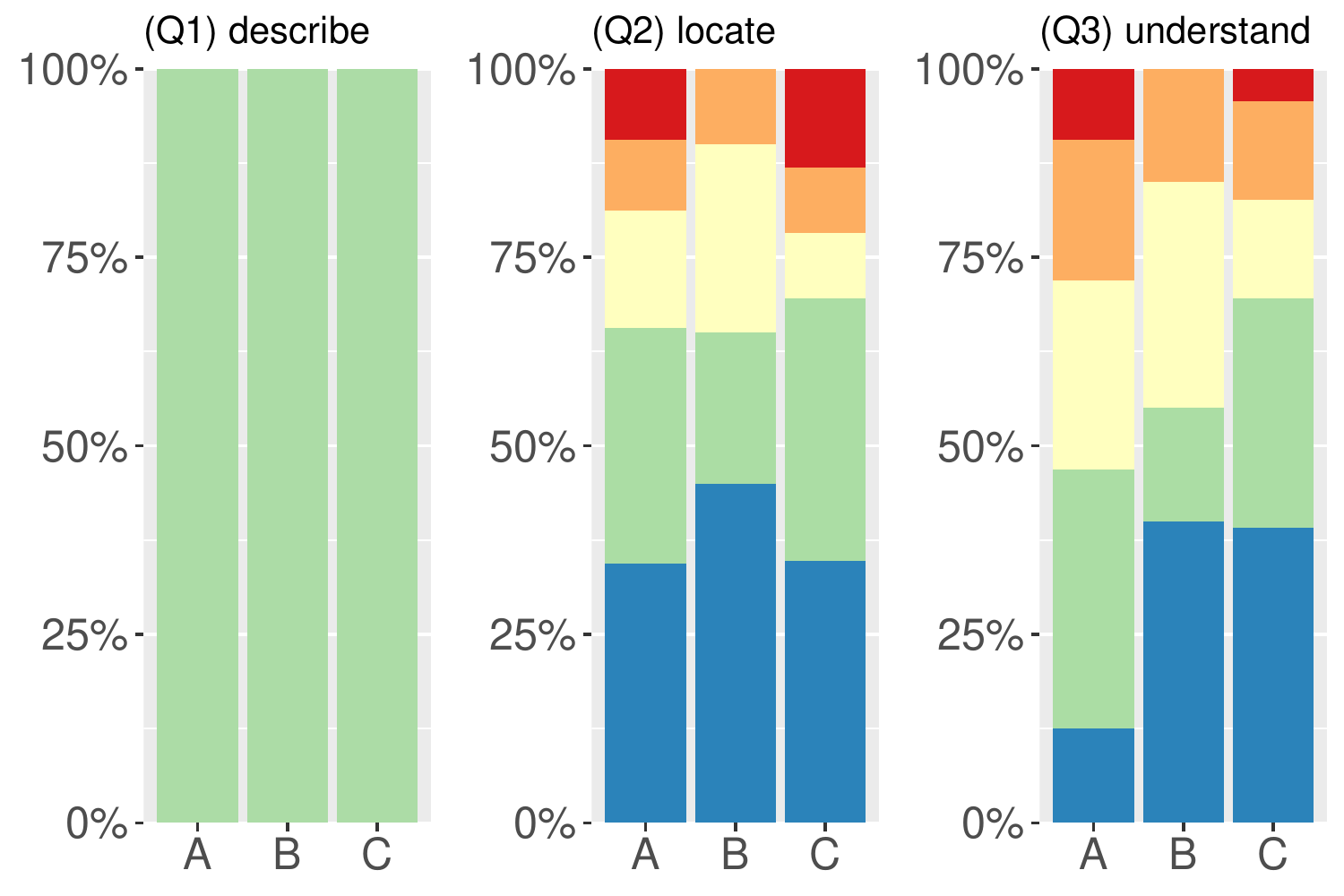}
\end{center}
\begin{lstlisting}[language=numsnippets]
let boolToBit x = if x then "0" else 1
\end{lstlisting}

\begin{Error}{Ocaml}
\vspace{0.2cm}
\begin{lstlisting}[language=error]
Line 1, characters 37-38:
1 | let boolToBit x = if x then "0" else 1
                                         ^
Error: This expression has type int but an expression was expected of type
         string
\end{lstlisting}
\end{Error}

\begin{Error}{Helium}
\vspace{0.2cm}
\begin{lstlisting}[language=error]
(1,38): Type error in else branch of conditional
 expression       : if x then "0" else 1
 term             : 1
   type           : int
   does not match : string
\end{lstlisting}
\end{Error}

\begin{Error}{\system{}}
\vspace{0.2cm}
\begin{lstlisting}[language=error]
[ERROR] Type `string` does not match `int`

        (string) ---> (?a) <--- (int)

◉ (string) comes from
│  - l.1:  let boolToBit x = if x then "0" else 1
▼                                      ^^^
◉ (?a) is assumed for
▲  - l.1:  let boolToBit x = if x then "0" else 1
│                            ^^^^^^^^^^^^^^^^^^^^
◉ (int) comes from
   - l.1:  let boolToBit x = if x then "0" else 1
                                                ^
\end{lstlisting}
\end{Error}

%%
%% Easy 2
%%
\subsection{Easy 2}
\label{subsec:appendix:easy2}
\begin{center}
  \includegraphics[width=6cm]{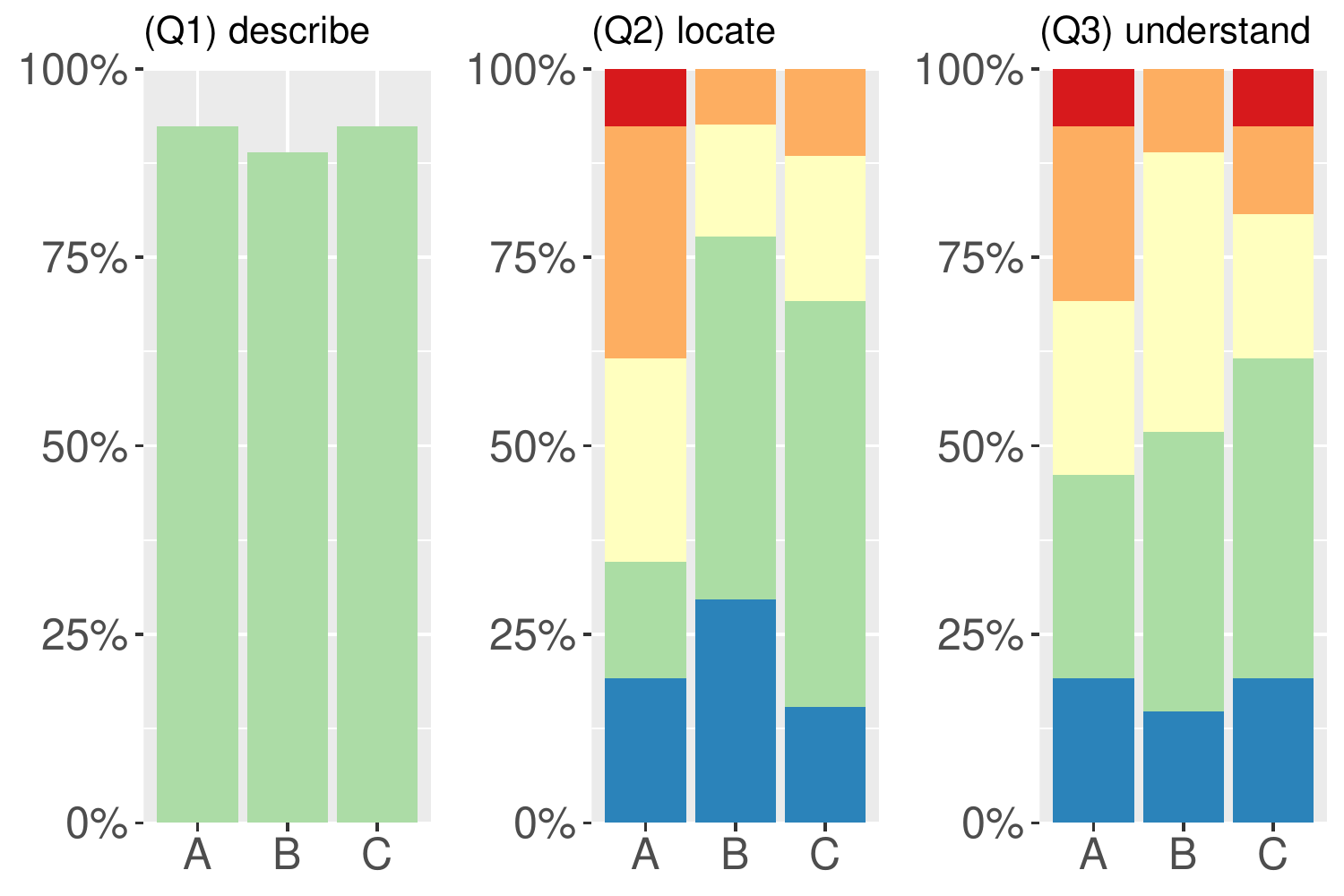}
\end{center}
\begin{lstlisting}[language=numsnippets]
let increment result = match result with
 | Left n -> n + 1
 | Right msg -> "ERROR: " ^ msg

(* (^): string -> string -> string
     is a string concatenation operator *)
(* Left and Right are constructors of the either type *)
\end{lstlisting}

\begin{Error}{Ocaml}
\vspace{0.2cm}
\begin{lstlisting}[language=error]
Line 3, characters 16-31:
3 |  | Right msg -> "ERROR: " ^ msg
                    ^^^^^^^^^^^^^^^
Error: This expression has type string but an expression was expected of type
         int
\end{lstlisting}
\end{Error}

\begin{Error}{Helium}
\vspace{0.2cm}
\begin{lstlisting}[language=error]
(3,17): Type error in right-hand side
 expression       : "ERROR: " ^ msg
   type           : string
   does not match : int
\end{lstlisting}
\end{Error}

\begin{Error}{\system{}}
\vspace{0.2cm}
\begin{lstlisting}[language=error]
[ERROR] Type `string` does not match `int`

        (string) ---> (?a) <--- (int)

◉ (string) comes from
│  - builtin:  let (^): string -> string -> string
│                                           ^^^^^^
│
│  - l.3:   | Right msg -> "ERROR: " ^ msg
▼                          ^^^^^^^^^^^^^^^
◉ (?a) is assumed for
▲  - l.1:  let increment result = match result with
│                                 ^^^^^^^^^^^^^^^^^
│           | Left n -> n + 1 ...
│           ^^^^^^^^^^^^^^^^^^^^^
│
│  - l.2:   | Left n -> n + 1
│                       ^^^^^
◉ (int) comes from
   - builtin:  let (+): int -> int -> int
                                      ^^^
\end{lstlisting}
\end{Error}

%%
%% Easy 3
%%
\subsection{Easy 3}
\label{subsec:appendix:easy3}

\begin{center}
  \includegraphics[width=6cm]{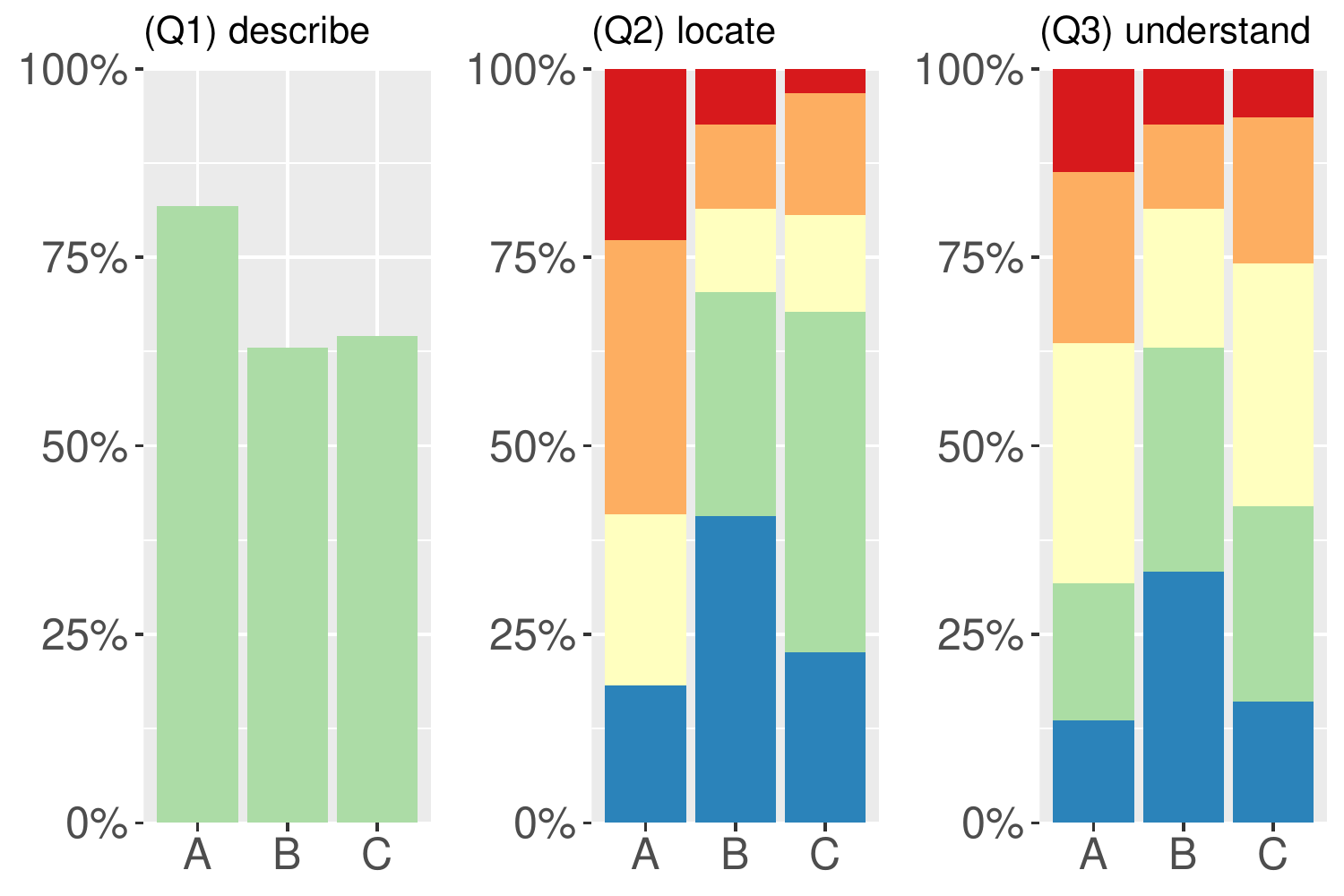}
\end{center}
\begin{lstlisting}[language=numsnippets]
let check result condition =
 if condition then
   match result with
    | Left n -> "OK"
    | Right msg -> "ERROR: " ^ msg
 else result

(* (^): string -> string -> string
     is string concatenation operator *)
(* Left and Right are constructors of the either type *)
\end{lstlisting}

\begin{Error}{Ocaml}
\vspace{0.2cm}
\begin{lstlisting}[language=error]
Line 6, characters 6-12:
6 |  else result
          ^^^^^^
Error: This expression has type ('a, string) either
       but an expression was expected of type string
\end{lstlisting}
\end{Error}

\begin{Error}{Helium}
\vspace{0.2cm}
\begin{lstlisting}[language=error]
(6,7): Type error in else branch of conditional
 expression       : if condition then ... else result
 term             : result
   type           : ('a, string) either
   does not match : string
\end{lstlisting}
\end{Error}

\begin{Error}{\system{}}
\vspace{0.2cm}
\begin{lstlisting}[language=error]
[ERROR] Type `(_, _) either` does not match `string`

        ((_, _) either) <--- (?a) ---> (?b) <--- (string)

◉ ((_, _) either) comes from
▲  - l.4:      | Left n -> "OK"
│                ^^^^
│
│  - l.3:     match result with
│                   ^^^^^^
◉ (?a) is assumed for
│  - l.1:  let check result condition =
│                    ^^^^^^
│
│  - l.6:   else result
▼                ^^^^^^
◉ (?b) is assumed for
▲  - l.2:   if condition then
│           ^^^^^^^^^^^^^^^^^
│             match result with ...
│             ^^^^^^^^^^^^^^^^^^^^^
│
│  - l.3:     match result with
│             ^^^^^^^^^^^^^^^^^
│              | Left n -> "OK" ...
│              ^^^^^^^^^^^^^^^^^^^^
│
│  - l.5:      | Right msg -> "ERROR: " ^ msg
│                             ^^^^^^^^^^^^^^^
◉ (string) comes from
   - builtin:  let (^): string -> string -> string
                                            ^^^^^^
\end{lstlisting}
\end{Error}

%%
%% Medium 1
%%
\subsection{Medium 1}
\label{subsec:appendix:medium1}

\begin{center}
  \includegraphics[width=6cm]{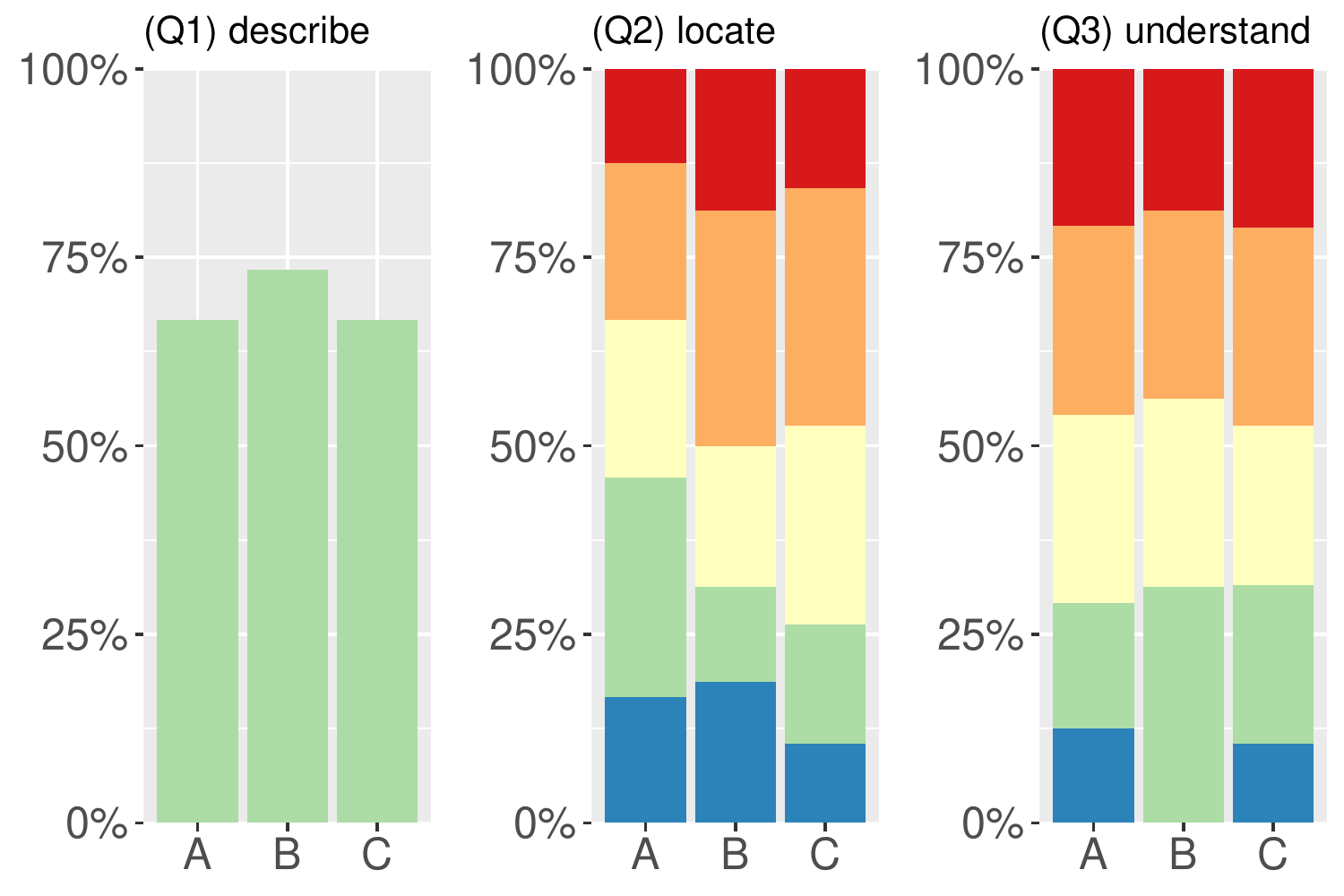}
\end{center}
\begin{lstlisting}[language=numsnippets]
let wrap x = x :: []
 
let test z cond = if cond
 then wrap z
 else wrap true
 
let rec check cond =
 test (if cond then false else check (not cond)) cond
\end{lstlisting}

\begin{Error}{Ocaml}
\vspace{0.2cm}
\begin{lstlisting}[language=error]
Line 8, characters 1-53:
8 |  test (if cond then false else check (not cond)) cond
     ^^^^^^^^^^^^^^^^^^^^^^^^^^^^^^^^^^^^^^^^^^^^^^^^^^^^
Error: This expression has type bool list
       but an expression was expected of type bool
\end{lstlisting}
\end{Error}

\begin{Error}{Helium}
\vspace{0.2cm}
\begin{lstlisting}[language=error]
(8,2): Type error in right-hand side
 expression       : test (if cond then false else check (not cond)) cond
   type           : bool list
   does not match : bool
\end{lstlisting}
\end{Error}

\begin{Error}{\system{}}
\vspace{0.2cm}
\begin{lstlisting}[language=error]
[ERROR] Type `bool` does not match `_ list`

        (bool) ---> (?a) <--- (_ list)

◉ (bool) comes from
│  - l.8:   test (if cond then false else check (not cond)) cond
▼                              ^^^^^
◉ (?a) is assumed for
▲  - l.8:   test (if cond then false else check (not cond)) cond
│                ^^^^^^^^^^^^^^^^^^^^^^^^^^^^^^^^^^^^^^^^^^
│
│  - l.8:   test (if cond then false else check (not cond)) cond
│                                         ^^^^^^^^^^^^^^^^
│
│  - l.8:   test (if cond then false else check (not cond)) cond
│           ^^^^^^^^^^^^^^^^^^^^^^^^^^^^^^^^^^^^^^^^^^^^^^^^^^^^^
│
│  - l.3:  let test z cond = if cond
│                            ^^^^^^^
│           then wrap z ...
│           ^^^^^^^^^^^^^^^
│
│  - l.5:   else wrap true
│                ^^^^^^^^^
◉ (_ list) comes from
   - l.1:  let wrap x = x :: []
                        ^^^^^^^
\end{lstlisting}
\end{Error}

%%
%% Medium 2
%%
\subsection{Medium 2}
\label{subsec:appendix:medium2}

\begin{center}
  \includegraphics[width=6cm]{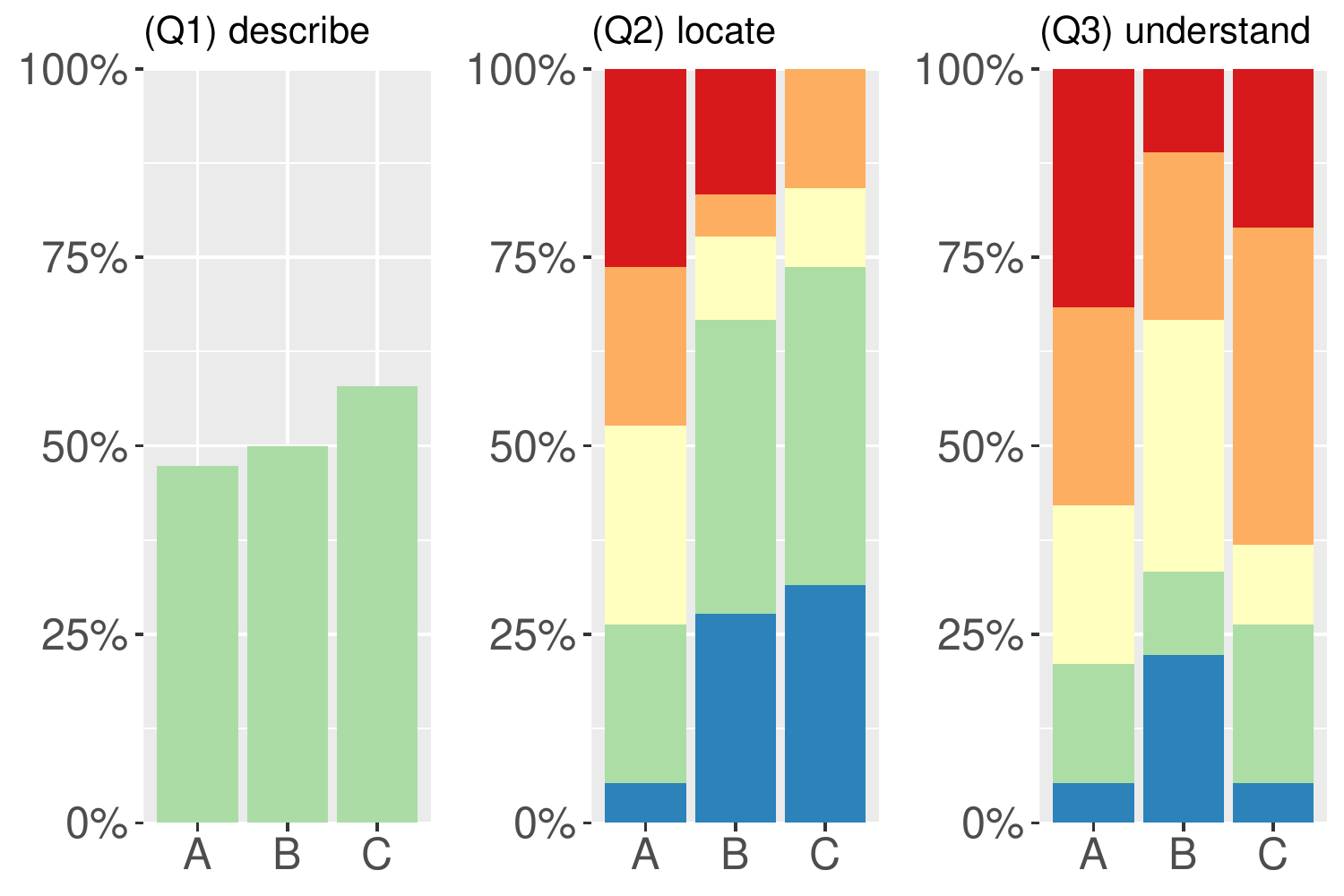}
\end{center}
\begin{lstlisting}[language=numsnippets]
let remainder x y = if (x * y) > 10 then (x * y) mod 10 else 0
  
let strings_of l = List.map string_of_int
  
let y = [1; 2; 3]
  
let rec mulByDigit i l =
  match List.rev l with
  | [] -> []
  | h::t -> [remainder strings_of y] @ (mulByDigit i t)
\end{lstlisting}

\begin{Error}{Ocaml}
\vspace{0.2cm}
\begin{lstlisting}[language=error]
Line 10, characters 23-33:
10 |   | h::t -> [remainder strings_of y] @ (mulByDigit i t)
                            ^^^^^^^^^^
Error: This expression has type 'a -> int list -> string list
       but an expression was expected of type int
\end{lstlisting}
\end{Error}

\begin{Error}{Helium}
\vspace{0.2cm}
\begin{lstlisting}[language=error]
(10,14): Type error in application
 expression       : remainder strings_of y
 term             : remainder
   type           : int -> int -> int
   does not match : ('a -> 'b list -> string list) -> int list -> 'c 
\end{lstlisting}
\end{Error}

\begin{Error}{\system{}}
\vspace{0.2cm}
\begin{lstlisting}[language=error]
[ERROR] Type `int` does not match `_ -> _ list -> _ list`

◉ `int` comes from
▲  - builtin:  let ( * ): int -> int -> int
│                         ^^^
│ 
│  - l.1:  let remainder x y = if (x * y) > 10 then (x * y) mod 10 else 0
│                                                    ^
│ 
│  - l.1:  let remainder x y = if (x * y) > 10 then (x * y) mod 10 else 0
│                        ^
│ 
│  - l.10:    | h::t -> [remainder strings_of y] @ (mulByDigit i t)
│                                  ^^^^^^^^^^
◉ `_ -> _ list -> _ list` comes from
   - l.3:  let strings_of l = List.map string_of_int
                          ^^^^^^^^^^^^^^^^^^^^^^^^^^
\end{lstlisting}
\end{Error}

%%
%% Medium 3
%%
\subsection{Medium 3}
\label{subsec:appendix:medium3}

\begin{center}
  \includegraphics[width=6cm]{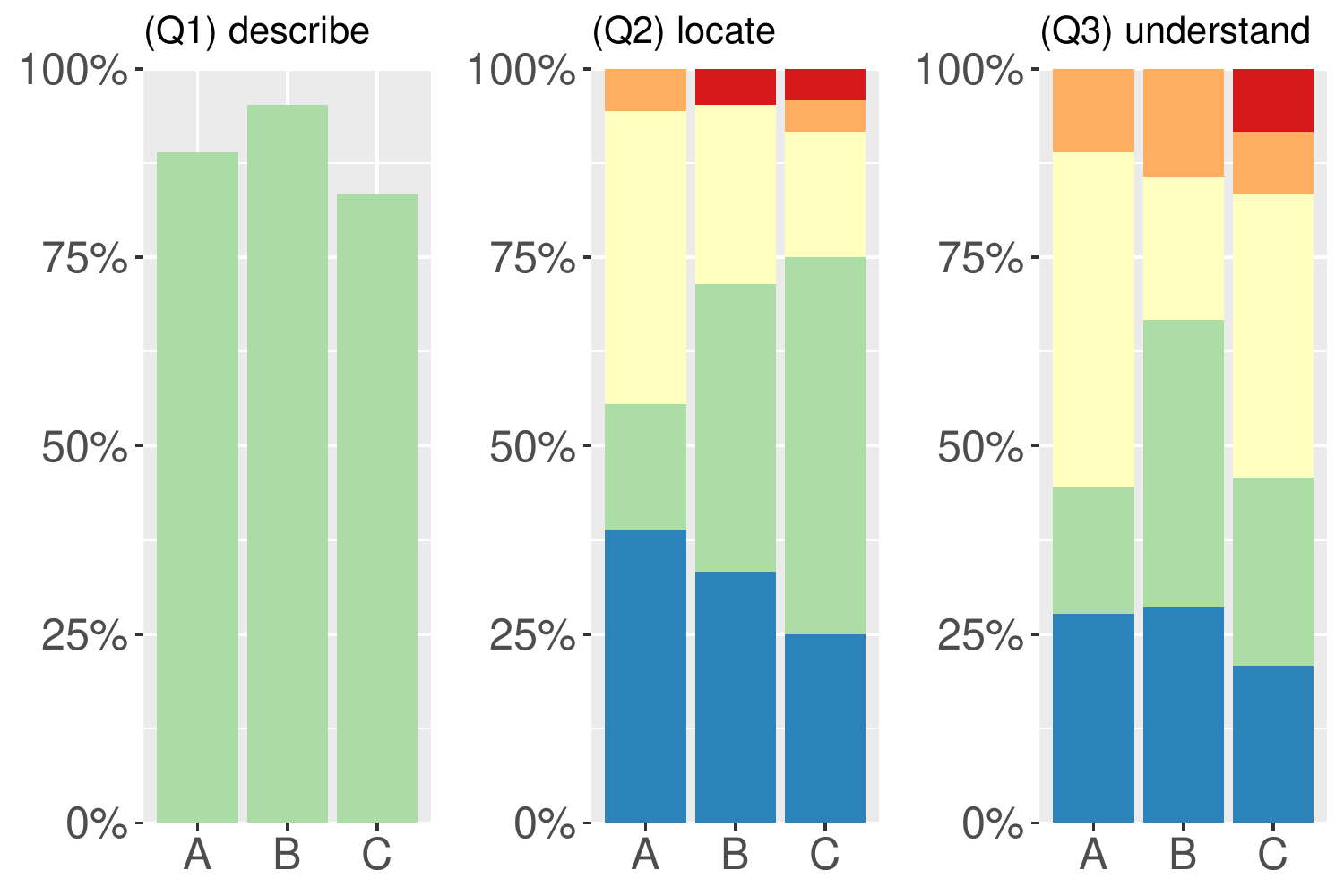}
\end{center}
\begin{lstlisting}[language=numsnippets]
let rec digitsOfInt n =
  if n <= 0 then [] else (digitsOfInt (n / 10)) @ [n mod 10]

let rec addNumbs n = match n with
 | [] -> 0
 | h::t -> h + (addNumbs t)

let digits n = digitsOfInt (abs n)

let rec additivePersistence n =
  match digits n with
  | [] -> 0
  | h::t -> if (addNumbs (h :: t)) >= 10 then false else true

(* (@): 'a list -> 'a list -> 'a list
     is a list concatenation operator *)
\end{lstlisting}

\begin{Error}{Ocaml}
\vspace{0.2cm}
\begin{lstlisting}[language=error]
Line 13, characters 46-51:
13 |   | h::t -> if (addNumbs (h :: t)) >= 10 then false else true
                                                   ^^^^^
Error: This expression has type bool but an expression was expected of type
         int
\end{lstlisting}
\end{Error}

\begin{Error}{Helium}
\vspace{0.2cm}
\begin{lstlisting}[language=error]
(13,13): Type error in right-hand side
 expression       : if (addNumbs (h :: t)) >= 10 then false else true
   type           : bool
   does not match : int
\end{lstlisting}
\end{Error}

\begin{Error}{\system{}}
\vspace{0.2cm}
\begin{lstlisting}[language=error]
[ERROR] Type `bool` does not match `int`

        (bool) ---> (?a) <--- (int)

◉ (bool) comes from
│  - l.13:    | h::t -> if (addNumbs (h :: t)) >= 10 then false else true
│                                                                    ^^^^
│
│  - l.13:    | h::t -> if (addNumbs (h :: t)) >= 10 then false else true
▼                       ^^^^^^^^^^^^^^^^^^^^^^^^^^^^^^^^^^^^^^^^^^^^^^^^^
◉ (?a) is assumed for
▲  - l.11:    match digits n with
│             ^^^^^^^^^^^^^^^^^^^
│             | [] -> 0 ...
│             ^^^^^^^^^^^^^
◉ (int) comes from
   - l.12:    | [] -> 0
                      ^
\end{lstlisting}
\end{Error}

%%
%% Medium 4
%%
\subsection{Medium 4}
\label{subsec:appendix:medium4}

\begin{center}
  \includegraphics[width=6cm]{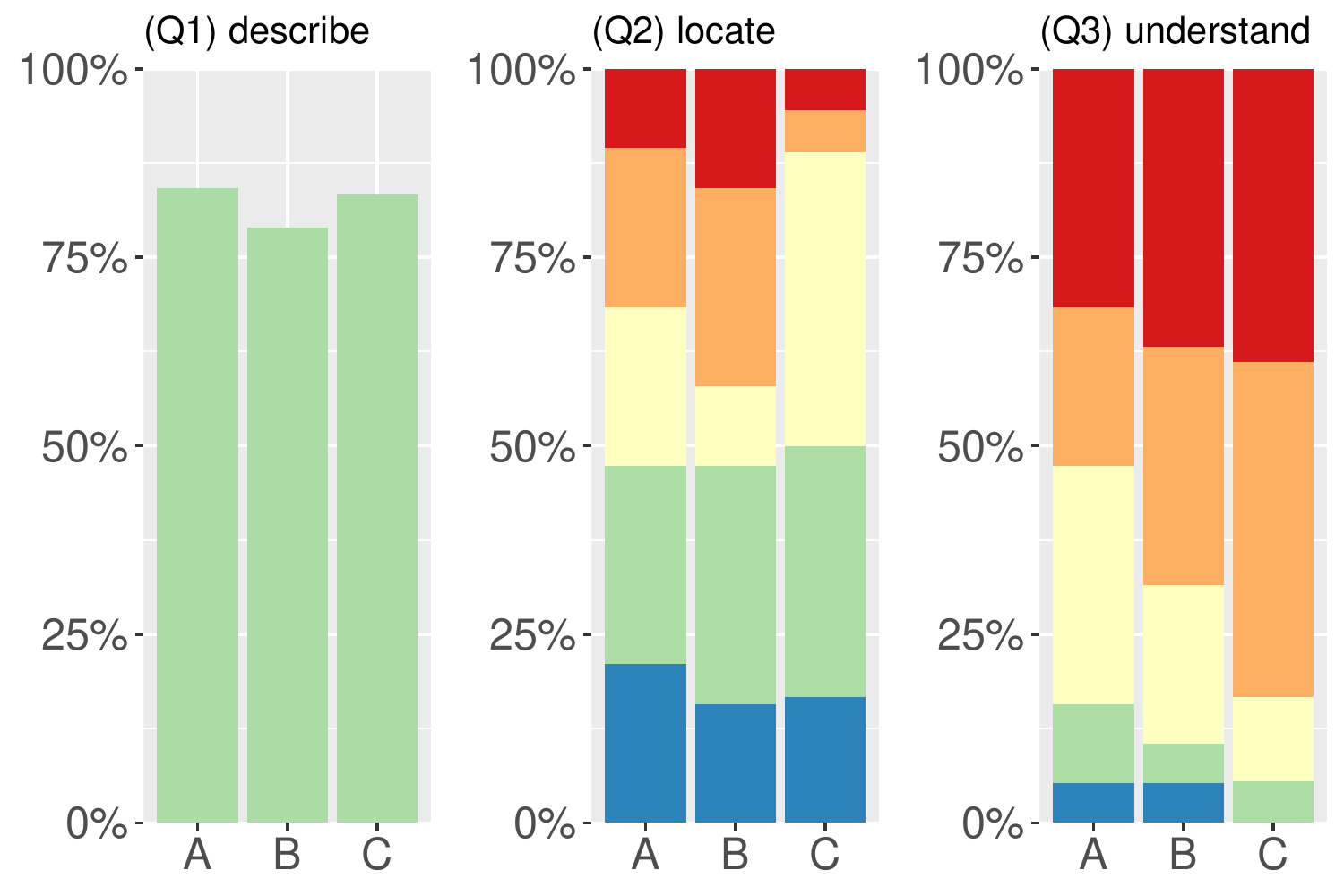}
\end{center}
\begin{lstlisting}[language=numsnippets]
let rec assoc (d,k,l) =
  match l with
  | [] -> d
  | h::t -> let (f,s) = h in if k = f then s else assoc d k t
\end{lstlisting}

\begin{Error}{Ocaml}
\vspace{0.2cm}
\begin{lstlisting}[language=error]
Line 4, characters 56-57:
4 |   | h::t -> let (f,s) = h in if k = f then s else assoc d k t
                                                            ^
Error: This expression has type 'a -> 'b -> 'c
       but an expression was expected of type
         ('a -> 'b -> 'c) * 'd * ('d * ('a -> 'b -> 'c)) list
\end{lstlisting}
\end{Error}

\begin{Error}{Helium}
\vspace{0.2cm}
\begin{lstlisting}[language=error]
(4,50): Type error in application
 expression       : assoc d k t
 term             : assoc
   type           : ('a * 'b * ('b * 'a) list) -> 'a                 
   does not match : 'a                         -> 'b -> ('b * 'a) list -> 'a
 because          : unification would give infinite type
\end{lstlisting}
\end{Error}

\begin{Error}{\system{}}
\vspace{0.2cm}
\begin{lstlisting}[language=error]
[ERROR] Type `_ * _ * _` does not match `_ -> _`

        (?a * _ * _) <--- (?a) ---> (_ -> _) 

◉ (?a * _ * _) comes from
▲  - l.1:  let rec assoc (d,k,l) =
│                        ^^^^^^^
│ 
│  - l.4:    | h::t -> let (f,s) = h in if k = f then s else assoc d k t
│                                                                  ^
◉ (?a) is assumed for
│  - l.1:  let rec assoc (d,k,l) =
│                         ^
│ 
│  - l.3:    | [] -> d
│                    ^
│ 
│  - l.2:    match l with
│            ^^^^^^^^^^^^
│            | [] -> d ...
▼            ^^^^^^^^^^^^^
◉ (_ -> _) comes from
   - l.4:    | h::t -> let (f,s) = h in if k = f then s else assoc d k t
                                                             ^^^^^^^
\end{lstlisting}
\end{Error}

%%
%% Hard 1
%%
\subsection{Hard 1}
\label{subsec:appendix:hard1}

\begin{center}
  \includegraphics[width=6cm]{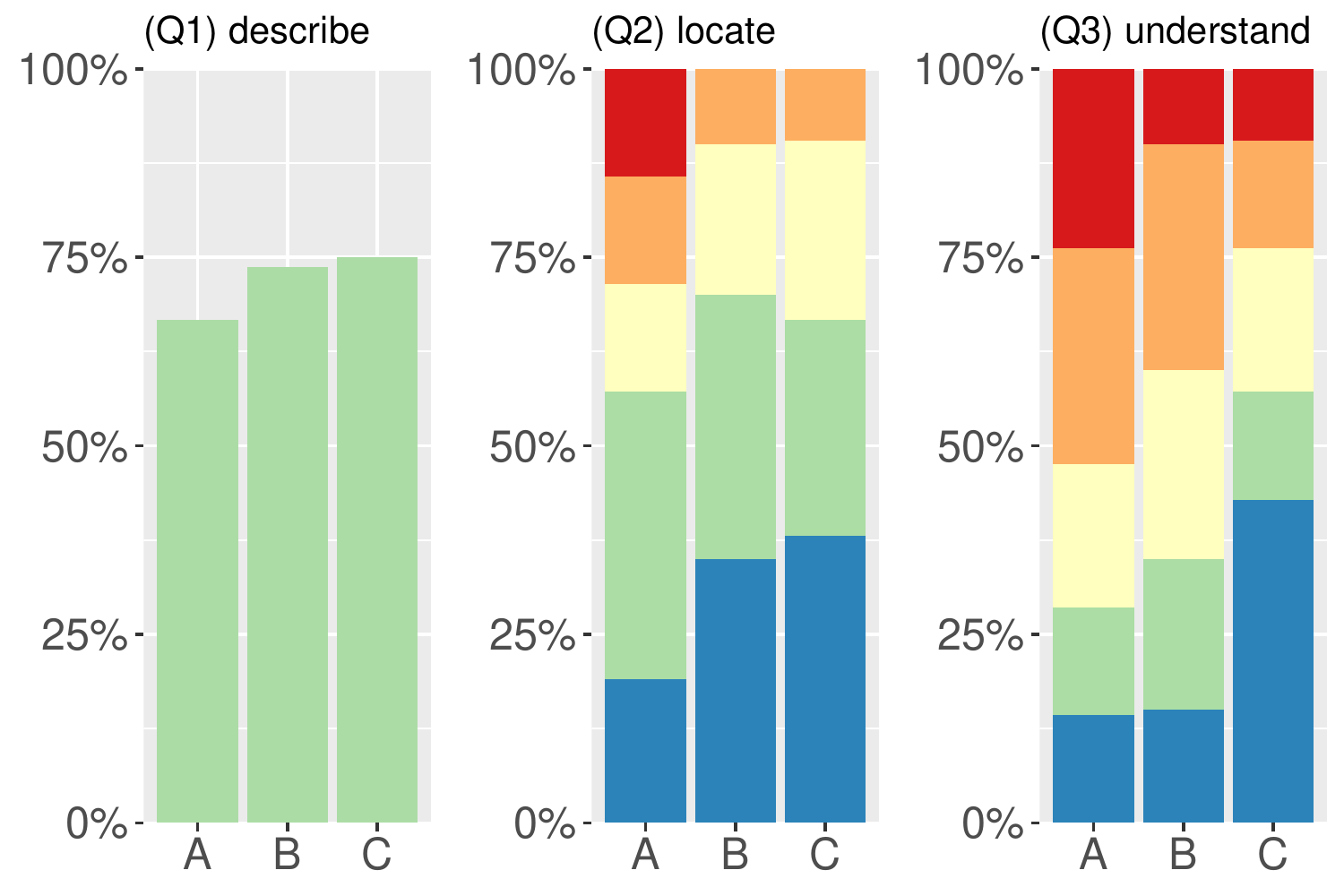}
\end{center}
\begin{lstlisting}[language=numsnippets]
let rec sepConcat sep sl =
  match sl with
  | [] -> ""
  | h::t ->
      let f a x = a ^ (sep ^ x) in
      let base = h in let l = t in List.fold_left f base l

let stringOfList f l = "[" ^ ((sepConcat "; " List.map (f, l)) ^ "]")

(* (^): string -> string -> string
     is a string concatenation operator *)
(* List.fold_left: ('a -> 'b -> 'a) -> 'a -> 'b list -> 'a *)
\end{lstlisting}

\begin{Error}{Ocaml}
\vspace{0.2cm}
\begin{lstlisting}[language=error]
Line 8, characters 31-40:
8 | let stringOfList f l = "[" ^ ((sepConcat "; " List.map (f, l)) ^ "]")
                                   ^^^^^^^^^
Error: This function has type string -> string list -> string
       It is applied to too many arguments; maybe you forgot a `;'.
\end{lstlisting}
\end{Error}

\begin{Error}{Helium}
\vspace{0.2cm}
\begin{lstlisting}[language=error]
(8,29): Type error in application
 expression       : sepConcat "; " List.map (f, l)
 term             : sepConcat
   type           : string -> string list -> string
   does not match : string -> (('a -> 'b) -> 'a list -> 'b list) -> ('c * 'd) -> string
 because          : too many arguments are given
\end{lstlisting}
\end{Error}

\begin{Error}{\system{}}
\vspace{0.2cm}
\begin{lstlisting}[language=error]
[ERROR] Type `_ list` does not match `(_ -> _) -> _ list -> _ list`

◉ `_ list` comes from
▲  - l.3:    | [] -> ""
│              ^^
│
│  - l.2:    match sl with
│                  ^^
│
│  - l.1:  let rec sepConcat sep sl =
│                                ^^
│
│  - l.8:  let stringOfList f l = "[" ^ ((sepConcat "; " List.map (f, l)) ^ "]")
│                                                        ^^^^^^^^
◉ `(_ -> _) -> _ list -> _ list` comes from
   - builtin:  let List.map: ('a -> 'b) -> 'a list -> 'b list
                              ^^^^^^^^^^^^^^^^^^^^^^^^^^^^^^^
\end{lstlisting}
\end{Error}

%%
%% Hard 2
%%
% \subsection{Hard 2}
% \label{subsec:appendix:hard2}

% \begin{center}
%   \includegraphics[width=6cm]{study/summary\string_hard2.pdf}
% \end{center}
% \lstinputlisting[language=numsnippets]{study/dataset/hard2.ml}

% \begin{Error}{Ocaml}
% \vspace{0.2cm}
% \lstinputlisting[language=error]{study/dataset/hard2.ocamlc}
% \end{Error}

% \begin{Error}{Helium}
% \vspace{0.2cm}
% \lstinputlisting[language=error]{study/dataset/hard2.helium}
% \end{Error}

% \begin{Error}{\system{}}
% \vspace{0.2cm}
% \lstinputlisting[language=error]{study/dataset/hard2.simplesub}
% \end{Error}

%%
%% Hard 3
%%
\subsection{Hard 3}
\label{subsec:appendix:hard3}

\begin{center}
  \includegraphics[width=6cm]{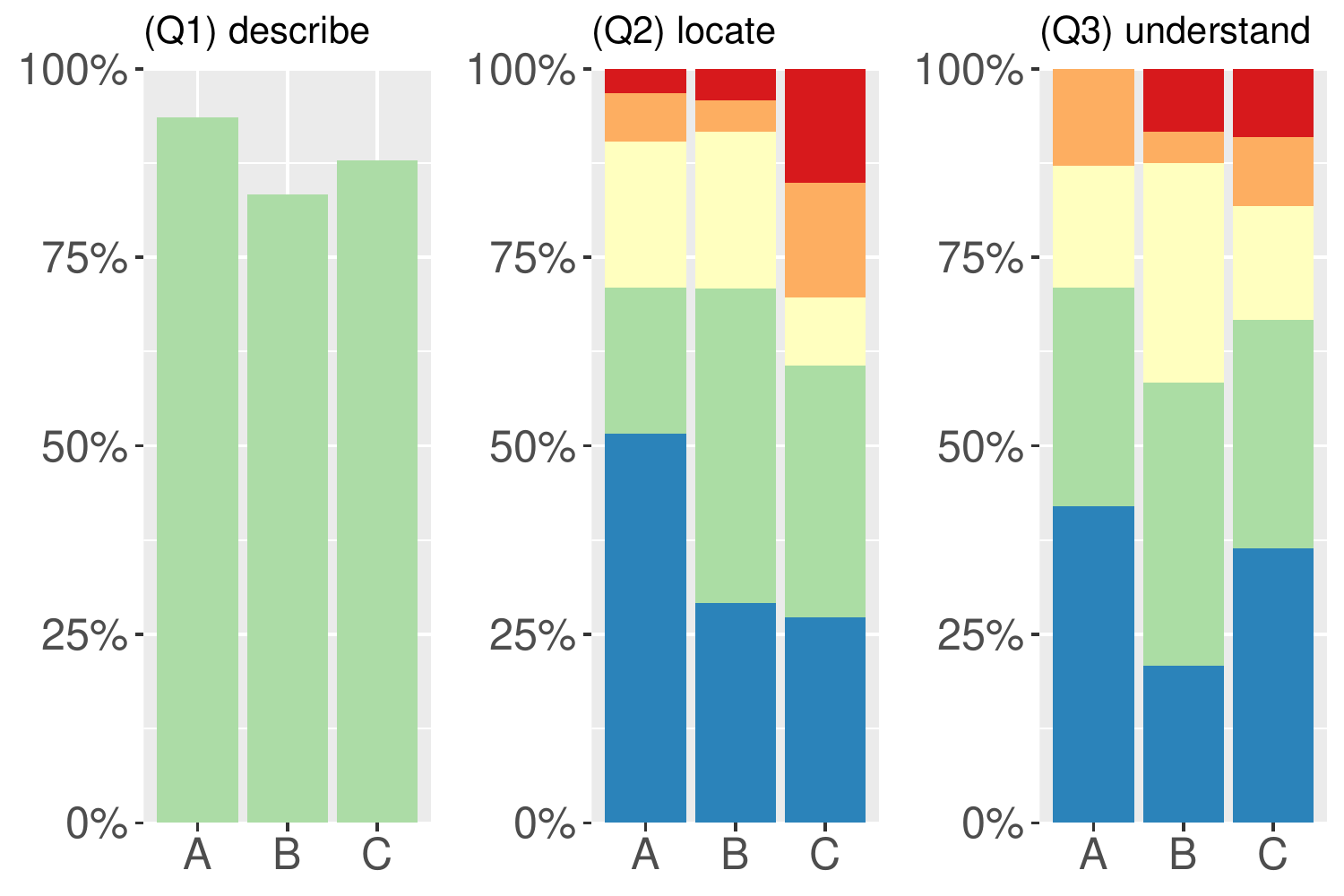}
\end{center}
\begin{lstlisting}[language=numsnippets]
let rec clone x n = if n <= 0 then [] else x :: (clone x (n - 1))

let padZero l1 l2 =
  let numZeros = (List.length l1) - (List.length l2) in
  let absNumZeros = abs numZeros in
  if numZeros = 0
  then (l1, l2)
  else
    (let listZeros = clone 0 absNumZeros in
     if numZeros > 0 then (l1, (listZeros @ l2)) else ((listZeros @ l1), l2))

let rec removeZero l =
  match l with | [] -> [] | 0::t -> removeZero t | h::t -> l

let bigAdd l1 l2 =
  let add (l1,l2) =
    let f a x =
      let (carry,currentSum) = a in
      if x = []
      then (0, (carry :: currentSum))
      else
        (let (toSum1,toSum2) = x in
         let intermediateValue = (toSum1 + toSum2) + carry in
         let valueToAddToArray = intermediateValue mod 10 in
         let carry = intermediateValue / 10 in
         (carry, (valueToAddToArray :: currentSum))) in
    let base = (0, []) in
    let args = List.rev (List.combine l1 l2) in
    let (_,res) = List.fold_left f base args in res in
  removeZero (add (padZero l1 l2))

(* (@): 'a list -> 'a list -> 'a list
     is a list concatenation operator *)
(* List.length: 'a list -> int *)
(* List.fold_left: ('a -> 'b -> 'a) -> 'a -> 'b list -> 'a *)
(* List.combine: 'a list -> 'b list -> ('a * 'b) list
     zips two lists together *)
(* List.rev: 'a list -> 'a list
     reverses a list *)
\end{lstlisting}

\begin{Error}{Ocaml}
\vspace{0.2cm}
\begin{lstlisting}[language=error]
Line 22, characters 31-32:
22 |         (let (toSum1,toSum2) = x in
                                    ^
Error: This expression has type 'a list
       but an expression was expected of type 'b * 'c
\end{lstlisting}
\end{Error}

\begin{Error}{Helium}
\vspace{0.2cm}
\begin{lstlisting}[language=error]
(22,32): Type error in right-hand side
 expression       : x
   type           : 'a list
   does not match : ('b, 'c)
\end{lstlisting}
\end{Error}

\begin{Error}{\system{}}
\vspace{0.2cm}
\begin{lstlisting}[language=error]
[ERROR] Type `_ list` does not match `_ * _`

        (_ list) <--- (?b) ---> (_ * _) 

◉ (_ list) comes from
▲  - l.19:        if x = []
│                        ^^
│ 
│  - l.19:        if x = []
│                    ^
◉ (?b) is assumed for
│  - l.17:      let f a x =
│                       ^
│ 
│  - l.22:          (let (toSum1,toSum2) = x in
▼                                          ^
◉ (_ * _) comes from
   - l.22:          (let (toSum1,toSum2) = x in
                         ^^^^^^^^^^^^^^^
\end{lstlisting}
\end{Error}

\section{More Example Programs}
\label{sec:appendix:verbose-examples}

\subsection{Verbose Version of Example From Introduction}
\label{sec:appendix-verbose-example}

Following is the additional result of running our type checker in \emph{verbose} mode:

\begin{Error}{\system{} (verbose mode)}
\begin{lstlisting}[language=error]
[ERROR] Type `float` does not match `string`

        (float) ~~~> (string * float) ---> (_ * ?a) ~~~> (?a) ---> (string)

◉ (float) comes from
   - l.4  let appInfo = ("My Application", 1.5)
                                           ^^^
  ◉ (string * float) comes from
  │  - l.4  let appInfo = ("My Application", 1.5)
  │                       ^^^^^^^^^^^^^^^^^^^^^^^
  │  - l.9  let test = process appInfo
  │                            ^^^^^^^
  ▼ 
  ◉ (_ * ?a) comes from
     - l.6  let process (name, vers) =
                        ^^^^^^^^^^^^
◉ (?a) is assumed for
│  - l.6  let process (name, vers) =
│                            ^^^^
│  - l.7    name ^ show_major (parse_version vers)
│                                            ^^^^
▼ 
◉ (string) comes from
   - l.1  val parse_version: string -> string
                             ^^^^^^
\end{lstlisting}
\end{Error}
Given these error reports,
we now understand exactly what the cause of the error is,
even without having seen the rest of the source code:
% the problem is that
version information passed into the @process@
function are expected to be strings,
since they are passed to the @parse_version@ function,
but the user provided a @float@ number instead while defining @appInfo@.

\end{document}